\documentclass[a4paper,11pt]{article}
\pdfoutput=1

\usepackage{jheppub}
\bibstyle{JHEP}

\usepackage[english]{babel}
\usepackage[utf8]{inputenc}
\usepackage[T1]{fontenc}
\usepackage{amsfonts}
\usepackage{amsmath}       
\usepackage{amsfonts}          
\usepackage{amssymb}
\usepackage{mathtools}
\usepackage{comment}
\usepackage{tikz}
\usetikzlibrary{decorations.markings}
\usepackage{physics}
\usepackage{dsfont}
\usepackage{tabu}
\usepackage{tikz}
\usepackage{xcolor}
\usetikzlibrary{arrows.meta}

\renewcommand{\d}{\text{d}}

\renewcommand\d{\delta}

\numberwithin{equation}{section}

\usepackage{amsmath}

\DeclareMathOperator\vol{Vol}

\usepackage{xcolor}

\renewcommand*\thesection{\arabic{section}}
\definecolor{mathematica1}{rgb}{0.368417, 0.506779, 0.709798}
\definecolor{mathematica2}{rgb}{0.880722, 0.611041, 0.142051}
\definecolor{mathematica3}{rgb}{0.560181, 0.691569, 0.194885}
\definecolor{mathematica4}{rgb}{0.922526, 0.385626, 0.209179}
\definecolor{mathematica6}{rgb}{0.772079, 0.431554, 0.102387}
\definecolor{pink}{rgb}{1, 0.5, 0.5}

\newcommand{\be}{\begin{equation}}
\newcommand{\ee}{\end{equation}}

\renewcommand{\d}{\mathrm{d}}

\usepackage{tabu}

\newcommand{\overbar}[1]{\mkern 1.5mu\overline{\mkern-1.5mu#1\mkern-1.5mu}\mkern 1.5mu}

\renewcommand{\tr}[1]{\mathrm{Tr}\,{#1}}

\title{3D Einstein action from 6D Kodaira--Spencer gravity}

\author[a]{Johanna Erdmenger,}
\author[a]{Jonathan Karl,}
\author[a]{Jani Kastikainen,}
\author[a]{and Henri Scheppach}

\affiliation[a]{Institute for Theoretical Physics and Astrophysics and Würzburg-Dresden Cluster of Excellence ct.qmat, Julius-Maximilians-Universität Würzburg, Am Hubland, 97074 Würzburg, Germany}

\emailAdd{jonathan.karl@uni-wuerzburg.de}
\emailAdd{jani.kastikainen@uni-wuerzburg.de}
\emailAdd{henri.scheppach@uni-wuerzburg.de}

\abstract{In view of embedding 3D gravity into topological string theory, 
we show that the dimensional reduction of 6D Kodaira--Spencer gravity on $\text{AdS}_3\times S^3$, i.e.~the low-energy description of twisted holography, contains a subsector that coincides with the action of chiral 3D gravity on $\text{AdS}_3$. Furthermore, we show that the full 3D Einstein action is obtained as the reduction of two independent complex-conjugate copies of 6D Kodaira--Spencer gravity. We perform the dimensional reduction explicitly at the level of the classical actions, extending our previous work that related the equations of motion of the two theories. Our reduction procedure is based on a novel rewriting of the non-local 6D Kodaira--Spencer action as a local 6D holomorphic Chern--Simons action for the gauge group $\mathrm{SL}(2,\mathbb{C})$, valid within a subsector of complex structure deformations that we identify. Our results provide a necessary step toward embedding the Euclidean path integral of 3D gravity into topological string theory.}

\begin{document} 
\maketitle
\flushbottom

\newpage

\section{Introduction}

Topological models of gravity, such as three-dimensional Einstein gravity (3D gravity), have proven to be a fruitful route for decoding holographic properties of gravity in a simplified setting. However, as bottom-up models, they lack control over the UV degrees of freedom. Within string theory, twisted holography~\cite{Costello:2018zrm} provides a top-down duality involving a six-dimensional topological theory, Kodaira--Spencer (KS) gravity~\cite{Bershadsky:1993cx}, with a well-defined two-dimensional holographic dual conformal field theory (CFT), known as the gauged beta-gamma system~\cite{Beem:2013sza}. There is ample evidence for this duality, including its derivation from brane backreaction, the matching of global symmetries, identification of single-trace operators with boundary conditions in KS gravity~\cite{Costello:2018zrm}, and identification of determinant operators with wrapped D1-branes~\cite{Budzik:2021fyh}. Nevertheless, the question how bottom-up $\text{AdS}_3\text{/CFT}_2$ duality, involving 3D gravity, is embedded into twisted holography has received little attention.

In this work, we continue the program initiated in~\cite{Erdmenger:2025lvv} to seek a UV completion of Euclidean 3D Einstein gravity with negative cosmological constant within twisted holography and topological string theory. That work provided evidence for the existence of such a 3D gravity subsector in twisted holography. On the other hand, it has been proposed that 3D gravity may describe an ensemble of CFTs rather than a single unitary 2D CFT~\cite{Maloney:2020nni}, with wormholes encoding statistical properties of the ensemble~\cite{Cotler:2020ugk,Cotler:2020hgz}. This is in apparent tension with top-down string-theoretic holography, which does not involve ensemble averaging, as discussed recently in~\cite{Kudler-Flam:2026nzz}. Twisted holography provides a controlled setting to study this tension: its topological nature makes sums over topologies transparent, while the absence of a 't Hooft coupling in the beta-gamma CFT allows $1\slash N$-corrections to be directly organized by the loop expansion of KS gravity, equivalently the worldsheet expansion of the topological B-model. In this context, we expect that the results of the present paper that embed 3D gravity into twisted holography may provide a new avenue towards resolving the tension between explicit top-down constructions and ensemble averaging.

In the work~\cite{Erdmenger:2025lvv}, the foundation for this program was laid out by relating the equations of motion of 6D KS gravity on $\text{AdS}_3\times S^3$ to 3D Einstein's equations on $\text{AdS}_3$. The main result of the present paper is to extend this relation to the level of classical actions: we show that dimensionally reducing a subsector of 6D KS gravity on $\text{AdS}_3\times S^3$ over the $S^3$, keeping only the $\mathrm{SU}(2)$-invariant zero-mode sector, yields a background dependent version of chiral 3D gravity on $\text{AdS}_3$ in the Chern--Simons (CS) formulation~\cite{Achucarro:1986uwr,Witten:1988hc,Witten:1989ip,Li:2008dq}. We show that the full 3D Einstein action, with two independent CS gauge fields, is obtained by starting with two independent, complex-conjugate copies of KS gravity in 6D.

Our results are based on the identification of a subsector of 6D complex structure deformations on $\text{AdS}_3\times S^3$ for which the non-local action of KS gravity becomes local and equal to a holomorphic Chern--Simons (hCS) action for the gauge group $\mathrm{SL}(2,\mathbb{C})$.\footnote{Here this hCS action really describes the closed-string sector of the B-model string theory and is thus different from the hCS theory arising as the open-string theory of space-filling D-branes~\cite{Witten:1992fb} that is also needed for anomaly cancellation purposes~\cite{Costello:2019jsy}. It is also different from the rewriting of KS gravity as CS theory for the infinite-dimensional Lie group of volume preserving diffeomorphisms~\cite{Bershadsky:1993cx}.} This subsector is characterized by a condition on complex structure deformations that we call the locality constraint. This condition is stronger than the one imposed in~\cite{Bershadsky:1993cx,Costello:2018zrm} in order for the variational principle of KS gravity to exist, since it restricts the field space even further. The existence of a hCS formulation of KS gravity is consistent with the earlier result that the integrability of the deformed complex structure, equivalent to the equation of motion of KS gravity, follows from the flatness of a certain 6D $\mathfrak{sl}(2,\mathbb{C})$-valued 1-form~\cite{zentner2013integrable,Herfray:2016std,Erdmenger:2025lvv}.

The strength of reformulating KS gravity as a local hCS theory is that it admits a direct dimensional reduction to 3D gravity: the six-dimensional $\mathfrak{sl}(2,\mathbb{C})$-valued $(0,1)$-form of hCS theory is completed into a connection of the principal $\mathrm{SU}(2)$-bundle $\text{AdS}_3\times S^3$ that straightforwardly reduces to an $\mathfrak{sl}(2,\mathbb{C})$-valued gauge field on the $\text{AdS}_3$ base. This 3D gauge field is exactly the dynamical field of Euclidean 3D gravity in its CS formulation. We show that the 6D hCS action directly reduces to a 3D CS action that we identify as a 3D gravity action, either chiral gravity or full Einstein gravity as mentioned above. Moreover, solutions of 6D Kodaira--Spencer equations reduce to 3D Einstein's equations, consistent with~\cite{Erdmenger:2025lvv}.

Our dimensional reduction to 3D gravity allows us to relate the 3D Newton's constant to the coefficient of the 6D KS action, the string coupling of the B-model. Using the Brown--Henneaux formula for 3D gravity, we then relate the string coupling to the central charge of the dual chiral $U(N)$ beta-gamma system, thereby fixing the proportionality constant in the expected large-$N$ scaling $g_{\text{s}} \sim \frac{1}{N}$~\cite{Costello:2018zrm} of the string coupling to be given by $g_{\text{s}}^2 = \frac{16\pi^3}{N^2}$. The resulting relative sign between the 3D Newton's constant and the 6D string coupling depends on the choice of orientation of $\text{AdS}_3$ inside $\text{AdS}_3\times S^3$. Since the central charge of the beta-gamma system is negative~\cite{Beem:2013sza}, requiring the string coupling to be positive fixes this orientation.

Since 3D gravity is a local theory, obtaining it from dimensional reduction is only possible within the subspace of 6D off-shell complex structure deformations on which the non-local KS action is rendered local. The identification of this constrained subspace is one of the main results of this work. The equations of motion of the constrained system are different from those of the original theory, however we show that they admit a non-trivial family of solutions that also solve the 6D Kodaira--Spencer and 3D Einstein's equations. These are the first steps toward relating the Euclidean path integral of 3D gravity to a UV complete path integral of 6D KS gravity.

\subsection{Summary of results}

Here we give a summary of the main results of this paper.

\paragraph{KS gravity as holomorphic CS theory.} Consider a complex manifold $P_6$ with six real dimensions that admits a holomorphic volume form $\Omega$. Deformations of the complex structure are described by a $(0,1)$-form valued $(1,0)$-vector field $\alpha$ which can be mapped to a $(2,1)$-form $\eta(\alpha) = \iota_\alpha\Omega$ by contracting with the holomorphic volume form. Assuming the cohomology condition $\partial\eta(\alpha) = 0$ such that locally $\eta(\alpha) = \partial\eta(\beta)$, Kodaira--Spencer gravity has the action
\begin{equation}
    I_{\text{KS}}[\beta;\Omega] = -\frac{1}{2g_{\text{s}}^2}\int_{P_6}\biggl( \eta(\alpha)\wedge\overbar{\partial}\eta(\beta)+\frac{1}{3}\,\eta(\alpha\wedge \alpha)\wedge\eta(\alpha)\biggr)\,,
\end{equation}
where $g_{\text{s}}$ is the string coupling of the B-model topological string theory. Solutions to its equation of motion describe integrable complex structure deformations.

In this work, we focus on 6D KS gravity on $P_6 = \mathbb{H}_3\times S^3$, where $\mathbb{H}_3$ denotes three-dimensional hyperbolic space (Euclidean AdS$_3$) which is the relevant background for twisted holography. The manifold $\mathbb{H}_3\times S^3$ admits a natural $\mathfrak{sl}(2,\mathbb{C})$-valued 1-form $\Sigma$, which is both holomorphic $\overbar{\partial}\Sigma = 0$ and flat $\d \Sigma + \Sigma\wedge\Sigma = 0$, coming from the identification $\mathbb{H}_3\times S^3\cong \mathrm{SL}(2,\mathbb{C})$~\cite{Erdmenger:2025lvv}. This $\Sigma$ defines both a background complex structure and a corresponding holomorphic $(3,0)$-form
\begin{equation}
    \Omega = \frac{1}{3}\,\tr{(\Sigma\wedge\Sigma\wedge\Sigma)}\,,
\end{equation}
where the trace is taken in the fundamental representation of $\mathfrak{su}(2)$. The complex structure deformation $\alpha$ defines the $\mathfrak{sl}(2,\mathbb{C})$-valued $(0,1)$-form $\mathcal{A}_{01} = \iota_\alpha\Sigma$ via contraction. Assuming $\alpha$ satisfies the locality constraint
\begin{equation}
    \partial \mathcal{A}_{01} +\mathcal{A}_{01}\wedge\Sigma + \Sigma\wedge\mathcal{A}_{01} = 0\,,
\end{equation}
which implies the cohomology condition $\partial\eta(\alpha) = 0$, although the converse does not hold, we show that the non-local Kodaira--Spencer action can be rewritten as local holomorphic Chern--Simons theory
\begin{equation}
    I_{\text{KS}}[\beta;\Omega] = -\frac{1}{2g_{\text{s}}^2}\int_{P_6}\Omega\wedge \biggl(\tr{(\mathcal{A}_{01}\wedge \overbar{\partial}\mathcal{A}_{01})}+\frac{2}{3}\,\tr{(\mathcal{A}_{01}\wedge \mathcal{A}_{01}\wedge \mathcal{A}_{01})}\biggr)\,.
\end{equation}
We show that the variation of the holomorphic CS action matches the variation of the KS action under the locality constraint,
\begin{equation}
    \tr{[(\overbar{\partial}\mathcal{A}_{01} + \mathcal{A}_{01}\wedge \mathcal{A}_{01})\wedge \Sigma\wedge\Sigma]} = \overbar{\partial}\eta(\alpha)+\frac{1}{2}\,\partial\eta(\alpha\wedge \alpha)\,.
\end{equation}
Furthermore, we derive the constrained equation of motion under the locality constraint and  show that it includes solutions of the original unconstrained KS equation.

Thus we have identified a subsector of Kodaira--Spencer gravity on $\mathbb{H}_3\times S^3$ which is local in $\alpha$ and described by holomorphic Chern--Simons theory.

\paragraph{Chiral 3D gravity from dimensional reduction.} The above reformulation of KS gravity as holomorphic CS theory provides a direct avenue for dimensional reduction over the $S^3$ to $\mathbb{H}_3$. Defining the $\mathfrak{sl}(2,\mathbb{C})$-valued 1-form
\begin{equation}
    \mathcal{A} = B_i\,\Sigma^i + \iota_\alpha\Sigma\,,
\end{equation}
where $B_i$ are arbitrary $\mathfrak{sl}(2,\mathbb{C})$-valued scalar fields and $\Sigma = \Sigma^i\,\tau_i$ with $\mathfrak{su}(2)$-generators $\tau_i$, the holomorphic CS action takes the form
\begin{equation}
    I_{\text{KS}}[\beta;\Omega] = -\frac{1}{g_{\text{s}}^2}\int_{P_6}\Omega\wedge\mathrm{CS}[\mathcal{A}]\,,
\end{equation}
where $\mathrm{CS}[\mathcal{A}]$ is the Chern--Simons 3-form. The dependence on $B_i$ drops out since the wedge product with $\Omega$ cancels all $(1,0)$-form parts of $\mathcal{A}$. It is convenient to fix $B_i$ such that $\mathcal{A}(\tau_i^{\#}) = \tau_i$ where $\tau_i^{\#}$ is the fundamental vector field associated with the $\mathfrak{su}(2)$-generator $\tau_i$. In this case, $\mathcal{A}$ is a connection and we have the dimensional reduction formulae
\begin{equation}
    \Sigma = G^{-1}\,A_0\,G + G^{-1}\d G\,,\quad \mathcal{A} = G^{-1}\,A\,G + G^{-1}\d G\,,
\end{equation}
where $A_0 = A_0^i\,\tau_i$ is a flat 3D gauge field determined by the background and $A$ is an off-shell 3D gauge field
\begin{equation}
    A = A_0 - c_i\,(A_0^i-\overbar{A}_0^i)\,,
\end{equation}
where $c_i $ is the pull-back of the 6D field $C_i = \Sigma_k^j\,\overbar{\sigma}^{\overbar{l}}_i\,\alpha_{\overbar{l}}^{\;\;\,k}\,\tau_j$ to the base $\mathbb{H}_3$ using a section.

As a result, we show that the dimensional reduction of KS gravity, on the local subspace described by holomorphic CS theory, is given by
\begin{equation}
    I_{\text{KS}}[\beta;\Omega] = \frac{2\vol{(S^3)}}{g_{\text{s}}^2}\int_{\mathbb{H}_3}\tr{\biggl(
\Delta A \wedge \d_{A_0} (\Delta A)+
\frac{2}{3}\,
\Delta A \wedge \Delta A \wedge \Delta A
\biggr)}\,,
\end{equation}
where $\Delta A \equiv A - A_0$ and $\d_{A_0} = \d + [A_0,\cdot\,]$ is the gauge covariant derivative relative to the background $A_0$. This is chiral 3D gravity for a single gauge field $A$ and the dependence on a background solution $A_0$ is the 3D manifestation of the background dependence of KS gravity on $\Omega$.

\paragraph{3D Einstein gravity from dimensional reduction.} To obtain full 3D Einstein gravity with two copies of the 3D CS action for complex-conjugate gauge fields $A$ and $\overbar{A} = (A^i)^*\tau_i$, we must dimensionally reduce two copies of 6D KS actions. The 6D action takes the form
\begin{equation}
    I[\beta,\overbar{\beta};\Omega,\overbar{\Omega}] = -\frac{i}{2}\,\bigl(I_{\text{KS}}[\beta;\Omega]-I_{\text{KS}}[\overbar{\beta};\overbar{\Omega}]\bigr)+\frac{i}{4 g_{\text{s}}^2}\int_{P_6} \Omega\wedge \overbar{\Omega}\,,
\end{equation}
where we have included a six-dimensional volume term involving the volume form $\Omega\wedge \overbar{\Omega}$ of the background. We show that its dimensional reduction gives
\begin{align}
    I[\beta,\overbar{\beta};\Omega,\overbar{\Omega}] &= -\frac{2i\vol{(S^3)}}{ g_{\text{s}}^2}\int_{\mathbb{H}_3} (\text{CS}[A]-\text{CS}[\overbar{A}])+ (\text{boundary terms})\,,
\end{align}
which is the action of 3D Einstein gravity in the CS formulation. Thus we are able to relate the 6D string coupling $g_{\text{s}}$ to the 3D Newton's constant $G_{\text{N}}$ as
\begin{equation}
    G_{\text{N}} = -\frac{g_{\text{s}}^2}{32\pi^3}\,.
\end{equation}
In particular, combining the Brown--Henneaux formula with the negative central charge $c=-3(N^2-1)$ of the beta-gamma system implies
\begin{equation}
    g_{\text{s}}^2 = \frac{16\pi^3}{N^2}+ \mathcal{O}(N^{-4})\,,\quad N\rightarrow \infty\,,
\end{equation}
which fixes the prefactor in the relation of the string coupling to the rank of the gauge group.

The paper is structured as follows. In Section~\ref{sec:KS_as_CS}, we review KS gravity and how it can be rewritten as local holomorphic CS theory on a subspace of complex structure deformations. Then, in Section~\ref{sec:3D_grav_red}, we explain how to dimensionally reduce KS gravity by reducing the holomorphic CS action. First, in Section~\ref{subapp:reduction_action}, we dimensionally reduce the KS action to chiral gravity, and in Section~\ref{sec:full_Einstein_reduce}, we reduce two copies of KS gravity to full 3D Einstein gravity. We conclude with a discussion of subtleties and future directions in Section~\ref{sec:discussion}. Technical details and complementary calculations are relegated to the appendices. Appendix~\ref{app:conventions} contains our differential geometry conventions and a proof of Tian's lemma; Appendix~\ref{app:KS_variation} contains derivations of equations of motion; details of the dimensional reduction of KS gravity are in Appendix~\ref{app:dim_red_details}; the CS formulation of 3D Einstein gravity is reviewed in detail in Appendix~\ref{app:3D_Einstein_CS_form}; and Appendix~\ref{app:flatness_integrability} contains a self-contained calculation relating flatness to the KS equation when the background volume form is not holomorphic.

\section{6D Kodaira--Spencer gravity as holomorphic Chern--Simons theory}\label{sec:KS_as_CS}

In this section, we rewrite the 6D KS action on $\mathbb{H}_3\times S^3$ as a 6D holomorphic CS action for the gauge group $\mathrm{SL}(2,\mathbb{C})$. To this end, we propose a locality constraint that renders the non-local KS action local. We discuss the resulting constrained field space and the constrained equations of motion.

\subsection{Definition of Kodaira--Spencer gravity}\label{subsec:KS_definition}

We consider a six-dimensional differential manifold $P_6$ equipped with a complex structure $J$ with real coordinates $x^a$, corresponding complex coordinates $(w^i,\overbar{w}^i)$ and a $(3,0)$-form $\Omega$ which is holomorphic $\overbar{\partial}\Omega = 0$. Let $\alpha = \alpha_{\overbar{i}}^{\;\;j}\, \d\overbar{w}^{\overbar{i}}\otimes \partial_j$ be a $(0,1)$-form valued $(1,0)$-vector field, a so-called $(1,1)$-polyvector, which is used to deform the background $J$ to an almost complex structure $J_\alpha$ via the formula\footnote{Strictly speaking, $J_\alpha$ is a complex endomorphism that deforms the space of $(1,0)$-forms but leaves the space of $(0,1)$-forms invariant. To recover a real automorphism, we need to deform the $(0,1)$-forms by the conjugate $\overbar{\alpha}$ of $\alpha$. Here, we work in the complexified cotangent space and set $\overbar{\alpha} = 0$.}
\begin{equation}
    J_\alpha = e^{\iota_\alpha}\,J\,e^{-\iota_\alpha} = J + 2i\,\alpha_{\overbar{i}}^{\;\;j}\, \d\overbar{w}^{\overbar{i}}\otimes \partial_j\,,
    \label{eq:J_alpha}
\end{equation}
where $\iota_\alpha = \alpha^i\wedge\iota_{\partial_i}$ denotes the contraction with the vector index of $\alpha$, $\iota_{\partial_i}$ is the contraction with the vector field $\partial_i$ and $e^{\iota_\alpha}$ is defined by the Taylor series of the exponential~\cite{Rao_2025}. It is defined so that for any $(1,0)$-form $X$ with respect to $J$, the form $X_\alpha = e^{\iota_\alpha} X = X+\iota_\alpha X$ has eigenvalue $+i$ of $J_\alpha$. In contrast, the space of $(0,1)$-forms defined by $J$ is not deformed. Since $\alpha$ defines a new almost complex structure in this way, it is called a complex structure deformation. It follows that $J_\alpha$ is integrable, i.e.~has a vanishing Nijenhuis tensor and defines a complex structure, if and only if $\alpha$ satisfies the Kodaira--Spencer equation~\cite{newlander1957complex}
\begin{equation}
    E_\alpha\equiv \overbar{\partial}\alpha - \frac{1}{2}\,[\alpha,\alpha] = 0\,,
    \label{eq:KS_equation}
\end{equation}
where $\overbar{\partial}$ is the Dolbeault operator acting on the form part of $\alpha$ and the bracket is the Nijenhuis bracket defined in \eqref{eq:Nijenhuis_bracket}. In components, the KS equation is (see Appendix~\ref{app:conventions} for our conventions)
\begin{equation}
    \frac{1}{2}\,(E_\alpha)_{\overbar{j}\overbar{k}}^i = \overbar{\partial}_{[\overbar{j}}\, \alpha_{\overbar{k}]}^{\;\;\,i}-\alpha_{[\overbar{j}}^{\;\;\,l}\partial_{\vert l \vert}\,\alpha_{\overbar{k}]}^{\;\;\,i}=0\,.
    \label{eq:KS_EOM}
\end{equation}
The background holomorphic $(3,0)$-form $\Omega$ can be used to define a map $\eta$ from polyvectors to forms by contracting the polyvector with $\Omega$ (see Appendix~\ref{app:conventions}). When applied to $\alpha$, the result is the $(2,1)$-form
\begin{equation}
    \eta(\alpha)\equiv \iota_\alpha \Omega\,,
    \label{eq:eta_alpha}
\end{equation}
where the holomorphic vector index of $\alpha$ is contracted with the holomorphic form index of $\Omega$. In components, we have $\Omega_{ijk} = e^{\upsilon}\varepsilon_{ijk}$, for some function $\upsilon$ and $\varepsilon_{ijk}$ is the Levi--Civita symbol, so that explicitly $\eta(\alpha)_{ij\overbar{k}} = \Omega_{ijl}\,\alpha_{\overbar{k}}^{\;\;\,l}$.

To ensure that the KS action exists, as given below, we assume that $\alpha$ satisfies the cohomology condition~\cite{Bershadsky:1993cx,Costello:2018zrm}
\begin{equation}
    \partial \eta(\alpha) = 0\,,
    \label{eq:cohomology_condition}
\end{equation}
which in components is equivalent to $ (\partial_{j} + \partial_j\upsilon)\,\alpha_{\overbar{i}}^{\;\;\,j} = 0$. Under this condition, the KS equation \eqref{eq:KS_equation} takes the form
\begin{equation}
    \eta(E_\alpha) = \overbar{\partial}\eta(\alpha)  + \frac{1}{2}\,\partial\eta(\alpha\wedge\alpha) = 0\,,
    \label{eq:KS_Tian}
\end{equation}
where $\alpha\wedge\alpha =  \alpha_{\overbar{i}}^{\;\;k}\alpha_{\overbar{j}}^{\;\;l}\, \d\overbar{w}^{\overbar{i}}\wedge \d\overbar{w}^{\overbar{j}}\otimes \partial_k\wedge \partial_l $ is a $(2,0)$-vector-valued $(0,2)$-form, called a $(2,2)$-polyvector, so that
\begin{equation}
    \eta(\alpha\wedge\alpha) = \iota_{\alpha\wedge\alpha}\Omega\equiv \iota_\alpha\iota_\alpha\Omega
    \label{eq:eta_alpha_alpha}
\end{equation}
is a $(1,2)$-form. This is the KS equation in Tian's form which follows from the cohomology condition \eqref{eq:cohomology_condition}, holomorphicity $\overbar{\partial}\Omega = 0$ and Tian's lemma\footnote{We adopt the convention $\iota_{\partial_i\wedge \partial_j} \equiv \iota_{\partial_j}\iota_{\partial_i}$ for the contraction with a bivector. With the opposite convention $\iota_{\partial_i\wedge \partial_j} \equiv \iota_{\partial_i}\iota_{\partial_j} = -\iota_{\partial_j}\iota_{\partial_i}$, Tian's lemma has a plus sign on the right-hand side as in~\cite{Bershadsky:1993cx}. See Appendix~\ref{app:conventions} for a complete summary of our conventions and for a proof of Tian's lemma.}
\begin{equation}
    \frac{1}{2}\,\eta([\alpha,\alpha]) = -\frac{1}{2}\,\partial\eta(\alpha\wedge\alpha)+\iota_\alpha\partial\eta(\alpha)\,.
    \label{eq:Tians_lemma_text}
\end{equation}
The KS equation and the cohomology condition have a natural interpretation from the perspective of the holomorphic volume form. The $(3,0)$-form with respect to the deformed almost complex structure $J_\alpha$ is given by~\cite{Bershadsky:1993cx,Kjelsnes:2026oox}
\begin{equation}\label{eq:volume_form_deformation}
    \Omega_\alpha = e^{\iota_\alpha}\,\Omega = \Omega + \eta(\alpha) + \frac{1}{2}\,\eta(\alpha\wedge \alpha) + \frac{1}{6}\,\eta(\alpha\wedge \alpha\wedge \alpha)\,,
\end{equation}
which follows from \eqref{eq:J_alpha} and where the second equality follows by expanding the exponential in a Taylor series. Together, the KS equation and the cohomology condition ensure that the deformed $(3,0)$-form is closed $ \d \Omega_\alpha = 0$~\cite{Bershadsky:1993cx}. As a result, $(J,\Omega)\rightarrow (J_\alpha,\Omega_\alpha)$ is a deformation that preserves integrability of the complex structure and closedness of the holomorphic top-form.

\paragraph{Kodaira--Spencer action.} Assuming the cohomology condition \eqref{eq:cohomology_condition} holds, locally there exists a $(2,1)$-polyvector $\beta = \frac{1}{2}\,\beta_{\overbar{i}}^{\;\;\,kl}\,\d\overbar{w}^{\overbar{i}}\otimes \partial_k\wedge\partial_l $ such that\footnote{We set the massless modes, denoted by $x$ in~\cite{Bershadsky:1993cx}, to zero. As explained in~\cite{Bershadsky:1993cx}, a non-zero massless mode is equivalent to changing the background $\Omega$ in \eqref{eq:KS_action_def} up to a classical on-shell action.}
\begin{equation}
    \eta(\alpha) = \partial \eta(\beta)\,,
    \label{eq:definition_beta}
\end{equation}
where $\eta(\beta) = \iota_\beta\, \Omega$ is a $(1,1)$-form with components $\eta(\beta)_{i\overbar{j}} = -\frac{1}{2}\,e^{\upsilon}\varepsilon_{imn}\,\beta_{\overbar{j}}^{\;\;\,mn}$. In components, the relation \eqref{eq:definition_beta} becomes $\alpha_{\overbar{i}}^{\;\;\,j} = (\partial_k+ \partial_k\upsilon)\,\beta_{\overbar{i}}^{\;\;\,kj}$.

The action of 6D KS gravity is a non-local action for $\alpha$ but a local action for $\beta$ given by\footnote{In~\cite{Bershadsky:1993cx}, the kinetic term has an opposite sign. Their kinetic term is $\frac{1}{\partial}\overbar{\partial}\eta(\alpha) = \frac{1}{\partial}\overbar{\partial}\partial\eta(\beta) = -\overbar{\partial}\eta(\beta)$ since $\overbar{\partial}\partial = -\partial\overbar{\partial}$. However, since they have a plus sign on the right-hand side of Tian's lemma \eqref{eq:Tians_lemma_text} they are using the alternative convention for the contraction with a bivector. Thus their cubic term differs from ours by a minus sign and their total action differs from ours by an overall minus sign.}~\cite{Bershadsky:1993cx}
\begin{equation}
    I_{\text{KS}}[\beta;\Omega] = -\frac{1}{2g_{\text{s}}^2}\int_{P_6}\biggl( \eta(\alpha)\wedge\overbar{\partial}\eta(\beta)+\frac{1}{3}\,\eta(\alpha\wedge \alpha)\wedge\eta(\alpha)\biggr)\,.
    \label{eq:KS_action_def}
\end{equation}
In KS gravity, $\Omega$ plays the role of a background field while $\beta$ is the dynamical field. The variation of the action with respect to $\beta$ is given by (see Appendix~\ref{app:KS_variation})
\begin{equation}
    \delta I_{\text{KS}}[\beta;\Omega] = -\frac{1}{g_{\text{s}}^2}\int_{P_6}\biggl(\overbar{\partial}\eta(\alpha)+\frac{1}{2}\,\partial\eta(\alpha\wedge \alpha)\biggr)\wedge\eta(\delta\beta) + (\text{boundary terms})\,,
    \label{eq:KS_action_variation}
\end{equation}
which gives the KS equation in Tian's form \eqref{eq:KS_Tian}. Now it is clear why it is necessary to impose the cohomology condition \eqref{eq:cohomology_condition}: it gives the $(1,1)$-form $\eta(\beta)$ whose variation can be wedged with the $(2,2)$-form $\eta(E_\alpha)$ in order to define a $(3,3)$-form $\eta(E_\alpha)\wedge \eta(\delta\beta)$ that can arise from the variation of an action.

\subsection{The 6D KS action as a local 6D holomorphic CS action}\label{subsec:hCS_rewriting}

Consider a six-dimensional principal $\mathrm{SU}(2)$-bundle $P_6$ over three-dimensional hyperbolic space $\mathbb{H}_3\subset \mathbb{R}^4$ defined by its standard embedding into $\mathbb{R}^4$. Since all such bundles are trivial~\cite{Dijkgraaf:1989pz}, the manifold is a global direct product $P_6 = \mathbb{H}_3\times S^3$. It turns out that $P_6$ is the group manifold of $\mathrm{SL}(2,\mathbb{C})$. This follows from the fact that any $M\in \mathrm{SL}(2,\mathbb{C})$ has the polar decomposition $M = S G$ where $G\in \mathrm{SU}(2)$ and $S \equiv \sqrt{MM^\dagger}>0$ is a positive-definite Hermitian matrix. Such Hermitian matrices are in one-to-one correspondence with $\mathbb{H}_3$ so that $S$ is the global section leading to the trivialization $\mathrm{SL}(2,\mathbb{C})\cong  \mathbb{H}_3\times S^3$; see~\cite{Erdmenger:2025lvv} for a more detailed discussion.

We now show that a local subsector of KS gravity on $P_6 = \mathbb{H}_3\times S^3$ can be written as holomorphic Chern--Simons theory.

\paragraph{Background structures on the deformed conifold.} The manifold $P_6 = \mathbb{H}_3\times S^3$ is naturally equipped with a flat $\mathfrak{sl}(2,\mathbb{C})$-valued 1-form
\begin{equation}
    \Sigma = M^{-1}\,\d M\,,\quad F_\Sigma = \d \Sigma + \Sigma\wedge\Sigma = 0\,.
    \label{eq:Sigma}
\end{equation}
In addition, there is a second $\mathfrak{sl}(2,\mathbb{C})$-valued 1-form given by $\overbar{\Sigma} \equiv -\Sigma^\dagger$ which is also flat $F_{\overbar{\Sigma}} = -F_\Sigma^\dagger = 0$. The forms can be decomposed into a basis of three 1-forms $\Sigma^i$ and $\overbar{\Sigma}^i$ as
\begin{equation}
    \Sigma  = \Sigma^i\,\tau_i\,,\quad \overbar{\Sigma} = \overbar{\Sigma}^i\tau_i\,,
\end{equation}
where the anti-Hermitian $\mathfrak{su}(2)$-generators $\tau_i = -\tau_i^\dagger$ satisfy
\begin{equation}
    [\tau_i,\tau_j] = \varepsilon_{ijk}\,\delta^{kl}\,\tau_l\,,\quad \tr{(\tau_i\tau_j)} = -\frac{1}{2}\,\delta_{ij}\,,\quad \tr{(\tau_i\tau_j\tau_k)} = -\frac{1}{4}\,\varepsilon_{ijk}\,.
\end{equation}
Therefore $\overbar{\Sigma}^i = (\Sigma^i)^*$ defines the complex conjugate basis.

Due to the presence of $\Sigma$ and $\overbar{\Sigma}$, the bundle $P_6 = \mathbb{H}_3\times S^3$ comes equipped with a natural triple consisting of a complex structure and a pair of volume forms $(J,\Omega,\overbar{\Omega})$ defined by
\begin{equation}
    J = -2i\,\tr{(\Sigma\otimes V - \overbar{\Sigma}\otimes \overbar{V})}\,,\quad \Omega = \frac{1}{3}\,\tr{(\Sigma\wedge\Sigma\wedge \Sigma)}\,,\quad \overbar{\Omega} = \frac{1}{3}\,\tr{(\overbar{\Sigma}\wedge\overbar{\Sigma}\wedge \overbar{\Sigma})}\,,
    \label{eq:Sigma_J_Omega}
\end{equation}
where we have defined the $\mathfrak{sl}(2,\mathbb{C})$-valued dual vector fields $V \equiv \sigma^{ij}\,\tau_i\,\partial_j$ and $\overbar{V} =\overbar{\sigma}^{i\overbar{j}}\,\tau_i\,\overbar{\partial}_{\overbar{j}}$ that satisfy $\Sigma^i(V^j) = \overbar{\Sigma}^i(\overbar{V}^j) = \delta^{ij}$ and $\Sigma^i(\overbar{V}^j) = \overbar{\Sigma}^i(V^j) = 0$. It follows that $\Sigma$ is a $(1,0)$-form and $\overbar{\Sigma}$ is a $(0,1)$-form with respect to $J$ so that we can decompose $\Sigma^i = \Sigma^i_j\,\d w^j$ and $\overbar{\Sigma}^i = \overbar{\Sigma}^i_{\overbar{j}}\,\d \overbar{w}^{\overbar{j}}$ with $\overbar{\Sigma}^i_{\overbar{j}} = (\Sigma^i_j)^*$. The existence of the dual vector fields $V, \overbar{V}$ implies that the components $\Sigma^i_j$ and $\overbar{\Sigma}^i_{\overbar{j}}$ form invertible matrices. Explicitly, from the duality relations, we find $\Sigma^i_k\, \sigma^k_{j}= \delta^{i}_j= \overbar{\Sigma}^{i}_{\overbar{k}}\,\overbar{\sigma}^{\overbar{k}}_j$, with $\sigma^i_j = \delta_{jk}\sigma^{ki}$ and $\overbar{\sigma}^{\overbar{i}}_j = \delta_{jk} \overbar{\sigma}^{k\overbar{i}}$.

Since $\Sigma$ is flat $(1,0)$-form, it is holomorphic $\overbar{\partial}\Sigma = 0$ with respect to $J$. Flatness also implies that \(J\) is integrable and that \(\Omega\) is closed~\cite{Erdmenger:2025lvv}
\begin{equation}
    \d \Omega = \tr{(\d\Sigma\wedge\Sigma\wedge\Sigma)} = -\tr{(\Sigma\wedge\Sigma\wedge\Sigma\wedge\Sigma)} = 0\,.
\end{equation}
Similarly, we have $\partial\overbar{\Sigma} = 0$ and $\d \overbar{\Omega} = 0 = \partial \overbar{\Omega}$.

\paragraph{Rewriting of the fields.} Consider the forms $\eta(\alpha)$ and $\eta(\beta)$ relevant for KS gravity on $P_6 = \mathbb{H}_3\times S^3$. For the above $\Omega$ \eqref{eq:Sigma_J_Omega}, we find
\begin{equation}
     \eta(\alpha) = \frac{1}{3}\,\iota_{\alpha}\tr{(\Sigma\wedge\Sigma\wedge \Sigma)} = \tr{(\mathcal{A}_{01}\wedge \Sigma\wedge \Sigma)}\,,
     \label{eq:eta_alpha_Sigma}
\end{equation}
where we have defined the $\mathfrak{sl}(2,\mathbb{C})$-valued $(0,1)$-form
\begin{equation}
    \mathcal{A}_{01} \equiv \iota_\alpha\Sigma
    \label{eq:A_01}
\end{equation}
and used that the contraction $\iota_\alpha$ with a $(1,1)$-polyvector satisfies the Leibniz rule without additional minus signs and the trace of the wedge product is invariant under cyclic permutations of forms whose total degree is one. The cohomology condition \eqref{eq:cohomology_condition} becomes
\begin{equation}
    \partial\eta(\alpha) = \tr{(\partial \mathcal{A}_{01}\wedge \Sigma\wedge\Sigma)} = 0\,,
    \label{eq:eta_alpha_A_cohomology}
\end{equation}
where we used that $\tr{(\mathcal{A}_{01}\wedge\Sigma\wedge\Sigma\wedge\Sigma)} = 0$ due to $\tr{(\tau_i\tau_j\tau_k\tau_l)} = \frac{1}{8}(\delta_{ij}\delta_{kl}-\delta_{ik}\delta_{jl} +\delta_{il}\delta_{jk})$ and antisymmetry of the wedge product $\Sigma^i\wedge \Sigma^j$.

The cohomology condition \eqref{eq:eta_alpha_A_cohomology} is not sufficient for writing $\eta(\beta)$ as a local function of $\mathcal{A}_{01}$. However, this turns out to be possible if we impose the condition
\begin{equation}
    (F_{\mathcal{B}})_{11} = \partial \mathcal{A}_{01} +\mathcal{A}_{01}\wedge\Sigma + \Sigma\wedge\mathcal{A}_{01} = 0 \,,
    \label{eq:A_cohomology}
\end{equation}
which can be identified as the vanishing of the $(1,1)$-component of the curvature $F_{\mathcal{B}} = \d \mathcal{B} + \mathcal{B}\wedge \mathcal{B}$ of $\mathcal{B}\equiv \Sigma + \iota_\alpha\Sigma$. The condition \eqref{eq:A_cohomology}, which we will henceforth call the locality constraint, implies the cohomology condition \eqref{eq:eta_alpha_A_cohomology} since
\begin{equation}
    \partial\eta(\alpha) = \tr{(\partial \mathcal{A}_{01}\wedge \Sigma\wedge\Sigma)} = \tr{[(F_{\mathcal{B}})_{11}\wedge \Sigma\wedge\Sigma]} = 0
    \label{eq:cohomology_F}
\end{equation}
due to $\tr{(\mathcal{A}_{01}\wedge\Sigma\wedge\Sigma\wedge\Sigma)} = 0$, but the reverse is not true.\footnote{The cohomology condition \eqref{eq:cohomology_F} is in components $\varepsilon_{ijk}\,\partial_{[l} (\mathcal{A}_{01})^i_{\vert \overbar{m}\vert} \Sigma_{n}^j\Sigma_{p]}^k = 0$. This can be inverted to yield $\sigma^j_k\,\partial_j(\mathcal{A}_{01})^k_{\overbar{i}} = 0$ which only constrains the trace of $\partial \mathcal{A}_{01}$ while the locality constraint \eqref{eq:A_cohomology} restricts all its form and $\mathfrak{su}(2)$ components.} Thus \eqref{eq:A_cohomology} is a stronger condition than the cohomology condition \eqref{eq:eta_alpha_A_cohomology}. Now using the locality constraint \eqref{eq:A_cohomology} together with $F_\Sigma = \partial\Sigma + \Sigma\wedge \Sigma = 0$, we see that
\begin{equation}
    \eta(\alpha) = -\tr{(\mathcal{A}_{01}\wedge \partial\Sigma)} = \partial\,\tr{(\mathcal{A}_{01}\wedge \Sigma)} + 2\,\tr{(\mathcal{A}_{01}\wedge \Sigma\wedge \Sigma)}\,.
\end{equation}
The last term is equal to $2\eta(\alpha)$ so that we obtain
\begin{equation}
    \eta(\alpha) = - \partial\,\tr{(\mathcal{A}_{01}\wedge \Sigma)}\,,
\end{equation}
which implies that
\begin{equation}
    \eta(\beta) = -\tr{(\mathcal{A}_{01}\wedge\Sigma)} + K
    \label{eq:eta_beta}
\end{equation}
for some $K$ satisfying $\partial K = 0$.

\paragraph{Rewriting of the KS action.} The locality constraint \eqref{eq:A_cohomology} is strong enough to allow a rewriting of the KS action as a local holomorphic Chern--Simons action for $\mathcal{A}_{01}$. First, we notice that $K$ in $\eta(\beta)$ \eqref{eq:eta_beta} contributes only to boundary terms in the KS action \eqref{eq:KS_action_def} since its contribution to the kinetic term $\eta(\alpha)\wedge\overbar{\partial}\eta(\beta)$ is
\begin{equation}
    \eta(\alpha)\wedge\overbar{\partial}K = \d(\eta(\beta)\wedge \overbar{\partial}K)
\end{equation}
due to the fact that $\partial K = 0$. Since $\overbar{\partial}\Sigma = 0$, the remaining contribution to the kinetic term is thus
\begin{equation}
    -\eta(\alpha)\wedge\tr{(\overbar{\partial}\mathcal{A}_{01}\wedge\Sigma)}  = -\frac{1}{8}\,\varepsilon_{ijk}\,\delta_{mn}\,\mathcal{A}_{01}^i\wedge \Sigma^j\wedge\Sigma^k\wedge \overbar{\partial}\mathcal{A}_{01}^m\wedge \Sigma^n\,.
\end{equation}
Now permuting the wedge product to move all $\Sigma$ to the left which flips the overall minus sign to a plus, using
\begin{equation}
    \Sigma^j\wedge \Sigma^k\wedge   \Sigma^n = \varepsilon^{jkn}\,\Sigma^1\wedge \Sigma^2\wedge   \Sigma^3 = -\frac{2}{3}\,\varepsilon^{jkn}\,\tr{(\Sigma\wedge\Sigma\wedge \Sigma)}
    \label{eq:Sigma_cubed}
\end{equation}
and that $\varepsilon_{ijk}\varepsilon^{jkn}=2\delta^n_i$, we obtain that the full kinetic term is
\begin{equation}
    \eta(\alpha)\wedge\overbar{\partial}\eta(\beta) = \frac{1}{3}\,\tr{(\Sigma\wedge \Sigma\wedge \Sigma)}\wedge\tr{(\mathcal{A}_{01}\wedge \overbar{\partial}\mathcal{A}_{01})} + \d(\eta(\beta)\wedge \overbar{\partial}K)\,.
\end{equation}
The total derivative contributes only a boundary term, and on a manifold with boundary, it can be canceled by adding an appropriate boundary term to the KS action. Such a term is also required to preserve the usual shift symmetry $\eta(\beta)\rightarrow \eta(\beta) + K$ with $\partial K = 0$ of KS gravity in the presence of a boundary~\cite{Dijkgraaf:2004te}. Thus, we assume that this boundary term has been included and neglect the resulting total derivative.

For the interaction term of the KS action, we need to compute $\eta(\alpha\wedge\alpha) = \iota_\alpha\eta(\alpha)$ in terms of $\mathcal{A}_{01}$. By using \eqref{eq:eta_alpha_Sigma} and that $\iota_\alpha\mathcal{A}_{01} = 0$ since the vector component of $\alpha$ is of type $(1,0)$, we obtain
\begin{equation}
    \eta(\alpha\wedge\alpha) = 2\,\tr{(\mathcal{A}_{01}\wedge \mathcal{A}_{01}\wedge \Sigma)}\,.
    \label{eq:eta_alpha_alpha_Sigma}
\end{equation}
Thus the interaction term becomes
\begin{equation}
    \frac{1}{3}\,\eta(\alpha\wedge \alpha)\wedge\eta(\alpha) 
    = \frac{2}{3}\frac{1}{16}\,\varepsilon_{ijk}\varepsilon_{lmn}\,\mathcal{A}_{01}^i\wedge \mathcal{A}_{01}^j\wedge \Sigma^k\wedge\mathcal{A}_{01}^l\wedge \Sigma^m\wedge \Sigma^n\,.
\end{equation}
Moving all $\Sigma$ to the left, which does not change the overall sign, using \eqref{eq:Sigma_cubed} for both the product of $\Sigma$ and for the product of $\mathcal{A}_{01}$, and that $\varepsilon_{ijk}\varepsilon_{lmn}\varepsilon^{kmn}\varepsilon^{ijl} = 12 $, we obtain
\begin{equation}
    \frac{1}{3}\,\eta(\alpha\wedge \alpha)\wedge\eta(\alpha) 
    = \frac{2}{9}\,\tr{(\Sigma\wedge \Sigma\wedge \Sigma)}\wedge\tr{(\mathcal{A}_{01}\wedge \mathcal{A}_{01}\wedge\mathcal{A}_{01})}\,.
\end{equation}
Thus, together with the expression \eqref{eq:Sigma_J_Omega} for the volume form, the KS action becomes up to the above boundary term
\begin{equation}
    I_{\text{KS}}[\beta;\Omega] = -\frac{1}{2g_{\text{s}}^2}\int_{P_6}\Omega\wedge \biggl(\tr{(\mathcal{A}_{01}\wedge \overbar{\partial}\mathcal{A}_{01})}+\frac{2}{3}\,\tr{(\mathcal{A}_{01}\wedge \mathcal{A}_{01}\wedge \mathcal{A}_{01})}\biggr)\,,
    \label{eq:KS_as_hCS}
\end{equation}
which is a 6D holomorphic Chern--Simons (hCS) action for $\mathcal{A}_{01} = \iota_\alpha \Sigma$. We see that the non-local KS action for $\alpha$ is rendered local by the locality constraint \eqref{eq:A_cohomology} which is stronger than the cohomology condition. We will discuss the implications of this condition in the next subsection.

\subsection{Constrained variation of the holomorphic CS action}\label{subsec:constrained_variation}

On the subset of complex structure deformations $\alpha$ satisfying the locality constraint \eqref{eq:A_cohomology}, the dynamics is governed by the local hCS action \eqref{eq:KS_as_hCS}. The variation of \eqref{eq:KS_as_hCS} with respect to $\mathcal{A}_{01}$ gives
\begin{equation}
    \delta I_{\text{KS}}[\beta;\Omega] = -\frac{1}{g_{\text{s}}^2}\int_{P_6}\Omega\wedge\tr{[(\overbar{\partial}\mathcal{A}_{01} + \mathcal{A}_{01}\wedge \mathcal{A}_{01}) \wedge \delta \mathcal{A}_{01}]} + (\text{boundary terms}) \,,
    \label{eq:hCS_variation}
\end{equation}
where we used $\d \Omega  = 0 = \overbar{\partial}\Omega$; see Appendix~\ref{app:CS_variation} for details. Here we can simplify
\begin{equation}
    \Omega\wedge\tr{[(\overbar{\partial}\mathcal{A}_{01} + \mathcal{A}_{01}\wedge \mathcal{A}_{01}) \wedge \delta \mathcal{A}_{01}]} = -\tr{[(\overbar{\partial}\mathcal{A}_{01} + \mathcal{A}_{01}\wedge \mathcal{A}_{01})\wedge\Sigma \wedge \Sigma]}\wedge\tr{(\delta\mathcal{A}_{01}\wedge\Sigma)}
\end{equation}
so that we obtain
\begin{equation}
    \delta I_{\text{KS}}[\beta;\Omega] = \frac{1}{g_{\text{s}}^2}\int_{P_6}\tr{[(\overbar{\partial}\mathcal{A}_{01} + \mathcal{A}_{01}\wedge \mathcal{A}_{01})\wedge \Sigma\wedge\Sigma]}\wedge \delta\,\tr{( \mathcal{A}_{01}\wedge \Sigma)} +(\text{bdy. terms}) \,.
    \label{eq:CS_action_variation}
\end{equation}
The second factor in the wedge product is equal to $\delta\,\tr{( \mathcal{A}_{01}\wedge \Sigma)} = -\delta\eta(\beta)$ by \eqref{eq:eta_beta}, where we neglect $\delta K $ as it is a shift symmetry once appropriate boundary terms are included as discussed above. For the first factor, we note that
\begin{equation}
    \tr{(\overbar{\partial}\mathcal{A}_{01}\wedge \Sigma\wedge\Sigma)} = \overbar{\partial}\,\tr{(\mathcal{A}_{01}\wedge \Sigma\wedge\Sigma)} = \overbar{\partial}\eta(\alpha)
\end{equation}
since $\overbar{\partial}\Sigma = 0$. Similarly, using $\partial\Sigma = -\Sigma\wedge\Sigma$, we obtain
\begin{equation}
    \tr{(\mathcal{A}_{01}\wedge  \mathcal{A}_{01}\wedge\Sigma\wedge\Sigma)} = -\partial\,\tr{(\mathcal{A}_{01}\wedge  \mathcal{A}_{01}\wedge\Sigma)} +2\,\tr{(\partial\mathcal{A}_{01}\wedge  \mathcal{A}_{01}\wedge\Sigma)}\,.
\end{equation}
Using the locality constraint \eqref{eq:A_cohomology}, the second term here becomes $2\,\tr{(\mathcal{A}_{01}\wedge  \mathcal{A}_{01}\wedge\Sigma\wedge\Sigma)}$. Thus we obtain
\begin{equation}
    \tr{(\mathcal{A}_{01}\wedge  \mathcal{A}_{01}\wedge\Sigma\wedge\Sigma)} = \partial\,\tr{(\mathcal{A}_{01}\wedge  \mathcal{A}_{01}\wedge\Sigma)} = \frac{1}{2}\,\partial\eta(\alpha\wedge\alpha)\,,
\end{equation}
where we used \eqref{eq:eta_alpha_alpha_Sigma}. Hence we find
\begin{equation}
    \tr{[(\overbar{\partial}\mathcal{A}_{01} + \mathcal{A}_{01}\wedge \mathcal{A}_{01})\wedge \Sigma\wedge\Sigma]} = \overbar{\partial}\eta(\alpha)+\frac{1}{2}\,\partial\eta(\alpha\wedge \alpha)
    \label{eq:KS_eq_A}
\end{equation}
so that the variation \eqref{eq:CS_action_variation} of the hCS action indeed coincides with the variation \eqref{eq:KS_action_variation} of the KS action obtained earlier. Since $\Sigma^i_j$ is invertible, it follows from \eqref{eq:KS_eq_A} that $\overbar{\partial}\mathcal{A}_{01} + \mathcal{A}_{01}\wedge \mathcal{A}_{01} = 0$ is equivalent to the KS equation (see Appendix~\ref{subapp:equivalence_hcs_KS}).

\paragraph{Equations of motion.} For general variations $\delta\mathcal{A}_{01}$, the equation of motion of hCS theory obtained from \eqref{eq:hCS_variation} is given by
\begin{equation}
    (F_{\mathcal{B}})_{02} =\overbar{\partial}\mathcal{A}_{01} + \mathcal{A}_{01}\wedge \mathcal{A}_{01} = 0\,,
    \label{eq:hCS_equation}
\end{equation}
with $\mathcal{B} = \Sigma + \iota_\alpha\Sigma$. However, since the variation $\delta\mathcal{A}_{01}$ is restricted by the locality constraint \eqref{eq:A_cohomology}, we do not get the full equation but a weaker equation, which is then supplemented with the constraint. This is because the constrained field space is smaller than the full field space. To derive the weaker equation, we impose the constraint via a Lagrange multiplier $Y$, which is an $\mathfrak{sl}(2,\mathbb{C})$-valued $(2,2)$-form, and consider the modified action
\begin{equation}
    \tilde{I}_{\text{KS}}[\beta;\Omega,Y] = I_{\text{KS}}[\beta;\Omega] - \frac{1}{g_{\text{s}}^2}\int_{P_6}\tr{[Y\wedge (F_{\mathcal{B}})_{11}]}\,.
\end{equation}
Varying the modified action with respect to $\mathcal{A}_{01}$ gives the weaker equation
\begin{equation}
    \Omega\wedge (\overbar{\partial}\mathcal{A}_{01} + \mathcal{A}_{01}\wedge \mathcal{A}_{01}) = \partial Y - Y\wedge\Sigma + \Sigma\wedge Y\,,
\end{equation}
while varying with respect to $Y$ imposes the locality constraint. We can also write the weaker version of the KS equation. Using the identity
\begin{equation}
    \Omega\wedge (\overbar{\partial}\mathcal{A}_{01} + \mathcal{A}_{01}\wedge \mathcal{A}_{01}) = \Sigma\wedge \tr{[(\overbar{\partial}\mathcal{A}_{01} + \mathcal{A}_{01}\wedge \mathcal{A}_{01})\wedge \Sigma\wedge\Sigma]}\,,
\end{equation}
we obtain
\begin{equation}
     \Sigma\wedge\biggl(\overbar{\partial}\eta(\alpha)+\frac{1}{2}\,\partial\eta(\alpha\wedge \alpha)\biggr)= \partial Y - Y\wedge\Sigma + \Sigma\wedge Y\,.
\end{equation}
Solutions of the original KS equation that also satisfy the locality constraint remain solutions of the constrained system with $Y=0$, while additional solutions with $Y\neq 0$ may exist.

\paragraph{Solutions in the constrained field space.} We are interested in solutions to the KS equation $(F_{\mathcal{B}})_{02}=0$ that also satisfy the locality constraint $(F_{\mathcal{B}})_{11}=0$ which renders the KS action local. It is not immediately obvious that such solutions exist, so we now construct them explicitly. First notice that the constraint \eqref{eq:A_cohomology} reduces to
\begin{equation}
    \partial\widetilde{\mathcal{A}}_{01} = M\,(F_{\mathcal{B}})_{11} M^{-1} = 0\,,
    \label{eq:constraint_tilde}
\end{equation}
where we have defined
\begin{equation}
\widetilde{\mathcal{A}}_{01} \equiv M\,\mathcal{A}_{01}\,M^{-1} \,.
\end{equation}
Then, $(F_{\mathcal{B}})_{02} =  0$ implies
\begin{equation}
    \overbar{\partial}\widetilde{\mathcal{A}}_{01} + \widetilde{\mathcal{A}}_{01}\wedge \widetilde{\mathcal{A}}_{01} = M\,(F_{\mathcal{B}})_{02}\,M^{-1} = 0\,.
    \label{eq:tilde_EOM}
\end{equation}
Locally, a class of solutions of \eqref{eq:constraint_tilde} and \eqref{eq:tilde_EOM} is given by
\begin{equation}
    \widetilde{\mathcal{A}}_{01} = H^{-1}\overbar{\partial}H\,,\quad \partial H = 0\,,
\end{equation}
with $H= H(\overbar{w}) \in \mathrm{SL}(2,\mathbb{C})$. Thus a general local solution $\mathcal{A}_{01}$ that also solves the equation of motion and the locality constraint is of the form
\begin{equation}
    \mathcal{A}_{01} = M^{-1}\,(H^{-1}\overbar{\partial}H)\,M\,,\quad \overbar{\partial} M = \partial H  = 0
    \label{eq:general_constraint_onshell}
\end{equation}
so there are infinitely many such solutions.

\section{Dimensional reduction to 3D gravity}\label{sec:3D_grav_red}

In this section, we show how the dimensional reduction of 6D KS gravity on $P_6 = \mathbb{H}_3\times S^3$ over the $S^3$ produces 3D gravity on the base $\mathbb{H}_3$. First, we demonstrate that a single KS action produces the action of chiral 3D gravity coupled to a flat background gauge field. Second, we show that including two copies of the KS action, together with a background volume term, dimensionally reduces to the full 3D Einstein action. Our approach is based on the above rewriting of 6D KS gravity as local 6D holomorphic CS theory whose dimensional reduction is straightforward.

\subsection{Reduction of the 6D Kodaira--Spencer action}\label{subapp:reduction_action}

We perform the dimensional reduction in parts by first reducing the background structures of KS gravity, then the complex structure deformation and last the action.

\paragraph{Reduction of the background.} Let us first dimensionally reduce the background 1-form $\Sigma$ defined in \eqref{eq:Sigma}. Substituting the polar decomposition $M = SG$, we obtain
\begin{equation}
    \Sigma = M^{-1}\,\d M = G^{-1}\,(A_0 +\Theta)\,G\,,
    \label{eq:Sigma_reduction}
\end{equation}
where $A_0= S^{-1}\d S$ is a flat 3D gauge field on the base\footnote{Strictly speaking, it is the pull-back of a 3D gauge field with the projection map of the bundle, but we leave the pull-back implicit.} $\mathbb{H}_3$ and the right-invariant Maurer--Cartan form $\Theta$ is given by
\begin{equation}
    \Theta =\Theta^i\,\tau_i = \d G\,G^{-1}\,.
    \label{eq:right_MC}
\end{equation}
Thus dimensional reduction of the background volume form $\Omega$ of 6D KS gravity on $\mathbb{H}_3\times S^3$ produces, through the reduction of $\Sigma$, a flat background 3D gauge field $A_0$ on $\mathbb{H}_3$. Similarly, $\overbar{\Sigma} = -\Sigma^\dagger$ is given by
\begin{equation}
    \overbar{\Sigma} = \overbar{M}^{-1}\d \overbar{M}\,,\quad \overbar{M} = (M^\dagger)^{-1}\,.
\end{equation}
From $M = SG$ it follows that $\overbar{M} = S^{-1}G$ and we obtain
\begin{equation}
    \overbar{\Sigma} = G^{-1}\,(\overbar{A}_0 +\Theta)\,G\,,
    \label{eq:Sigma_bar_reduction}
\end{equation}
where $\overbar{A}_0 \equiv -S\,A_0\,S^{-1}$; see Appendix~\ref{subapp:dim_red_Sigmas} for details. Note that $\overbar{A}_0 = -A_0^\dagger$ which implies $\overbar{A}_0^i = (A_0^i)^*$.

\paragraph{Reduction of the complex structure deformation.} Consider the 6D KS action on $\mathbb{H}_3\times S^3$. To dimensionally reduce it over the $S^3$, we start with its reformulation as 6D holomorphic CS theory \eqref{eq:KS_as_hCS} valid when $\alpha$ satisfies the constraint \eqref{eq:A_cohomology} written in terms of $\mathcal{A}_{01}$. This constraint makes the KS action local and equal to a holomorphic CS action making dimensional reduction to 3D Einstein gravity possible.

The holomorphic CS action is invariant under the addition of a $(1,0)$-form to $\mathcal{A}_{01}$ since the wedge product with $\Omega$ projects out any $(1,0)$-component. Thus if we define
\begin{equation}\label{eq:A_param}
    \mathcal{A} = B_i\,\Sigma^i + \iota_\alpha\Sigma\,,
\end{equation}
where $B_i = B_i^j\,\tau_j$ are three $\mathfrak{sl}(2,\mathbb{C})$-valued scalar fields, the KS action \eqref{eq:KS_as_hCS} becomes
\begin{equation}
    I_{\text{KS}}[\beta;\Omega] = -\frac{1}{g_{\text{s}}^2}\int_{P_6}\Omega\wedge\mathrm{CS}[\mathcal{A}]\,,
    \label{eq:KS_action_CS_action}
\end{equation}
where the Chern--Simons 3-form is given by
\begin{equation}
    \text{CS}[\mathcal{A}]\equiv \frac{1}{2}\,\tr{\biggl(\mathcal{A}\wedge \d \mathcal{A}+\frac{2}{3}\,\mathcal{A}\wedge \mathcal{A}\wedge \mathcal{A}\biggr)}\,.
    \label{eq:CS_3_form}
\end{equation}
Thus instead of dimensionally reducing $\alpha$ directly, we will reduce $\mathcal{A}$.

To this end, we introduce a basis $(\d y^\mu,\Theta^i)$ for 6D 1-forms, where $y^\mu$ with $\mu = 1,2,3$ are coordinates on the base $\mathbb{H}_3$, and $\Theta^i$ are $\mathfrak{su}(2)$-components of $\Theta$. Note that $\d y^\mu$ is horizontal, $\d y^\mu(\tau_i^{\#}) = 0$. In this basis, we can decompose a general $\mathfrak{sl}(2,\mathbb{C})$-valued 1-form as
\begin{equation}
    \mathcal{A} = G^{-1}\,(A + \Phi_i\,\Theta^i)\,G\,,
\end{equation}
where $A = A_\mu(x)\,\d y^\mu$ is a horizontal $\mathfrak{sl}(2,\mathbb{C})$-valued 1-form, $\Phi_i = \Phi^j_i(x)\,\tau_j$ are three $\mathfrak{sl}(2,\mathbb{C})$-valued scalars and $G = G(x)\in \mathrm{SU}(2)$ is defined by the polar decomposition $M = SG$ of the background. By horizontality of $\d y^\mu$, the scalars $\Phi_i$ determine the action of $\mathcal{A}$ on the fundamental vector field $\tau_i^{\#}$. However, since the KS action is independent of the scalars $B_i$, we are free to fix them in a way that makes the dimensional reduction simple.\footnote{Note the 1-form $\mathcal{B} =\Sigma + \iota_\alpha\Sigma$ introduced in Section~\ref{subsec:hCS_rewriting} does not admit a dimensional reduction as a connection $\mathcal{B}\neq G^{-1}\,b\, G + G^{-1}\d G$ with a horizontal 1-form $b$ when $\alpha\neq 0$, because imposing $\mathcal{B}(\tau_i^{\#}) = \tau_i$ implies $\alpha = 0$. To see this, we use $\Sigma(\tau_i^{\#}) = \tau_i$ to obtain $ (\iota_\alpha\Sigma)(\tau_i^{\#})  = 0$. Now since $\overbar{\Sigma}^i(\tau_j^{\#}) = \delta^i_j$, this implies $C_i = 0$ for $C_i$ defined below in \eqref{eq:C_equation}. Thus $\alpha = 0$. It is only $\mathcal{A} = B_i\Sigma^i + \iota_\alpha\Sigma$ which admits a dimensional reduction as a connection.} In particular, we fix $B_i$ such that $\Phi_i = \tau_i$ which implies $\mathcal{A}(\tau_i^{\#}) = \tau_i$. To do this, we write $\mathcal{A}_{01}$ in the basis of $(0,1)$-forms $\overbar{\Sigma}^i$ as
\begin{equation}
    \mathcal{A}_{01} = \iota_\alpha\Sigma = C_i\,\overbar{\Sigma}^i\,,
    \label{eq:C_equation}
\end{equation}
where the $\mathfrak{sl}(2,\mathbb{C})$-valued scalars $C_i = C^j_i\,\tau_j $ are given by
\begin{equation}
    C_i^j = \Sigma_k^j\,\overbar{\sigma}^{\overbar{l}}_i\,\alpha_{\overbar{l}}^{\;\;\,k}\,,
    \label{eq:C_scalar}
\end{equation}
where $\overbar{\sigma}^{\overbar{j}}_i$ is the inverse of $\overbar{\Sigma}^i_{\overbar{j}}$ satisfying $\overbar{\sigma}^{\overbar{k}}_i\,\overbar{\Sigma}^j_{\overbar{k}} = \delta^j_i$. From the formulae \eqref{eq:Sigma_reduction} and \eqref{eq:Sigma_bar_reduction} it follows together with $A_0(\tau_i^{\#}) = \overbar{A}_0(\tau_i^{\#}) = 0$ that $\Sigma^i(\tau_j^{\#}) = \overbar{\Sigma}^i(\tau_j^{\#}) = \delta^i_j$. Hence we obtain
\begin{equation}
    \mathcal{A}_{01}(\tau_{i}^{\#}) = C_i
\end{equation}
so that $\Phi_i = \tau_i$ corresponds to $B_i = \tau_i - C_i $. With this choice, we obtain the simple decomposition into horizontal and vertical parts
\begin{equation}
    \mathcal{A} = G^{-1}\,(A + \Theta)\,G = G^{-1}AG + G^{-1}\d G\,.
    \label{eq:A_dim_red}
\end{equation}
In general, the horizontal components $A_\mu = A_\mu(y,\theta)$ depend on both the base $y^\mu$ and fiber coordinates $\theta^i$. We restrict to the zero-mode of $A$ by imposing that $\mathcal{A}$ is $\mathrm{SU}(2)$-equivariant
\begin{equation}
    \mathcal{R}^*_u\,\mathcal{A} = u^{-1}\mathcal{A}\,  u\,,
    \label{eq:A_equivariance}
\end{equation}
where the diffeomorphism $\mathcal{R}_u\colon P_6\rightarrow P_6$ is defined by $\mathcal{R}_u(p) = pu$ with $p\in \mathrm{SL}(2,\mathbb{C})$ and constant $u\in \mathrm{SU}(2)$. The right-action with a constant $u\in \mathrm{SU}(2)$ preserves horizontal 1-forms $\mathcal{R}^*_u\d y^\mu = \d y^\mu$, and since $G\circ \mathcal{R}_u = Gu$, it follows that the Maurer--Cartan forms are also preserved $\mathcal{R}^*_u\Theta^i = \Theta^i$. Thus equivariance \eqref{eq:A_equivariance} implies that $A_\mu(y,\theta) = A_\mu(y)$ is independent of the $S^3$-directions and $A$ is a true 3D gauge field on $\mathbb{H}_3$.

We can obtain an explicit formula for the 3D gauge field $A$ in terms of $\alpha$ as follows. By using $B_i = \tau_i - C_i$, we obtain
\begin{equation}
    \mathcal{A} = \Sigma - C_i\,(\Sigma^i - \overbar{\Sigma}^i)\,,
\end{equation}
where the second term is horizontal since $(\Sigma^i - \overbar{\Sigma}^i)(\tau_j^{\#}) = 0$. Thus equivariance \eqref{eq:A_equivariance} of $\mathcal{A}$ requires
\begin{equation}
    \mathcal{R}_u^* [C_i\,(\Sigma^i - \overbar{\Sigma}^i)] = u^{-1}\,C_i\,(\Sigma^i - \overbar{\Sigma}^i)\,u\,,
    \label{eq:equivariance_horizontal_part}
\end{equation}
which also imposes a constraint on the transformation of $C_i$ under right-action since $\Sigma$ and $\overbar{\Sigma}$ are equivariant themselves. From \eqref{eq:equivariance_horizontal_part} it follows that the 3D gauge field $A$ in \eqref{eq:A_dim_red} becomes (see Appendix~\ref{subapp:dim_red_Sigmas} for details)
\begin{equation}
    A = A_0 - c_i\,(A_0^i - \overbar{A}_0^i)\,.
    \label{eq:A_explicit}
\end{equation}
where $c_i = C_i\circ S$ is the pull-back of $C_i$ to the base $\mathbb{H}_3$ using the section $S = \sqrt{MM^\dagger}$. This is the explicit relation between the dimensionally reduced 3D gauge field $A$ and the 6D complex structure deformation $\alpha$.

\paragraph{Reduction of the KS action.} We will dimensionally reduce the local hCS form of the KS action using the horizontal-vertical decompositions \eqref{eq:Sigma_reduction} and \eqref{eq:A_dim_red} of $\Sigma$ and $\mathcal{A}$ respectively. In Appendix~\ref{subapp:CS_reduction}, we show that the dimensional reduction of the Lagrangian of 6D hCS theory is given by
\begin{equation}
   \Omega\wedge \text{CS}[\mathcal{A}]  = 4\,\epsilon_{S^3}\wedge(\mathrm{CS}[A] -\mathrm{CS}[A_0]) -2\,\d [\epsilon_{S^3}\wedge\tr{(A_0\wedge A)}]\,,
    \label{eq:hCS_reduction_1}
\end{equation}
where $\mathrm{CS}[A]$ is the CS form of the three-dimensional gauge field $A$ involving the 3D exterior derivative (since $A$ depends only on the base coordinate by equivariance), we have used that the background is flat $F_{A_0} = G\,F_\Sigma\,G^{-1} = 0 $ and defined the volume form $\epsilon_{S^3}$ on the $S^3$ as
\begin{equation}
    \epsilon_{S^3} = \frac{1}{12}\,\tr{(\Theta\wedge \Theta\wedge \Theta)}\,,\quad \vol{(S^3)} \equiv \int_{S^3}\epsilon_{S^3} = 2\pi^2\,.
\end{equation}
Substituting \eqref{eq:hCS_reduction_1} into \eqref{eq:KS_action_CS_action}, we obtain for the action
\begin{equation}
    I_{\text{KS}}[\beta;\Omega] = \frac{4\vol{(S^3)}}{g_{\text{s}}^2}\int_{\mathbb{H}_3}(\mathrm{CS}[A] -\mathrm{CS}[A_0]) + \frac{2\vol{(S^3)}}{g_{\text{s}}^2}\int_{\partial \mathbb{H}_3}\tr{(A_0\wedge A)}\,,
    \label{eq:KS_as_3D_CS}
\end{equation}
where there is a sign flip in the bulk term due to our choice of orientation.\footnote{We have already fixed the orientation of $P_6$ to be the one induced by the background $J$ and here we now fix the orientation of $\mathbb{H}_3$ such that $\int_{P_6}\epsilon_{S^3}\wedge C_3 = -\int_{S^3}\epsilon_{S^3}\,\int_{\mathbb{H}_3}C_3$, where $C_3$ is a horizontal 3-form. This implies Stokes' theorem with a plus sign $\int_{P_6} \d(\epsilon_{S^3}\wedge C_2) = -\int_{P_6} \epsilon_{S^3}\wedge \d C_2 = \int_{S^3}\epsilon_{S^3}\,\int_{\partial \mathbb{H}_3} C_2$, where $C_2$ is a horizontal 2-form.\label{footnote:orientation}} This is a background dependent version of chiral 3D gravity with a negative cosmological constant~\cite{Li:2008dq} for an $\mathfrak{sl}(2,\mathbb{C})$-valued 1-form $A$ with $A_0$ being the background.

The action \eqref{eq:KS_action_CS_action} may further be written in the following more suggestive way. By writing $A = A_0 + \Delta A$, where $\Delta A \equiv A - A_0$, we obtain
\begin{equation}
    \mathrm{CS}[A] - \mathrm{CS}[A_0] = \mathrm{CS}[\Delta A; A_0] - \frac{1}{2}\,\d \,\tr{(A_0\wedge A)}\,,
\end{equation}
where we used $F_{A_0} = 0$, defined the relative CS 3-form
\begin{equation}
    \mathrm{CS}[\Delta A; A_0] \equiv \frac{1}{2}\, \tr{\biggl(
\Delta A \wedge \d_{A_0} (\Delta A)+
\frac{2}{3}\,
\Delta A \wedge \Delta A \wedge \Delta A
\biggr)}
\end{equation}
and the gauge covariant derivative $\d_{A_0} = \d + [A_0,\cdot\,]$ relative to the background $A_0 = S^{-1}\d S $. Thus we can also write the hCS Lagrangian \eqref{eq:hCS_reduction_1} as (see also Appendix~\ref{subapp:CS_reduction})
\begin{equation}
   \Omega\wedge \text{CS}[\mathcal{A}]  = 4\,\epsilon_{S^3}\wedge\mathrm{CS}[\Delta A; A_0]\,.
    \label{eq:hCS_reduction_relative}
\end{equation}
Substituting into \eqref{eq:KS_action_CS_action}, we obtain the alternative form
\begin{equation}
    I_{\text{KS}}[\beta;\Omega] = \frac{2\vol{(S^3)}}{g_{\text{s}}^2}\int_{\mathbb{H}_3}\tr{\biggl(
\Delta A \wedge \d_{A_0} (\Delta A)+
\frac{2}{3}\,
\Delta A \wedge \Delta A \wedge \Delta A
\biggr)}
\label{eq:background_CS_action}
\end{equation}
without any boundary terms. From \eqref{eq:A_explicit} we find explicitly in terms of the 6D complex structure deformation
\begin{equation}
    \Delta A^i = -c_j^i\,(S^{-1}\d S + \d S S^{-1})^j\,,
\end{equation}
where the 3D scalar field $c^i_j = C^i_j\circ S$ is the pull-back of the 6D scalar field $C^i_j = \Sigma_k^i\,\overbar{\sigma}^{\overbar{l}}_j\,\alpha_{\overbar{l}}^{\;\;\,k} $ onto the $\mathbb{H}_3$ base with the section $S = \sqrt{MM^\dagger}$, see Appendix~\ref{subapp:dim_red_Sigmas} for a more detailed discussion.

\paragraph{3D Einstein's equations.} The unconstrained variation of the 3D chiral gravity action \eqref{eq:background_CS_action} produces the equation of motion
\begin{equation}
    \d_{A_0}(\Delta A) + \Delta A\wedge \Delta A = 0\,,
\end{equation}
which by $F_{A_0} = 0$ reduces to
\begin{equation}
    F_A = \d A + A\wedge A = 0\,,
    \label{eq:F_A_0}
\end{equation}
whose real and imaginary parts are equivalent to 3D Einstein's equations when the metric and the connection are independent fields: the curvature equation and the torsion-free condition. However, since the 6D $(0,1)$-form $\mathcal{A}_{01}$ must satisfy the locality constraint \eqref{eq:A_cohomology} in order for the KS action to be local, the 3D gauge field is also restricted. Thus the variation $\delta A$ is constrained and $A$ satisfies a weaker equation than \eqref{eq:F_A_0}. This reflects the fact that variations $\delta\mathcal{A}_{01}$ obeying the locality constraint \eqref{eq:A_cohomology} give a weaker equation than the usual hCS equation as discussed in Section~\ref{subsec:constrained_variation}. Regardless, solutions of 3D Einstein's equations \eqref{eq:F_A_0} are still solutions of the weaker equation.

Now solutions of the 3D Einstein's equations $F_A = 0$ uplift to solutions of the 6D KS equation. This follows from the fact that by the dimensional reduction formula \eqref{eq:A_dim_red}, we have
\begin{equation}
    F_{\mathcal{A}} = G^{-1}\,F_{A}\,G \, ,
    \label{eq:FA_reduction}
\end{equation}
such that $F_{\mathcal{A}}$ is horizontal. Now $\Sigma^i-\overbar{\Sigma}^i$ form a basis of horizontal 1-forms so that we can expand
\begin{equation}
    F_{\mathcal{A}} = \frac{1}{2}\,G^{-1}\,(F_{A})_{ij}\,G\,(\Sigma^i-\overbar{\Sigma}^i)\wedge(\Sigma^j-\overbar{\Sigma}^j)\,.
\end{equation}
Since $(\Sigma^i-\overbar{\Sigma}^i)_{01} = -\overbar{\Sigma}^i$, we obtain
\begin{equation}
    (F_{\mathcal{A}})_{02} =\frac{1}{2}\,G^{-1}\,(F_{A})_{ij}\,G\,\overbar{\Sigma}^i\wedge\overbar{\Sigma}^j\,.
    \label{eq:F02_F_A}
\end{equation}
Thus $F_A = 0$ is equivalent to $(F_{\mathcal{A}})_{02} = 0$, which is equivalent to the KS equation via \eqref{eq:KS_eq_A}. Thus we have the equivalence
\begin{equation}
    F_A = 0\quad \Leftrightarrow \quad \overbar{\partial}\eta(\alpha)+\frac{1}{2}\,\partial\eta(\alpha\wedge \alpha) = 0
\end{equation}
between 3D Einstein's equations and 6D KS equation for complex structure deformations which satisfy the locality constraint and which are dimensionally reducible via an equivariant connection $\mathcal{A}$.

The question is whether there are solutions of chiral 3D gravity which uplift to $\mathcal{A}_{01}$ which also satisfy the 6D locality constraint. The answer is yes. To see this, consider a 3D solution
\begin{equation}
    A = q^{-1}\,dq\,,
    \label{eq:explicit_3D_solutions}
\end{equation}
where $q = q(y)\in \mathrm{SL}(2,\mathbb{C})$. Its uplift to 6D is given by
\begin{equation}
    \mathcal{A} = L^{-1}\,\d L\,,\quad L = q G
\end{equation}
and its $(0,1)$-component is
\begin{equation}
    \mathcal{A}_{01} = L^{-1}\,\overbar{\partial} L\,.
\end{equation}
When $\overbar{\partial}\mathcal{A}_{01} + \mathcal{A}_{01}\wedge \mathcal{A}_{01} = 0$, the general solution to the locality constraint \eqref{eq:A_cohomology} is given by \eqref{eq:general_constraint_onshell}. Thus the uplift satisfies the locality constraint \eqref{eq:A_cohomology} if and only if $L$ factorizes as
\begin{equation}
    L = QHM\,,\quad \overbar{\partial} Q = \overbar{\partial} M  = 0\,,\quad \partial H = 0\,,
\end{equation}
where $M(x)\in \mathrm{SL}(2,\mathbb{C})$ is defined by the background $\Sigma = M^{-1}\partial M$ and $Q(x),H(x)\in \mathrm{SL}(2,\mathbb{C})$. Indeed, in this case we obtain the uplift
\begin{equation}
    \mathcal{A}_{01} = M^{-1} H^{-1}Q^{-1}\,\overbar{\partial}(QHM) = M^{-1} (H^{-1}\overbar{\partial}H)\,M\,.
\end{equation}
Thus $q = QH S$, with $QH$ right-invariant, defines an explicit family of 3D gravity solutions \eqref{eq:explicit_3D_solutions} which uplift to an $\mathcal{A}_{01}$ satisfying $\overbar{\partial}\mathcal{A}_{01} + \mathcal{A}_{01}\wedge \mathcal{A}_{01}$ and the locality constraint.

\subsection{3D Einstein gravity from two copies of 6D Kodaira--Spencer gravity}\label{sec:full_Einstein_reduce}

Let $y^\mu$ with $\mu = 1,2,3$ be coordinates on $\mathbb{H}_3$ and $\hat{y}^{\hat{\mu}}$ with $\hat{\mu} = 1,2$ be coordinates on the boundary $\partial \mathbb{H}_3$. The 3D Einstein gravity in the Einstein--Palatini formulation in the Euclidean signature is described by the action
\begin{equation}
    I_{\text{EP}}[g,\Gamma] = -\frac{1}{16\pi G_{\text{N}}}\int_{\mathbb{H}_3}d^3y\sqrt{g}\,(R+2) - \frac{1}{8\pi G_{\text{N}}}\int_{\partial \mathbb{H}_3}d^2\hat{y}\sqrt{\gamma}\,(K-1)\,,
    \label{eq:Einstein_Palatini}
\end{equation}
where $ G_{\text{N}}$ is the 3D Newton's constant, $g_{\mu\nu}$ is the 3D metric, $R^{\rho}_{\;\;\sigma\mu\nu}=\partial_\mu\Gamma^\rho_{\nu\sigma}-\partial_\nu\Gamma^\rho_{\mu\sigma}+\Gamma^\rho_{\mu \lambda}\Gamma^\lambda_{\nu\sigma}-\Gamma^\rho_{\nu \lambda}\Gamma^\lambda_{\mu\sigma}$ is the Riemann tensor of the connection $\Gamma$ which is treated as an independent degree of freedom from the metric, $\gamma_{\hat{\mu}\hat{\nu}}$ is the induced metric of the boundary and $K = \gamma^{\hat{\mu}\hat{\nu}}K_{\hat{\mu}\hat{\nu}} $ is the trace of the extrinsic curvature $K_{\hat{\mu}\hat{\nu}} = \partial_{\hat{\mu}}Q^\rho\,\partial_{\hat{\nu}}Q^\sigma\,\nabla_{\rho}n_\sigma\vert_{\partial \mathbb{H}_3}$ of $\partial \mathbb{H}_3$ with $\nabla_{\mu}n_\nu = \partial_\mu n_\nu - \Gamma^\rho_{\mu\nu}\,n_\rho$ being the covariant derivative of the unit normalized outward-pointing normal vector $n^\mu$ associated to the connection $\Gamma$ and $Q^\mu$ being the embedding of $\partial \mathbb{H}_3$. When $\Gamma = \mathring{\Gamma}(g)$ is the Levi--Civita connection determined by $g$, the action reduces to the 3D Einstein action $I_{\text{EP}}[g,\mathring{\Gamma}(g)] = I_{\text{E}}[g]$.

We introduce an invertible dreibein $E^i_\mu$, with inverse $E^\mu_i$, and the spin connection $(\omega^i_{\;\; j})_\mu$ in terms of which
\begin{equation}
    g_{\mu\nu} = \delta_{ij}\, E^i_{\mu}E^j_{\nu} \,,\quad \Gamma^\mu_{\nu\rho} = E^\mu_i\,E^j_\rho\,(\omega^i_{\;\;\,j})_\nu + E^\mu_i\,\partial_\nu E^i_\rho\,.
    \label{eq:g_Gamma}
\end{equation}
We package them into $\mathfrak{su}(2)$-valued 1-forms as
\begin{equation}
    E = E^i \tau_i, \qquad W = W^i \tau_i,
    \qquad W^i \equiv -\frac{1}{2}\varepsilon^{i}{}_{jk}\,\omega^{jk}\,,\quad \omega^i_{\;\; j} = -\varepsilon^i{}_{jk}\,W^k\,.
    \label{eq:E_W_defs}
\end{equation}
The normal vector can also be packaged into an $\mathfrak{su}(2)$-valued scalar field $N\equiv n^i\tau_i$ where $n^i \equiv E^i_\mu\, n^\mu$. With these definitions, the Einstein--Palatini action \eqref{eq:Einstein_Palatini} takes the form (see Appendix~\ref{app:3D_Einstein_CS_form})
\begin{align}
    I_{\text{EP}}[g,\Gamma] &= -\frac{2}{8\pi G_{\text{N}}}\int_{\mathbb{H}_3}\tr{\biggl(E\wedge F_{W} - \frac{1}{3}\,E\wedge E\wedge E\biggr)}\nonumber\\
    &\quad+\frac{4}{8\pi G_{\text{N}}}\int_{\partial \mathbb{H}_3}\tr{(NE\wedge \d_{W} N)}-\frac{2}{8\pi G_{\text{N}}}\int_{\partial \mathbb{H}_3}\tr{(NE\wedge E)}\,.
    \label{eq:Einstein_Palatini_action}
\end{align}
By introducing the $\mathfrak{sl}(2,\mathbb{C})$-valued gauge field
\begin{equation}
    A = W + iE\,,\quad \overbar{A} = W - iE\,,
\end{equation}
the action takes the form
\begin{align}
    I_{\text{EP}}[g,\Gamma]
    &= \frac{i}{8\pi G_{\text{N}}}\int_{\mathbb{H}_3}(\text{CS}[A]-\text{CS}[\overbar{A}])\label{eq:EP_CS}\\
    &+ \frac{i}{8\pi G_{\text{N}}}\int_{\partial \mathbb{H}_3}\tr{\biggl(\frac{1}{2}\,A\wedge \overbar{A} -2\, N\,(A-\overbar{A})\wedge \d N -\frac{i}{2}\, N\,(A-\overbar{A})\wedge(A-\overbar{A})\biggr)}\,.\nonumber
\end{align}
Local $\mathrm{SO}(3)$ rotations of the frame translate to $\mathrm{SU}(2)$ gauge transformations
\begin{equation}
    A\rightarrow U^{-1}AU + U^{-1}\d U\,,\quad N\rightarrow U^{-1}NU\,,
\end{equation}
under which the action \eqref{eq:EP_CS} is invariant. There is a gauge transformation which only acts near the boundary $\partial \mathbb{H}_3$ and sets the normal vector to a constant $N\vert_{\partial \mathbb{H}_3} = \tau_3$. In this case, $\d N = 0$ and boundary terms simplify to the ones derived in~\cite{Ebert:2022cle}.

\paragraph{Two copies of KS gravity.} Let us now consider two copies of 6D KS gravity. We consider the action
\begin{equation}
    I[\beta,\overbar{\beta};\Omega,\overbar{\Omega}] \equiv -\frac{i}{2}\,\bigl(I_{\text{KS}}[\beta;\Omega]-I_{\text{KS}}[\overbar{\beta};\overbar{\Omega}]\bigr)+\frac{i}{4 g_{\text{s}}^2}\int_{P_6} \Omega\wedge \overbar{\Omega}\,,
    \label{eq:two_copies_of_KS}
\end{equation}
where $\overbar{\beta} = \beta^*$ is the complex conjugate polyvector and the KS action is defined in \eqref{eq:KS_action_CS_action}. The complex conjugate field $\overbar{\beta}$ defines the deformation $\overbar{\alpha} = \alpha^*$ as $\overbar{\eta}(\overbar{\alpha}) = \overbar{\partial}\overbar{\eta}(\overbar{\beta})$ where $\overbar{\eta}$ is the map from polyvectors to forms defined by contractions of $\overbar{\Omega}$. The 6D volume involving the volume form $\Omega\wedge\overbar{\Omega}$ of the background is added to cancel the dependence on the background $A_0$ arising from the dimensional reduction of the KS actions.

Now the KS action reduces to the 3D CS action as in \eqref{eq:KS_as_3D_CS} and produces a CS 3-form $\mathrm{CS}[A_0]$ of the background $A_0$. The reduction of the volume form is performed in Appendix~\ref{subapp:volume_reduction} and given by
\begin{equation}
    \Omega\wedge\overbar{\Omega} = 8\,\epsilon_{S^3}\wedge(\mathrm{CS}[A_0]- \mathrm{CS}[\overbar{A}_0]) +4\,\d[\epsilon_{S^3}\wedge\tr{(A_0\wedge \overbar{A}_0)}]\,,
    \label{eq:volume_via_CS}
\end{equation}
where we have used that the background gauge field is flat $F_{A_0} = 0$. Thus we obtain
\begin{equation}
    \frac{i}{4 g_{\text{s}}^2}\int_{P_6} \Omega\wedge \overbar{\Omega} = -\frac{2i\vol{(S^3)}}{ g_{\text{s}}^2}\int_{\mathbb{H}_3}(\mathrm{CS}[A_0]- \mathrm{CS}[\overbar{A}_0]) +\frac{i\vol{(S^3)}}{ g_{\text{s}}^2}\int_{\partial \mathbb{H}_3}\tr{(A_0\wedge \overbar{A}_0)}\,.
\end{equation}
The $\mathrm{CS}[A_0]$ and $\mathrm{CS}[\overbar{A}_0]$ terms coming from the volume form are canceled by the same contributions coming from the KS actions. Thus substituting into \eqref{eq:two_copies_of_KS} together with \eqref{eq:KS_as_3D_CS}, we obtain the dimensional reduction
\begin{align}
    I[\beta,\overbar{\beta};\Omega,\overbar{\Omega}] &= -\frac{2i\vol{(S^3)}}{ g_{\text{s}}^2}\int_{\mathbb{H}_3} (\text{CS}[A]-\text{CS}[\overbar{A}])\nonumber\\
    &\quad +\frac{i\vol{(S^3)}}{ g_{\text{s}}^2}\int_{\partial \mathbb{H}_3}\tr{(A\wedge A_0-\overbar{A}\wedge \overbar{A}_0+A_0\wedge\overbar{A}_0)}\,,
\end{align}
which is proportional to the 3D Einstein action \eqref{eq:EP_CS}. We see that all dependence on the background gauge field $A_0$ coming from the 6D KS actions is canceled by the 6D volume and pushed to the boundary terms. Comparing the coefficients, we find that the 3D Newton's constant is given by the coefficient of the 6D KS action as
\begin{equation}
    G_{\text{N}} = -\frac{g_{\text{s}}^2}{16\pi\vol{(S^3)}}= -\frac{g_{\text{s}}^2}{32\pi^3}\,,
    \label{eq:GN_gs}
\end{equation}
which is negative under our choices of orientations and signs for the actions. With this choice, we obtain the explicit relation
\begin{align}
I[\beta,\overbar{\beta};\Omega,\overbar{\Omega}]
= I_{\mathrm{EP}}[g,\Gamma] + I_{\mathrm{bdy}}\,,
\end{align}
where the boundary term is
\begin{align}
I_{\mathrm{bdy}}
=
\frac{i\vol{(S^3)}}{g_{\mathrm{s}}^2}
\int_{\partial\mathbb{H}_3}
&\tr{\biggl(
A\wedge A_0-\overbar{A}\wedge\overbar{A}_0
+A_0\wedge\overbar{A}_0+A\wedge\overbar{A}\biggr.}
\nonumber\\
&\qquad\quad\biggl.-4N(A-\overbar{A})\wedge\d N
-iN\,(A-\overbar{A})\wedge(A-\overbar{A})
\biggr)\,.
\end{align}
In order for these boundary terms to cancel, $I[\beta,\overbar{\beta};\Omega,\overbar{\Omega}]$ must be supplemented with 6D boundary terms which give them via dimensional reduction. These boundary terms are not important for the analysis of this paper and we leave their study for future work.

\paragraph{Brown--Henneaux formula for KS gravity.} The relative minus sign in \eqref{eq:GN_gs} can appear strange, but it is natural, because the dual theory has actually a negative central charge. It is known that 3D Einstein gravity describes a dual CFT with the Brown--Henneaux central charges
\begin{equation}
    c = \overbar{c} = \frac{3}{2G_{\text{N}}}\,,
\end{equation}
where $c$ and $\overbar{c}$ are the central charges of the chiral and anti-chiral Virasoro algebras respectively. We obtained 3D Einstein gravity as the dimensional reduction of two 6D KS gravities with the Newton's constant given by \eqref{eq:GN_gs}. Thus we find the relation
\begin{equation}
    c = \overbar{c} = -\frac{24\pi\vol{(S^3)}}{g_{\text{s}}^2} = -\frac{48\pi^3}{g_{\text{s}}^2}\,.
    \label{eq:BH_KS}
\end{equation}
On the other hand, two complex-conjugate copies of 6D KS gravities are holographically dual to two commuting copies of the chiral $\mathrm{U}(N)$ beta-gamma system via twisted holography~\cite{Costello:2018zrm} with central charges $c = \overbar{c} = -3\,(N^2-1)$~\cite{Beem:2013sza}. Substituting into \eqref{eq:BH_KS}, we obtain in the large-$N$ limit
\begin{equation}
    g_{\text{s}}^2 = \frac{16\pi^3}{N^2} + \mathcal{O}(N^{-4})\,,\quad N\rightarrow \infty\,.
\end{equation}

\paragraph{On-shell action of the background.} 

Let us compute the on-shell action of the undeformed background on $P_6=\mathbb{H}_3\times S^3$ without complex structure deformations $\alpha = \overbar{\alpha} = 0$. We assume that the conformal boundary $\partial\mathbb{H}_3$ is a torus $\partial \mathbb{H}_3 = S^1\times S^1_\beta$.  In three dimensions, this corresponds to the Euclidean version of empty $\text{AdS}_3$ with torus boundary and computes the vacuum contribution to the thermal trace of the dual 2D CFT at inverse temperature $\beta$ in the large-$N$ limit. Since we have two copies of KS gravity in this setup, the dual CFT is no longer chiral and is expected to involve two commuting holomorphic and antiholomorphic beta-gamma CFTs of~\cite{Beem:2013sza,Costello:2018zrm}.

In this case, the 6D action \eqref{eq:two_copies_of_KS} is simply
\begin{equation}
    I[0,0;\Omega,\overbar{\Omega}] =\frac{i}{4 g_{\text{s}}^2}\int_{P_6} \Omega\wedge \overbar{\Omega}\,.
\end{equation}
In Appendix~\ref{subapp:volume_reduction}, we show that by defining $ E_0 = \frac{1}{2i}\,(A_0 - \overbar{A}_0) $, the dimensional reduction of the background volume form can be written as (see also~\cite{Herfray:2016std})
\begin{equation}
    \Omega\wedge\overline{\Omega}=\frac{32 i}{3}\,\epsilon_{S^3}\wedge \tr{(E_0\wedge E_0\wedge E_0)}
    \label{eq:volume_conifold}
\end{equation}
so that the action becomes
\begin{equation}
    I[0,0;\Omega,\overbar{\Omega}] =\frac{8\vol{(S^3)}}{ 3g_{\text{s}}^2}\int_{\mathbb{H}_3} \tr{(E_0\wedge E_0\wedge E_0)}\,,
    \label{eq:I_0_0_reduction}
\end{equation}
where an additional minus sign comes from our orientation convention (see footnote~\ref{footnote:orientation}).

For our calculation here, we will write the dimensional reduction of the background as
\begin{equation}
    \Sigma = \widetilde{G}^{-1}\,(\widetilde{A}_0 + \widetilde{\Theta})\,\widetilde{G}\,,
\end{equation}
where $\widetilde{G}(y,\theta) = U(y)^{-1}\,G(\theta)$, with $U(y) \in \mathrm{SU}(2)$, and $\widetilde{A}_0 = \widetilde{S}^{-1}\d \widetilde{S} $, with $\widetilde{S} = SU$, is a flat 3D gauge field related to $A_0 = S^{-1}\d S$ by a 3D gauge transformation
\begin{equation}
    \widetilde{A}_0 = U^{-1}\,A_0\,U + U^{-1}\d U\,.
\end{equation}
The 6D action is diffeomorphism invariant, and since 3D gauge transformations descend from certain 6D diffeomorphisms, the dimensionally reduced 3D action is gauge invariant. Thus we are free to compute the on-shell action in any gauge and we choose
\begin{equation}
    \widetilde{S}(\rho,\phi,\tau) = e^{i\tau\tau_2}\,e^{i\rho\,(\cos{\phi}\,\tau_1+\sin{\phi}\,\tau_3)}\,,
\end{equation}
where $\rho \geq 0$, $\phi \sim \phi + 2\pi$ and $\tau \sim \tau + \beta$ are global coordinates of $\mathbb{H}_3$ with the conformal boundary at $\rho = \infty$. This yields the flat 3D gauge field
\begin{equation}
\begin{aligned}
    \widetilde{A}^1_0
    &=i\cos\phi\,\d\rho
      -\sinh\rho\,\sin\phi\,(\d\tau+i\,\d\phi)\,, \\
    \widetilde{A}^2_0
    &=i\cosh\rho\,\d\tau+(1-\cosh\rho)\d\phi\,, \\
    \widetilde{A}^3_0
    &=i\sin\phi\,\d\rho
      +\sinh\rho\,\cos\phi\,(\d\tau+i\,\d\phi)\,,
\end{aligned}
\end{equation}
which vanishes at the origin $\rho = 0$ and is therefore regular there. Decomposing
\begin{equation}
    \widetilde{A}_0 = \widetilde{W}_0 + i \widetilde{E}_0\,,\quad \overbar{\widetilde{A}}_0 = \widetilde{W}_0 - i \widetilde{E}_0\,,
\end{equation}
where $\overbar{\widetilde{A}}_0 = (\widetilde{A}_0^i)^*\,\tau_i$, we obtain the dreibein
\begin{equation}
\begin{aligned}
\widetilde{E}^1_0
&=\cos\phi\,\d\rho
  -\sinh\rho\,\sin\phi\,\d\phi\,, \\
\widetilde{E}^2_0
&=\cosh\rho\,\d\tau\,, \\
\widetilde{E}^3_0
&=\sin\phi\,\d\rho
  +\sinh\rho\,\cos\phi\,\d\phi\,,
  \label{eq:regular_dreibein}
\end{aligned}
\end{equation}
so that the corresponding 3D metric $g_{\mu\nu}\,\d y^\mu \d y^\nu=\delta_{ij}\,\widetilde{E}_0^i\widetilde{E}_0^j$ takes the form
\begin{equation}
    g_{\mu\nu}\,\d x^\mu \d x^\nu = \d\rho^2+ \cosh^2{\rho}\,\d \tau^2 + \sinh^2{\rho}\,\d \phi^2  \,,
    \label{eq:3D_metric}
\end{equation}
which is the standard hyperbolic metric on $\mathbb{H}_3$ in global coordinates. The same metric is obtained from $A_0 = W_0 + iE_0 $ since the metric is gauge invariant. Using \eqref{eq:regular_dreibein}, we obtain
\begin{align}
    \tr{(E_0\wedge E_0\wedge E_0)} =  \tr{(\widetilde{E}_0\wedge \widetilde{E}_0\wedge \widetilde{E}_0)}=-\frac{3}{2}\cosh\rho\,\sinh\rho \,\d\rho\wedge \d\tau\wedge \d\phi\,.
\end{align}
Substituting into the action \eqref{eq:I_0_0_reduction}, we find
\begin{equation}
    I[0,0;\Omega,\overbar{\Omega}] = -\frac{4\pi\vol{(S^3)}}{ g_{\text{s}}^2}\,\beta\sinh^2{\rho_{\text{c}}}\,,
\end{equation}
where $\rho_{\text{c}}\rightarrow \infty$ is an IR cut-off. Using the Brown--Henneaux formula \eqref{eq:BH_KS} for KS gravity, we obtain
\begin{equation}
    I[0,0;\Omega,\overbar{\Omega}] = \frac{c\beta}{24}\,(e^{2\rho_{\text{c}}} -2 + \mathcal{O}(e^{-2\rho_{\text{c}}}))\,,\quad \rho_{\text{c}}\rightarrow \infty\,.
\end{equation}
The finite term
\begin{equation}
    I_{\text{ren}}[0,0;\Omega,\overbar{\Omega}] = -\frac{c \beta}{12}
\end{equation}
gives the correct Casimir energy $-\frac{c}{12}$ of the dual beta-gamma system on a spatial unit circle $S^1$ via $\tr{e^{-\beta \mathcal{H}}} = e^{-I_{\text{ren}}} = e^{\frac{c \beta}{12}}$, where $\mathcal{H}$ is the Hamiltonian of the CFT. Note that the Casimir energy is negative in the beta-gamma CFT since the central charge is negative.

\section{Discussion}\label{sec:discussion}

In this work, we have identified a subsector of 6D KS gravity on $\mathbb{H}_3\times S^3$ described by a local holomorphic CS action with gauge group $\mathrm{SL}(2,\mathbb{C})$ and which dimensionally reduces to the action of 3D gravity on $\mathbb{H}_3$. This embeds 3D gravity into twisted holography and topological string theory opening multitude possibilities for future study. In this section, we discuss subtleties with our results and possible future directions.

\paragraph{Rendering the KS action local.} The KS action is a non-local action for the complex structure deformation $\alpha$ with the cohomology condition $\partial\eta(\alpha) = 0$ ensuring that the non-local operator $\frac{1}{\partial}$ is well defined. We discovered a subsector of KS gravity whose action is local in $\alpha$. In particular, the action is given by a subsector of holomorphic CS theory. Both subsectors are defined by the locality constraint \eqref{eq:A_cohomology}, which is slightly stronger than the cohomology condition. Rendering the KS action local is necessary in order to identify a local 3D gravity subsector from dimensional reduction as in Section~\ref{sec:3D_grav_red}.

Even though the dimensional reduction of $\alpha$ works irrespectively of the locality constraint \eqref{eq:A_cohomology}, the unfortunate fact of \eqref{eq:A_cohomology} is that it also constrains the 3D gravity gauge field $A$. Thus, the field space of zero-modes of complex structure deformations $\alpha$, for which the KS action is local, does not coincide with the full field space of 3D gravity gauge fields $A$. Within our construction, the 6D KS action of a deformation $\alpha$ that violates the locality constraint is not related to the local 3D gravity action of the corresponding 3D gauge field $A$. It is an interesting question how much of the dual beta-gamma CFT is captured by this local subspace of KS gravity in twisted holography. We will return to this question in future work.

\paragraph{Kaluza--Klein modes and path integrals.} We have focused on equivariant 6D 1-forms $\mathcal{A}$ which restrict to the zero-mode of the horizontal gauge field $A$. This is a restriction on the 6D complex structure deformation $\alpha$ making it dimensionally reducible. These higher KK modes play an important role in UV completion of 3D gravity via twisted holography. In future work, we will study KK modes and derive their associated 3D actions.

Furthermore, it will be very interesting to extend the relationship of 3D gravity and 6D KS gravity to the quantum level and study the relations between the path integrals of 6D KS gravity, 6D holomorphic CS theory, chiral 3D gravity and full 3D Einstein gravity. The quantum extension is particularly important when comparing with other approaches towards a quantization of 3D gravity with a negative cosmological constant~\cite{Cotler:2018zff,Collier:2023fwi}. The quantum extension is complicated because of the fact that the full off-shell field spaces of the theories are not equal, but only a restricted subsector is, as discussed above. This might help uncover a multitude of subtleties with the path integral of 3D gravity in the presence of black holes~\cite{Bac:2026eqj}. We plan to return to these issues in future work.

\paragraph{Connection to Hitchin gravity and related theories.} In~\cite{Herfray:2016std}, the action of 3D Einstein gravity in the pure connection formulation is obtained as the dimensional reduction of 6D Hitchin gravity. Similarly to KS gravity, Hitchin gravity can be interpreted as a theory of deformations of a holomorphic volume form $\Omega$, but the origin of the deformation is different from the one induced by a complex structure deformation in \eqref{eq:volume_form_deformation}. It will be interesting to relate our 6D action \eqref{eq:two_copies_of_KS}, which reduces to 3D Einstein gravity, to the Hitchin gravity reduction in~\cite{Herfray:2016std}. Another closely related theory, introduced in~\cite{Costello:2019jsy} and recently given an action formulation in~\cite{Boffo:2025jpx}, describes arbitrary closed deformations of $\Omega$. Moreover, these theories might be related to a 3D real analog of KS gravity derived in~\cite{Ben-Shahar:2026dgh} which also has a 6D origin. We leave a systematic comparison of our work to these formulations for future work.

\paragraph{Relation to our earlier work.} In our previous work~\cite{Erdmenger:2025lvv}, it was shown that a $\mathfrak{sl}(2,\mathbb{C})$-valued 1-form $\mathcal{A}$ and its conjugate $\overbar{\mathcal{A}}$ as well as their dual vector fields define an almost complex structure $J'$. Then $\mathcal{A}$ is a $(1,0)$-form with respect to $J'$. Setting $\mathcal{\overbar{A}}=\overbar{\Sigma}$, one might expect that $J'$ is equal to the almost complex structure $J_\alpha$ defined in \eqref{eq:J_alpha}. However, this is not true, since $\mathcal{A}$ is not a $(1,0)$-form with respect to $J_\alpha$. Thus, the almost complex structure $J_\alpha$ considered in the present work is a priori different from $J'$ considered in~\cite{Erdmenger:2025lvv}, however, they are related. Using the parametrization $\mathcal{A} = \Sigma'+\iota_{\alpha'}\Sigma'$ of~\cite{Erdmenger:2025lvv}, which is different from \eqref{eq:A_param}, we observe that $J' = J_{\alpha'} = e^{\iota_{\alpha'}} J e^{-\iota_{\alpha'}}$. Here, $\Sigma'$ is a general $\mathfrak{sl}(2,\mathbb{C})$-valued $(1,0)$-form with respect to the background, which is not in general holomorphic $\overbar{\partial} \Sigma'\neq 0$. In order for the dual vector fields to exists, the components $\Sigma' = {\Sigma'}_i^j\,\tau_j\, \d w^i$ have to form an invertible matrix, $\det \Sigma'^j_i \neq 0$.

Comparing the parametrization of $\mathcal{A}$ in~\cite{Erdmenger:2025lvv} with \eqref{eq:A_param}, we find the relations
\begin{equation}
    \Sigma'^i_j = B_k^i\,\Sigma^k_j  \,,\quad \alpha'^{\;\;\,j}_{\overbar{i}} = (B^{-1})^k_l\,\sigma_k^j\,\Sigma^l_n\,\alpha^{\;\;\,n}_{\overbar{i}}\,,\quad \alpha^{\;\;\,j}_{\overbar{i}} = (B'^{-1})^{\overbar{k}}_{\overbar{i}}\,\alpha'^{\;\;\,j}_{\overbar{k}}\,.
    \label{eq:primed_from_nonprimes}
\end{equation}
Here, we use that that $B^i_j = \delta^i_j - \Sigma_k^i\,\overbar{\sigma}^{\overbar{l}}_j\,\alpha_{\overbar{l}}^{\;\;\,k} $ is invertible, which follows from $\det \Sigma'^j_i\neq 0$ and define $B'^{\overbar{i}}_{\overbar{j}} \equiv \delta^{\overbar{i}}_{\overbar{j}} + \Sigma_l^k\,\overbar{\sigma}^{\overbar{i}}_k\,\alpha'^{\;\;\,l}_{\overbar{j}} $, which is also invertible. Thus we see that the complex structure deformation $\alpha'$ in~\cite{Erdmenger:2025lvv} is a non-linear function of the deformation $\alpha$ considered in this paper. In Appendix~\ref{app:flatness_integrability}, we improve on~\cite{Erdmenger:2025lvv} and show that the KS equation for $\alpha'$ is precisely the $(0,2)$-component of flatness $F_{\mathcal{A}} = 0$ even when $\overbar{\partial}\Sigma' \neq 0$. If $\mathcal{A}$ is $\mathrm{SU}(2)$-equivariant, flatness is equivalent to $(F_{\mathcal{A}})_{02} = 0$. As discussed in \eqref{eq:KS_eq_A}, this condition is also equivalent to the KS equation for $\alpha$. Thus, even though $\alpha$ and $\alpha'$ are complicated non-linear functions of one another, they simultaneously satisfy the KS equation if and only if the $(0,2)$-component $(F_{\mathcal{A}})_{02} = 0$ of the curvature of the connection $\mathcal{A} $ vanishes.

\acknowledgments{We would like to thank Jakob Hollweck and Jitendra Pal for useful discussions and comments. We acknowledge financial support by the Deutsche Forschungsgemeinschaft (DFG, German Research Foundation) through the German-Israeli Project Cooperation (DIP) grant ‘Holography and the Swampland’, as well as under Germany’s Excellence Strategy through the W\"{u}rzburg-Dresden Cluster of Excellence on Complexity and Topology in Quantum Matter - ct.qmat (EXC 2147, project-id 390858490). Several OpenAI language models were used to assist with calculations throughout this project. In particular, GPT-5.6 Sol was used to verify the calculations presented in the final version of the manuscript. All outputs were independently reviewed and validated by the authors, who take full responsibility for the accuracy and integrity of this work.}

\appendix

\section{Conventions and Tian's lemma}\label{app:conventions}

We write $(p,q)$-forms as
\begin{equation}
    C = \frac{1}{p!q!}\,C_{i_1\ldots i_p \overbar{j}_1\ldots \overbar{j}_q}\,\d w^{i_1}\wedge \cdots \wedge \d w^{i_p}\wedge \d \overbar{w}^{j_1}\wedge \cdots \wedge \d \overbar{w}^{j_q}\,.
\end{equation}
For a $(p,q)$-polyvectors, we use the convention
\begin{equation}
    C = \frac{1}{p!q!}\,C_{\overbar{i}_1\ldots \overbar{i}_q}^{\quad\quad j_1\ldots j_p}\;\d \overbar{w}^{\overbar{i}_1}\wedge \cdots \wedge \d \overbar{w}^{\overbar{i}_q}\otimes \partial_{j_1}\wedge \cdots\wedge \partial_{j_p}\,.
\end{equation}
The Dolbeault operators act as
\begin{equation}
    \partial C = \frac{1}{p!q!}\,\partial_k C_{i_1\ldots i_p \overbar{j}_1\ldots \overbar{j}_q}\,\d w^k\wedge \d w^{i_1}\wedge \cdots \wedge \d w^{i_p}\wedge \d \overbar{w}^{\overbar{j}_1}\wedge \cdots \wedge \d \overbar{w}^{\overbar{j}_q}
\end{equation}
and as
\begin{equation}
     \overbar{\partial} C = \frac{1}{p!q!}\,\overbar{\partial}_{\overbar{k}}C_{i_1\ldots i_p \overbar{j}_1\ldots \overbar{j}_q}\,\d\overbar{w}^{\overbar{k}}\wedge \d w^{i_1}\wedge \cdots \wedge \d w^{i_p}\wedge \d \overbar{w}^{\overbar{j}_1}\wedge \cdots \wedge \d \overbar{w}^{\overbar{j}_q}\,.
\end{equation}
In components, we obtain
\begin{equation}
    (\partial C)_{ki_1\ldots i_p \overbar{j}_1\ldots \overbar{j}_q} = (p+1)\,\partial_{[k}C_{i_1\ldots i_p \overbar{j}_1\ldots \overbar{j}_q]}\,,\quad (\overbar{\partial} C)_{i_1\ldots i_p \overbar{k}\overbar{j}_1\ldots \overbar{j}_q} = (-1)^p(q+1)\,\overbar{\partial}_{[\overbar{k}}C_{i_1\ldots i_p \overbar{j}_1\ldots \overbar{j}_q]}\,,
\end{equation}
where the factor $(-1)^p$ comes from commuting $\d\overbar{w}^{\overbar{k}}$ to the right. 

\paragraph{Wedge products.} For a $(q,m)$-polyvector $\zeta$ and a $(p,l)$-polyvector $\kappa$, we define the wedge product by
\begin{equation}
    \kappa  \wedge\zeta  = \kappa  ^{i_1\dots i_p} \wedge \zeta^{j_1\dots j_q} \otimes \partial_{i_1}\wedge\dots \wedge\partial_{i_p}\wedge\partial_{j_1}\wedge\dots \wedge \partial_{j_q}\,.
\end{equation}
Here, we have defined $\zeta  =\zeta ^{i_1\dots i_q}\otimes \partial_{i_1}\wedge\dots \partial_{i_q}$ and similarly for $\kappa$. Since the wedge product for both the vector and the form parts are antisymmetric, we find
\begin{equation}
    \kappa   \wedge \zeta  = (-1)^{qp+ml} \zeta \wedge \kappa \,.
\end{equation}

\paragraph{Contractions.} For a $(q,m)$-polyvector $\zeta$, we define its contraction with a form $C$ by
\begin{equation}
    \iota_\zeta C = \zeta^{i_1\dots i_q} \wedge\iota_{\partial_{i_1}\wedge\dots\wedge \partial_{i_q}} C\,,
\end{equation}
where the contraction with a multivector is defined as $\iota_{\partial_{j_1}\wedge \cdots \wedge\partial_{j_q}}\equiv \iota_{\partial_{j_q}}\cdots \iota_{\partial_{j_1}}$ and the contraction with a vector field $V$ is defined by $\iota_{V}\d w^i = \d w^i(V)$ and $\iota_{V}\d \overbar{w}^{\overbar{i}} = \d \overbar{w}^{\overbar{i}}(V)$. Contraction with a vector $V$ satisfies the Leibniz rule
\begin{equation}
    \iota_V(C\wedge D) = \iota_V C \wedge D+(-1)^{\vert C \vert} C \wedge \iota_V D\,.
\end{equation}
For a $(q,m)$-polyvector $\zeta $ and a $(p,l)$-polyvector $\kappa$, by explicit calculation, we find
\begin{equation}
    \iota_\kappa   \iota_\zeta = 
(-1)^{(q+m)(p+l)}  \iota_\zeta \iota_\kappa\,.
\label{eq:iota_commutation}
\end{equation}
For a $(1,s)$-polyvector $\gamma$, there is the identity
\begin{equation}
    \iota_\gamma(A\wedge B) = \iota_\gamma A\wedge B + (-1)^{(s+1)\vert A\vert}\,A\wedge \iota_\gamma B\,.
\end{equation}

\paragraph{Nijenhuis bracket.} Let $\alpha$ and $\gamma$ be two $(1,1)$-polyvectors
\begin{equation}
    \alpha = \alpha_{\overbar{i}}^{\;\;j}\, \d\overbar{w}^{\overbar{i}}\otimes \partial_j \,,\quad \gamma = \gamma_{\overbar{i}}^{\;\;j}\, \d\overbar{w}^{\overbar{i}}\otimes \partial_j\,.
\end{equation}
The Nijenhuis bracket is defined as
\begin{equation}
    [\alpha,\gamma] \equiv \d \overbar{w}^{\overbar{i}}\wedge \d \overbar{w}^{\overbar{j}} \otimes   [\alpha_{\overbar{i}},\gamma_{\overbar{j}}] = \frac{1}{2}\,[\alpha,\gamma]_{\overbar{i}\overbar{j}}^k\;\d \overbar{w}^{\overbar{i}}\wedge \d \overbar{w}^{\overbar{j}}\otimes \partial_k\,,
    \label{eq:Nijenhuis_bracket}
\end{equation}
where $\alpha_{\overbar{i}} = \alpha_{\overbar{i}}^{\;\;j}\partial_j$ and $\gamma_{\overbar{i}} = \gamma_{\overbar{i}}^{\;\;j}\partial_j$ are $(1,0)$-vector fields, $[\alpha_{\overbar{i}},\gamma_{\overbar{j}}]$ is their Lie bracket and 
\begin{equation}
    [\alpha,\gamma]_{\overbar{i}\overbar{j}}^k = [\alpha_{\overbar{i}},\gamma_{\overbar{j}}]^k+[\gamma_{\overbar{i}},\alpha_{\overbar{j}}]^k = \alpha_{\overbar{i}}^{\;\;l}\,\partial_l \gamma_{\overbar{j}}^{\;\;k} - \alpha_{\overbar{j}}^{\;\;l}\,\partial_l \gamma_{\overbar{i}}^{\;\;k}+\gamma_{\overbar{i}}^{\;\;l}\,\partial_l \alpha_{\overbar{j}}^{\;\;k} - \gamma_{\overbar{j}}^{\;\;l}\,\partial_l \alpha_{\overbar{i}}^{\;\;k}\,.
\end{equation}
When $\gamma=\alpha$ this reduces to twice the Lie bracket of vector fields
\begin{equation}
    [\alpha,\alpha]_{\overbar{i}\overbar{j}}^k = 2\,[\alpha_{\overbar{i}},\alpha_{\overbar{j}}]^k = 2\,(\alpha_{\overbar{i}}^{\;\;l}\,\partial_l \alpha_{\overbar{j}}^{\;\;k} - \alpha_{\overbar{j}}^{\;\;l}\,\partial_l \alpha_{\overbar{i}}^{\;\;k})\,.
    \label{eq:twice_the_lie_bracket}
\end{equation}
The contraction with the Nijenhuis bracket is given by
\begin{equation}
    \iota_{[\alpha,\gamma]}= \d\overbar{w}^{\overbar{i}}\wedge \d \overbar{w}^{\overbar{j}} \wedge   \iota_{[\alpha_{\overbar{i}},\gamma_{\overbar{j}}]} \,.
    \label{eq:Nijenhuis_contraction}
\end{equation}
Here contraction with the Lie bracket is given by
\begin{equation}
    \iota_{[\alpha_{\overbar{i}},\gamma_{\overbar{j}}]} = [\pounds_{\alpha_{\overbar{i}}},\iota_{\gamma_{\overbar{j}}}] = [\{\d, \iota_{\alpha_{\overbar{i}}}\},\iota_{\gamma_{\overbar{j}}}] = [\{\partial, \iota_{\alpha_{\overbar{i}}}\},\iota_{\gamma_{\overbar{j}}}]+[\{\overbar{\partial}, \iota_{\alpha_{\overbar{i}}}\},\iota_{\gamma_{\overbar{j}}}]\,,
    \label{eq:Lie_bracket_contraction}
\end{equation}
where we used the Cartan formula for the Lie derivative $\pounds_X = \{\d, \iota_{X}\}$, $\d = \partial + \overbar{\partial}$, $[\,,]$ denotes the commutator and $\{\,,\}$ the anticommutator. Now here we have
\begin{equation}
    \{\overbar{\partial}, \iota_{\alpha_{\overbar{i}}}\}=\overbar{\partial} (\alpha_{\overbar{i}}^{\;\;\,j}\,\iota_{\partial_j}) + \alpha_{\overbar{i}}^{\;\;\,j}\,\iota_{\partial_j}\overbar{\partial} = \overbar{\partial} \alpha_{\overbar{i}}^{\;\;\,j}\wedge \iota_{\partial_j}+\alpha_{\overbar{i}}^{\;\;\,j}\,\{\overbar{\partial},\iota_{\partial_j}\}\,.
\end{equation}
Since $\{\overbar{\partial},\iota_{\partial_j}\} = 0$, we obtain
\begin{equation}
    \{\overbar{\partial}, \iota_{\alpha_{\overbar{i}}}\}=\overbar{\partial} \alpha_{\overbar{i}}^{\;\;\,j}\wedge \iota_{\partial_j}\,.
\end{equation}
Thus we find that the second term in \eqref{eq:Lie_bracket_contraction} vanishes
\begin{equation}
    [\{\overbar{\partial}, \iota_{\alpha_{\overbar{i}}}\},\iota_{\gamma_{\overbar{j}}}]  = \overbar{\partial} \alpha_{\overbar{i}}^{\;\;\,k}\wedge \iota_{\partial_k}\iota_{\gamma_{\overbar{j}}}-\iota_{\gamma_{\overbar{j}}}(\overbar{\partial}\alpha_{\overbar{i}}^{\;\;\,k}\wedge \iota_{\partial_k}) =  \overbar{\partial} \alpha_{\overbar{i}}^{\;\;\,k}\wedge (\iota_{\partial_k}\iota_{\gamma_{\overbar{j}}}+\iota_{\gamma_{\overbar{j}}}\iota_{\partial_k}) = 0\,,
\end{equation}
where we used $\iota_{\gamma_{\overbar{j}}}\overbar{\partial}\alpha_{\overbar{i}}^{\;\;\,k} = 0$ since $\gamma_{\overbar{j}}$ is a $(1,0)$-vector field and that contractions with two vector fields anticommute. Thus we are left with
\begin{equation}
    \iota_{[\alpha_{\overbar{i}},\gamma_{\overbar{j}}]} = [\{\partial, \iota_{\alpha_{\overbar{i}}}\},\iota_{\gamma_{\overbar{j}}}] \,.
\end{equation}
Substituting to \eqref{eq:Nijenhuis_contraction}, we obtain
\begin{equation}
    \iota_{[\alpha,\gamma]}= \d\overbar{w}^{\overbar{i}}\wedge \d \overbar{w}^{\overbar{j}} \wedge   [\{\partial, \iota_{\alpha_{\overbar{i}}}\},\iota_{\gamma_{\overbar{j}}}] = \d\overbar{w}^{\overbar{i}}\wedge \d \overbar{w}^{\overbar{j}} \wedge (\partial\iota_{\alpha_{\overbar{i}}}\iota_{\gamma_{\overbar{j}}} -\iota_{\gamma_{\overbar{j}}}\partial\iota_{\alpha_{\overbar{i}}}+\iota_{\alpha_{\overbar{i}}}\partial\iota_{\gamma_{\overbar{j}}} -\iota_{\gamma_{\overbar{j}}}\iota_{\alpha_{\overbar{i}}}\partial) \,.
\end{equation}
In the first term, we have to use
\begin{equation}
    \d\overbar{w}^{\overbar{i}}\wedge\d\overbar{w}^{\overbar{j}}\wedge \partial \iota_{\alpha_{\overbar{i}}}\iota_{\gamma_{\overbar{j}}}C = \partial (\d\overbar{w}^{\overbar{i}}\wedge\d\overbar{w}^{\overbar{j}}\wedge \iota_{\alpha_{\overbar{i}}}\iota_{\gamma_{\overbar{j}}}C) = -\partial (\d\overbar{w}^{\overbar{i}}\wedge\iota_{\alpha_{\overbar{i}}}(\d\overbar{w}^{\overbar{j}}\wedge \iota_{\gamma_{\overbar{j}}}C)) = -\partial\iota_\alpha\iota_\gamma C\,.
\end{equation}
In the second term, we write
\begin{equation}
    \d\overbar{w}^{\overbar{i}}\wedge \d \overbar{w}^{\overbar{j}} \wedge\iota_{\gamma_{\overbar{j}}}\partial\iota_{\alpha_{\overbar{i}}}C = -\d\overbar{w}^{\overbar{j}}\wedge \d \overbar{w}^{\overbar{i}} \wedge\iota_{\gamma_{\overbar{j}}}\partial\iota_{\alpha_{\overbar{i}}}C =\d\overbar{w}^{\overbar{j}}\wedge \iota_{\gamma_{\overbar{j}}}(\d \overbar{w}^{\overbar{i}} \wedge\partial\iota_{\alpha_{\overbar{i}}}C) = -\iota_\gamma\partial(\d \overbar{w}^{\overbar{i}} \wedge\iota_{\alpha_{\overbar{i}}}C)\,, 
\end{equation}
which is equal to $-\iota_\gamma\partial \iota_\alpha C$. The third term simplifies in the same way as a second term, but there is no need to swap $\d\overbar{w}^{\overbar{i}}\wedge\d\overbar{w}^{\overbar{j}} = -\d\overbar{w}^{\overbar{j}}\wedge\d\overbar{w}^{\overbar{i}}$ so that it becomes
\begin{equation}
    \d\overbar{w}^{\overbar{i}}\wedge \d \overbar{w}^{\overbar{j}} \wedge\iota_{\alpha_{\overbar{i}}}\partial\iota_{\gamma_{\overbar{j}}}C = \iota_\alpha\partial\iota_\gamma C
\end{equation}
with a plus sign. In the fourth term, we use
\begin{equation}
    \d\overbar{w}^{\overbar{i}}\wedge \d \overbar{w}^{\overbar{j}} \wedge\iota_{\gamma_{\overbar{j}}}\iota_{\alpha_{\overbar{i}}}\partial C = -\d\overbar{w}^{\overbar{j}}\wedge \d \overbar{w}^{\overbar{i}} \wedge\iota_{\gamma_{\overbar{j}}}\iota_{\alpha_{\overbar{i}}}\partial C = \d\overbar{w}^{\overbar{j}}\wedge \iota_{\gamma_{\overbar{j}}}(\d \overbar{w}^{\overbar{i}} \wedge\iota_{\alpha_{\overbar{i}}}\partial C) = \iota_\gamma\iota_\alpha \partial C\,.
\end{equation}
Thus we obtain
\begin{equation}
     \iota_{[\alpha,\gamma]}=  -\partial\iota_{\alpha}\iota_{\gamma} +\iota_{\gamma}\partial\iota_{\alpha}+\iota_{\alpha}\partial\iota_{\gamma} -\iota_{\gamma}\iota_{\alpha}\partial\,,
\end{equation}
which simplifies to the useful formula
\begin{equation}
    \iota_{[\alpha,\gamma]}= -[[\partial,\iota_\alpha],\iota_\gamma]\,.
    \label{eq:useful_nijenhuis}
\end{equation}

\paragraph{Tian's lemma.} Using the formula \eqref{eq:useful_nijenhuis}, we obtain
\begin{equation}
    \eta([\alpha,\gamma]) =  -[[\partial,\iota_\alpha],\iota_\gamma]\,\Omega = -(\partial \iota_\alpha\iota_\gamma-\iota_\gamma\partial \iota_\alpha - \iota_\alpha\partial \iota_\gamma + \iota_\gamma\iota_\alpha\partial  )\,\Omega\,.
\end{equation}
Here $\partial\Omega = 0$ since $\Omega$ is a $(3,0)$-form so that
\begin{equation}
    \eta([\alpha,\gamma]) =-\partial (\iota_\alpha\iota_\gamma\Omega)+\iota_\gamma\partial \eta(\alpha) + \iota_\alpha\partial \eta(\gamma)\,,
    \label{eq:eta_nijenhuis_midstep}
\end{equation}
which matches~\cite{kasuya2025properties}. Here we have
\begin{equation}
    \iota_\alpha\iota_\gamma\Omega = \iota_\alpha(\gamma^i\wedge \iota_{\partial_i}\Omega) = \alpha^j\wedge \iota_{\partial_j}(\gamma^i\wedge \iota_{\partial_i}\Omega) = -\alpha^j\wedge \gamma^i\wedge \iota_{\partial_j}\iota_{\partial_i}\Omega \,,
\end{equation}
where we used $\iota_\alpha\gamma^i = 0$. Using $\iota_{\partial_j}\iota_{\partial_i} = \iota_{\partial_i\wedge \partial_j}$ and that $\alpha^j\wedge \gamma^i = -\gamma^i\wedge \alpha^j$ we obtain
\begin{equation}
    \iota_\alpha\iota_\gamma\Omega = \gamma^i\wedge \alpha^j\wedge \iota_{\partial_i\wedge \partial_j}\Omega = \iota_{\gamma\wedge\alpha}\Omega = \eta(\gamma\wedge\alpha)\,.
    \label{eq:11_PV_product_contraction}
\end{equation}
Thus \eqref{eq:eta_nijenhuis_midstep} is equal to
\begin{equation}
    \eta([\alpha,\gamma]) = -\partial\eta(\gamma\wedge \alpha) +\iota_{\gamma} \partial\eta(\alpha)+\iota_{\alpha} \partial\eta(\gamma)\,,
\end{equation}
which is Tian's lemma. For $\gamma = \alpha$, this reduces to
\begin{equation}
    \frac{1}{2}\,\eta([\alpha,\alpha]) =  -\frac{1}{2}\,\partial\eta(\alpha\wedge\alpha) + \iota_\alpha\partial\eta(\alpha)\,.
    \label{eq:Tians_lemma_alpha}
\end{equation}
Note that since contractions anticommute, the opposite convention $\iota_{\partial_i\wedge \partial_j}^{\text{opp}} = \iota_{\partial_i}\iota_{\partial_j} = -\iota_{\partial_j}\iota_{\partial_i} = -\iota_{\partial_i\wedge \partial_j}$ would give a minus sign $\eta^{\text{opp}}(\alpha\wedge\alpha) = -\eta(\alpha\wedge\alpha) = -\iota_\alpha\iota_\alpha\Omega$.

\section{Variations of actions}\label{app:KS_variation}

In this appendix, we compute the variations of KS and holomorphic CS actions in detail.

\subsection{Variation of the Kodaira--Spencer action}

Let us consider the variation of the KS action assuming the cohomology condition $\partial\eta(\alpha) = 0$ so that locally $\eta(\alpha) = \partial\eta(\beta)$. Note that $\delta\eta(\alpha) = \eta(\delta\alpha) = \partial \delta\eta(\beta) = \partial\eta(\delta\beta)$ since $\delta\Omega = 0$. We obtain
\begin{align}    &\delta\biggl[\eta(\alpha)\wedge\overbar{\partial}\eta(\beta)+\frac{1}{3}\,\eta(\alpha\wedge \alpha)\wedge\eta(\alpha)\biggr]\nonumber\\
    &= \eta(\alpha)\wedge\overbar{\partial}\eta(\delta\beta)+\eta(\delta\alpha)\wedge\overbar{\partial}\eta(\beta) +\frac{1}{3}\,[2\,\eta(\delta\alpha\wedge \alpha)\wedge\eta(\alpha)+\eta(\alpha\wedge \alpha)\wedge\eta(\delta\alpha)]\,,\label{eq:first_KS_variation}
\end{align}
where we used that
\begin{equation}
    \eta(\alpha\wedge\delta\alpha) = \eta(\delta\alpha\wedge\alpha)
\end{equation}
since $\iota_\alpha\iota_{\delta \alpha} = \iota_{\delta\alpha}\iota_\alpha$ for $(1,1)$-polyvectors, see \eqref{eq:iota_commutation}.
We have
\begin{equation}
    \eta(\delta\alpha\wedge \alpha)\wedge\eta(\alpha) = \iota_{\delta\alpha\wedge \alpha}\,\Omega\wedge \iota_{\alpha}\Omega = \iota_{\alpha}\iota_{\delta\alpha}\,\Omega\wedge \iota_{\alpha}\Omega\,,
\end{equation}
where in the last equation we used \eqref{eq:11_PV_product_contraction}. Now we can use that
\begin{align}
    0=\iota_{\alpha}(\iota_{\delta\alpha}\Omega\wedge \iota_{\alpha}\Omega) &= \iota_{\alpha}\iota_{\delta\alpha}\,\Omega\wedge \iota_{\alpha}\Omega+\iota_{\delta\alpha}\,\Omega\wedge \iota_{\alpha}\iota_{\alpha}\Omega\nonumber\\
    &= \eta(\delta\alpha\wedge \alpha)\wedge\eta(\alpha)-\eta(\alpha\wedge\alpha)\wedge \eta(\delta\alpha)\,,
\end{align}
where the first equality follows from the fact that $\iota_{\delta\alpha}\Omega\wedge \iota_{\alpha}\Omega$ is a $(4,2)$-form and in the last line we used $\iota_{\delta\alpha}\,\Omega\wedge \iota_{\alpha}\iota_{\alpha}\Omega = -\iota_{\alpha}\iota_{\alpha}\Omega\wedge \iota_{\delta\alpha}\,\Omega$ since both factors have total form degree three. Thus we obtain
\begin{equation}
    \eta(\delta\alpha\wedge \alpha)\wedge\eta(\alpha) =\eta(\alpha\wedge\alpha)\wedge \eta(\delta\alpha)\,. 
\end{equation}
Substituting into \eqref{eq:first_KS_variation}, we obtain
\begin{align}    &\delta\biggl[\eta(\alpha)\wedge\overbar{\partial}\eta(\beta)+\frac{1}{3}\,\eta(\alpha\wedge \alpha)\wedge\eta(\alpha)\biggr]\\
    &=\eta(\alpha)\wedge\overbar{\partial}\eta(\delta\beta)+\partial\eta(\delta\beta)\wedge\overbar{\partial}\eta(\beta)+\eta(\alpha\wedge \alpha)\wedge\partial\eta(\delta\beta)\,.
\end{align}
We have (note $\eta(\alpha)$ is a $(2,1)$-form of total degree three)
\begin{equation}
    \d[\eta(\alpha)\wedge\eta(\delta\beta)]  = \overbar{\partial}[\eta(\alpha)\wedge\eta(\delta\beta)] =  \overbar{\partial}\eta(\alpha)\wedge\eta(\delta\beta)-\eta(\alpha)\wedge\overbar{\partial}\eta(\delta\beta)\,,
\end{equation}
which gives
\begin{equation}
    \eta(\alpha)\wedge\overbar{\partial}\eta(\delta\beta) = \overbar{\partial}\eta(\alpha)\wedge\eta(\delta\beta)-\d[\eta(\alpha)\wedge\eta(\delta\beta)]\,.
\end{equation}
Similarly (note that $\eta(\delta\beta)$ is a $(1,1)$-form of total degree two, $\overbar{\partial}\eta(\beta)$ is a $(1,2)$-form of total degree three)
\begin{equation}
    \d[\eta(\delta\beta)\wedge\overbar{\partial}\eta(\beta)] = \partial[\eta(\delta\beta)\wedge\overbar{\partial}\eta(\beta)] = \partial\eta(\delta\beta)\wedge\overbar{\partial}\eta(\beta)+\eta(\delta\beta)\wedge\partial\overbar{\partial}\eta(\beta)\,.
\end{equation}
Here in the last term we can use $\partial\overbar{\partial}\eta(\beta) = -\overbar{\partial}\partial\eta(\beta) = -\overbar{\partial}\eta(\alpha)$
which gives
\begin{equation}
    \partial\eta(\delta\beta)\wedge\overbar{\partial}\eta(\beta) = \eta(\delta\beta)\wedge\overbar{\partial}\eta(\alpha)+\d[\overbar{\partial}\eta(\beta)\wedge \eta(\delta\beta)] = \overbar{\partial}\eta(\alpha)\wedge \eta(\delta\beta)+\d[\overbar{\partial}\eta(\beta)\wedge \eta(\delta\beta)]\,.
\end{equation}
Similarly (note that $\eta(\alpha\wedge \alpha)$ is a $(1,2)$-form of total degree three)
\begin{equation}
    \d[\eta(\alpha\wedge \alpha)\wedge\eta(\delta\beta)] =\partial[\eta(\alpha\wedge \alpha)\wedge\eta(\delta\beta)] = \partial\eta(\alpha\wedge \alpha)\wedge\eta(\delta\beta) -\eta(\alpha\wedge \alpha)\wedge\partial\eta(\delta\beta)\,,
\end{equation}
which gives
\begin{equation}
    \eta(\alpha\wedge \alpha)\wedge\partial\eta(\delta\beta) = \partial\eta(\alpha\wedge \alpha)\wedge\eta(\delta\beta)-\d[\eta(\alpha\wedge \alpha)\wedge\eta(\delta\beta)]\,.
\end{equation}
Substituting to the variation, we obtain
\begin{align}    
&\delta\biggl[\eta(\alpha)\wedge\overbar{\partial}\eta(\beta)+\frac{1}{3}\,\eta(\alpha\wedge \alpha)\wedge\eta(\alpha)\biggr]\nonumber\\
&= 2\,\biggl(\overbar{\partial}\eta(\alpha)+\frac{1}{2}\,\partial\eta(\alpha\wedge \alpha)\biggr)\wedge\eta(\delta\beta)\nonumber\\
    &+ \d[-\eta(\alpha)\wedge\eta(\delta\beta)+\overbar{\partial}\eta(\beta)\wedge \eta(\delta\beta)-\eta(\alpha\wedge \alpha)\wedge\eta(\delta\beta)]\,.
\end{align}
Thus, we see that the equations of motion are
\begin{equation}
    \overbar{\partial}\eta(\alpha)+\frac{1}{2}\,\partial\eta(\alpha\wedge \alpha) = 0\,.
\end{equation}

\subsection{Variation of the holomorphic Chern--Simons action}\label{app:CS_variation}

The variation of the
CS 3-form is
\begin{equation}
    \delta\mathrm{CS}[\mathcal A]
    =
    \tr{(F_{\mathcal A}\wedge\delta\mathcal A)}
    -\frac{1}{2}\,\d\,\tr{(\mathcal A\wedge\delta\mathcal A)}\,.
    \label{eq:delta_CS}
\end{equation}
We define the holomorphic CS 3-form as
\begin{equation}
    \mathrm{hCS}[\mathcal A_{01}]
    \equiv
    \frac{1}{2}\,\tr{\biggl(
        \mathcal{A}_{01}\wedge\overbar{\partial}\mathcal{A}_{01}
        +\frac{2}{3}\,
        \mathcal{A}_{01}\wedge\mathcal{A}_{01}\wedge\mathcal{A}_{01}
    \biggr)}\,,
\end{equation}
whose variation is
\begin{equation}
    \delta\,\mathrm{hCS}[\mathcal{A}_{01}]
    =
    \tr{[
        (\overbar{\partial}\mathcal{A}_{01}
        +\mathcal{A}_{01}\wedge\mathcal{A}_{01})
        \wedge\delta\mathcal{A}_{01}
    ]}
    -\frac{1}{2}\,\overbar{\partial}\,
        \tr{(\mathcal{A}_{01}\wedge\delta\mathcal{A}_{01})}\,.
    \label{eq:delta_hCS}
\end{equation}
which gives
\begin{equation}
    \delta(\Omega\wedge\text{CS}[\mathcal{A}]) = \Omega\wedge\tr{[(F_{\mathcal{A}})_{02}\wedge\delta\mathcal{A}_{01} ]} +\frac{1}{2}\,\d[\Omega\wedge\tr{(\mathcal{A}\wedge \delta\mathcal{A})}]\,,
    \label{eq:CS_var_midstep}
\end{equation}
where we have used $\d\Omega  = 0 = \overbar{\partial}\Omega$, that $\Omega \wedge \overbar{\partial} X = -\d (\Omega\wedge X)$ and that $\Omega$ projects to the antiholomorphic component $(F_{\mathcal{A}})_{02} = \overbar{\partial}\mathcal A_{01}
        +\mathcal A_{01}\wedge\mathcal A_{01}$. Now we notice that
\begin{align}
    \tr{[(F_{\mathcal{A}})_{02}\wedge\Sigma \wedge \Sigma]}\wedge\tr{(\delta\mathcal{A}_{01}\wedge\Sigma)} 
    &= \frac{1}{8}\,(F_{\mathcal{A}})_{02}^i\wedge\delta\mathcal{A}^m_{01}\wedge\Sigma^j\wedge\Sigma^k\wedge \Sigma^n\,\delta_{mn}\,\varepsilon_{ijk}
\end{align}
where we have moved $\delta\mathcal{A}^m_{01}$ to the left. Using $\Sigma^j\wedge\Sigma^k\wedge \Sigma^n = \Sigma^{1}\wedge\Sigma^2\wedge \Sigma^3\,\varepsilon^{jkn}$, $\varepsilon^{jkn}\varepsilon_{ijk} = 2\delta^n_i$ and $\Sigma^{1}\wedge\Sigma^2\wedge \Sigma^3 = \frac{1}{6}\,\Sigma^{i}\wedge\Sigma^j\wedge\Sigma^k\,\varepsilon_{ijk} = -\frac{2}{3}\,\tr{(\Sigma\wedge\Sigma\wedge \Sigma)}$, we obtain
\begin{equation}
    \tr{[(F_{\mathcal{A}})_{02}\wedge\Sigma \wedge \Sigma]}\wedge\tr{(\delta\mathcal{A}_{01}\wedge\Sigma)} 
    = \frac{1}{3}\,\tr{[(F_{\mathcal{A}})_{02}\wedge\delta\mathcal{A}_{01} ]}\wedge\tr{(\Sigma\wedge\Sigma\wedge \Sigma)}\,.
\end{equation}
Substituting into \eqref{eq:CS_var_midstep}, we obtain
\begin{equation}
    \delta(\Omega\wedge\text{CS}[\mathcal{A}]) = -\tr{[(F_{\mathcal{A}})_{02}\wedge\Sigma\wedge\Sigma ]}\wedge\tr{(\delta\mathcal{A}_{01}\wedge\Sigma)} +\frac{1}{2}\,\d[\tr{(\Omega\wedge\mathcal{A}\wedge \delta\mathcal{A})} ]\,,
\end{equation}
which gives the formula for the variation of the holomorphic CS action used in the main text.

\subsection{Equivalence of hCS equation and the KS equation}\label{subapp:equivalence_hcs_KS}

In the main text we showed that $\tr{[(F_{\mathcal{B}})_{02}\wedge \Sigma\wedge \Sigma]} = \eta(E_\alpha) $ where $E_\alpha = 0$ is the KS equation. Now we can expand in a basis
\begin{align}
    \tr{[(F_{\mathcal{B}})_{02}\wedge \Sigma\wedge \Sigma]} &= -\frac{1}{8}\,\varepsilon_{ijk}\,[(F_{\mathcal{A}})_{02}]_{\overbar{m}\overbar{n}}^i\, \Sigma^j_p\, \Sigma^k_q\,\d \overbar{w}^{\overbar{m}}\wedge \d \overbar{w}^{\overbar{n}}\wedge \d w^p\wedge \d w^q\,,\nonumber\\
    \eta(E_\alpha) &= \frac{1}{4}\,\eta(E_\alpha)_{pq\overbar{m}\overbar{n}}\,\d w^p\wedge \d w^q\wedge \d \overbar{w}^{\overbar{m}}\wedge \d \overbar{w}^{\overbar{n}}\,.
\end{align}
Thus in components, this equation becomes
\begin{equation}
    \varepsilon_{ijk}\, \Sigma^j_p\, \Sigma^k_q\,[(F_{\mathcal{B}})_{02}]_{\overbar{m}\overbar{n}}^i = -2\,\eta(E_\alpha)_{pq\overbar{m}\overbar{n}}\,.
\end{equation}
Since $\Sigma^i_j$ is invertible with inverse $\sigma^i_j$ as $\Sigma^i_k\,\sigma^k_j  = \delta^i_j$, we obtain
\begin{equation}
    \varepsilon_{ijk}\, [(F_{\mathcal{B}})_{02}]_{\overbar{m}\overbar{n}}^i = -2\,\sigma^p_j \sigma^q_k\,\eta(E_\alpha)_{pq\overbar{m}\overbar{n}}\,.
\end{equation}
Contracting with a Levi--Civita symbol, this simplifies to
\begin{equation}
    [(F_{\mathcal{B}})_{02}]_{\overbar{m}\overbar{n}}^i = -\varepsilon^{ijk}\,\sigma^p_j \sigma^q_k\,\eta(E_\alpha)_{pq\overbar{m}\overbar{n}}\,.
\end{equation}
Thus we see that $\eta(E_\alpha) = 0$ implies that all components of $(F_{\mathcal{B}})_{02}$, both form- and $\mathfrak{su}(2)$-components, vanish $[(F_{\mathcal{B}})_{02}]_{\overbar{m}\overbar{n}}^i = 0$. Thus, when $\mathcal{A}_{01}$ satisfies the locality constraint, the KS equation is equivalent to $(F_{\mathcal{B}})_{02} = 0$.

\section{Details of the dimensional reduction}\label{app:dim_red_details}

In this appendix, we give details on the dimensional reduction of the 6D KS action to three dimensions.

\subsection{Dimensional reduction to a 3D gauge field}\label{subapp:dim_red_Sigmas}

We have
\begin{equation}
    \mathcal{A} = B_i\,\Sigma^i + \iota_\alpha\Sigma = B_i\,\Sigma^i + C_i\overbar{\Sigma}^i\,,
\end{equation}
where the $\mathfrak{sl}(2,\mathbb{C})$-valued scalar fields are
\begin{equation}
    B_i = B^j_i\,\tau_j\,,\quad C_i = C^j_i\,\tau_j
\end{equation}
such that
\begin{equation}
    C^j_i = \Sigma_k^j\,\overbar{\sigma}^{\overbar{l}}_i\,\alpha_{\overbar{l}}^{\;\;\,k}\,.
\end{equation}
Since
\begin{equation}
    \Sigma = M^{-1}\,\d M\,,\quad \overbar{\Sigma} = \overbar{M}^{-1}\,\d \overbar{M}\,,
    \label{eq:Sigma_M_Mbar}
\end{equation}
where $M = M(w)$ and $\overbar{M} = \overbar{M}(\overbar{w})$ are holomorphic and antiholomorphic respectively. We expand
\begin{equation}
    \Sigma = \Sigma^i\tau_i\,,\quad \overbar{\Sigma}= \overbar{\Sigma}^i\tau_i\,,
\end{equation}
where by definition
\begin{equation}
    \overbar{\Sigma}^i = (\Sigma^i)^*\,.
\end{equation}
Since $\tau_i^\dagger = -\tau_i$, we obtain
\begin{equation}
    \overbar{\Sigma} = -\Sigma^\dagger\,.
\end{equation}
Substituting \eqref{eq:Sigma_M_Mbar}, we obtain
\begin{equation}
    \overbar{M}^{-1}\,\d \overbar{M} =-\d M^\dagger\,(M^{-1})^\dagger = -\d M^\dagger\,(M^\dagger)^{-1} = M^\dagger\,\d (M^\dagger)^{-1} \equiv \widetilde{M}^{-1}\,\d \widetilde{M}\,,
\end{equation}
where $\widetilde{M}\equiv (M^\dagger)^{-1}$ and we used $0 = \d[M^\dagger\,(M^\dagger)^{-1} ]$. Thus it follows that
\begin{equation}
    \overbar{M} = b\,\widetilde{M} = b\,(M^\dagger)^{-1}\,,
    \label{eq:M_bar}
\end{equation}
where $b$ is a constant $\mathrm{SL}(2,\mathbb{C})$ matrix that we set to one as it drops out from $\overbar{\Sigma}$. Now we have the polar decomposition $M = SG$ and obtain
\begin{equation}
     \Sigma = G^{-1}\,(A_0+\Theta)\,G\,,\quad A_0 \equiv S^{-1}\d S\,.
     \label{eq:Sigma_red_app}
\end{equation}
Similarly, we obtain from the polar decompositon of $M$ that
\begin{equation}
    \overbar{M} = (G^{\dagger}S^\dagger)^{-1} = (G^{-1}S)^{-1} = S^{-1}\, G\,.
\end{equation}
Thus we obtain
\begin{equation}
     \overbar{\Sigma} = G^{-1}\,(\overbar{A}_0+\Theta)\,G\,,
     \label{eq:Sigma_bar_red_app}
\end{equation}
where we have defined
\begin{equation}
    \overbar{A}_0\equiv (S^{-1})^{-1}\d S^{-1} = -S\,(S^{-1}\,\d S)\,S^{-1} = -S\,A_0\,S^{-1}\,,
    \label{eq:A_0_bar_app}
\end{equation}
which follows from the identity $\d S^{-1} = -S^{-1}\,\d S\,S^{-1}$. The isomorphism of positive definite matrices $S$ to $\mathbb{H}_3$ naturally defines a section which is invariant under projection~\cite{Erdmenger:2025lvv}.

\paragraph{Action on fundamental vector fields.} Let $\tau_i^{\#}$ be the fundamental vector field of $\tau_i$ defined by
\begin{equation}
    \tau_i^{\#}(p) = \frac{\d}{\d t} \mathcal{R}_{e^{t\tau_i}}(p)\bigg\vert_{t = 0}\,,
\end{equation}
where $p\in P_6 = \mathbb{H}_3\times S^3$. Since $A_0$ and $\overbar{A}_0$ are horizontal 1-forms they satisfy
\begin{equation}
    A_0(\tau_i^{\#}) = \overbar{A}_0(\tau_i^{\#}) = 0\,.
\end{equation}
which follows from
\begin{equation}
    \d S(\tau_i^{\#}) = \tau_i^{\#}(S) = \frac{\d}{\d t}S\circ \mathcal{R}_{e^{t\tau_i}}\bigg\vert_{t = 0} = \frac{\d}{\d t}S\bigg\vert_{t = 0} = 0\,,
\end{equation}
since $\mathcal{R}_{e^{t\tau_i}}(M) = Me^{t\tau_i}$ where $e^{t\tau_i}$ cancels in $S = \sqrt{MM^\dagger}$ by $\tau_i^\dagger = -\tau_i$. Similarly, we have
\begin{equation}
    \d G(\tau_i^{\#}) = \frac{\d}{\d t}Ge^{t\tau_i}\bigg\vert_{t = 0} =G\,\tau_i\,.
\end{equation}
Thus we obtain
\begin{equation}
    G^{-1}\,\d G(\tau_i^{\#})  = \tau_i= G^{-1}\,\Theta(\tau_i^{\#})\,G\,.
\end{equation}
Thus it follows that
\begin{equation}\label{eq:Sigma_action_fund_vector_field}
    \Sigma(\tau_i^{\#}) = \overbar{\Sigma}(\tau_i^{\#}) = \tau_i\,.
\end{equation}
Thus the action of $\mathcal{A}$ on a fundamental vector field is given by
\begin{equation}
    \mathcal{A}(\tau_i^{\#}) = B_i + C_i = (B^j_i + C_i^j)\,\tau_j\,.
\end{equation}
Imposing $\mathcal{A}(\tau_i^{\#}) = \tau_i$ gives
\begin{equation}
    B_i = \tau_i - C_i
\end{equation}
so that
\begin{equation}
    \mathcal{A} = \Sigma + C_i\,(\overbar{\Sigma}^i-\Sigma^i)\,.
    \label{eq:A_form_app}
\end{equation}
Note that the second term is horizontal,
\begin{equation}
    C_i\,(\overbar{\Sigma}^i-\Sigma^i)(\tau_j^\#) = C_i\, \delta_j^i - C_i\, \delta_j^i = 0\,,
\end{equation}
where we used \eqref{eq:Sigma_action_fund_vector_field}.

\paragraph{Equivariance.} Right-action with a constant element $u\in \mathrm{SU}(2)$ is given by $\mathcal{R}^*_uM = Mu$ which by \eqref{eq:M_bar} implies $\mathcal{R}^*_u\overbar{M} = \overbar{M} u$ since $u^\dagger = u^{-1}$. It follows that
\begin{equation}
    \mathcal{R}^*_u\,\Sigma = u^{-1}\,\Sigma\,u\,,\quad \mathcal{R}^*_u\,\overbar{\Sigma} = u^{-1}\,\overbar{\Sigma}\,u
\end{equation}
are equivariant. In components, equivariance amounts to
\begin{equation}
    \mathcal{R}^*_u\,\Sigma^i = \Lambda(u)^{i}_{\;\;\,j}\,\Sigma^j\,,\quad \mathcal{R}^*_u\,\overbar{\Sigma}^i = \Lambda(u)^{i}_{\;\;\,j}\,\overbar{\Sigma}^j\,,
\end{equation}
where $u^{-1}\,\tau_i\,u = \Lambda(u)^{j}_{\;\;\,i}\,\tau_j $. We require that $\mathcal{A}$ is also equivariant
\begin{equation}
    \mathcal{R}^*_u\,\mathcal{A} = u^{-1}\,\mathcal{A}\,u\,,
\end{equation}
which restricts to the zero-mode as explained in the main text. Since $\Sigma$ is equivariant, \eqref{eq:A_form_app} implies that $\mathcal{A}$ is equivariant if and only if $\Delta\mathcal{A}\equiv \mathcal{A}-\Sigma = -C_i\,(\Sigma^i-\overbar{\Sigma}^i)$ is equivariant. Any equivariant horizontal Lie algebra valued $1$-form is of the form
\begin{equation}
    \Delta\mathcal{A} = G^{-1}\, \Delta A\, G\,,
\end{equation}
where $\Delta A$ is the pull-back along the section $S$. Defining $c_i = C_i \circ S$
\begin{equation}\label{eq:dimensional_red_horizontal}
    \Delta A = -c_i\,(A_0^i-\overbar{A}_0^i)\,.
\end{equation}
Thus, plugging \eqref{eq:dimensional_red_horizontal} and \eqref{eq:Sigma_red_app} into \eqref{eq:A_form_app}, we find the dimensional reduction formula
\begin{equation}
    \mathcal{A} = G^{-1} (A + \Theta) G\,,
\end{equation}
where $A$ is horizontal and given by
\begin{equation}
    A = A_0 - c_i\,(A_0^i-\overbar{A}_0^i)\,.
\end{equation}
Using \eqref{eq:A_0_bar_app}, this can also be written as
\begin{equation}
    \Delta A^i = -c_j^i\,(S^{-1}\d S + \d S S^{-1})^j\,. 
\end{equation}

\subsection{Dimensional reduction of the holomorphic CS action}\label{subapp:CS_reduction}

In this appendix, we provide the derivations for the identities necessary to dimensionally reduce 6D KS gravity to three dimensions.

The Lagrangian we will reduce is $\Omega\wedge \mathrm{CS}[\mathcal{A}]$. Explicitly, we have
\begin{align}
    6\,\Omega\wedge \mathrm{CS}[\mathcal{A}]&=2\,\tr{(\Sigma\wedge \Sigma\wedge \Sigma)}\wedge\text{CS}[\mathcal{A}]\nonumber\\
    &= \tr{(\Sigma\wedge \Sigma\wedge \Sigma)}\wedge\biggl(\tr{(\mathcal{A}\wedge F_{\mathcal{A}})}-\frac{1}{3}\,\tr{(\mathcal{A}\wedge \mathcal{A}\wedge \mathcal{A})}\biggr)\,.
    \label{eq:CS_Omega_app}
\end{align}
Using $\mathcal{A} = G^{-1}\,(A+\Theta)\,G$, we have $F_{\mathcal{A}} = G^{-1}F_{A}G$ so that
\begin{equation}
    \tr{(\mathcal{A}\wedge F_{\mathcal{A}})} = \tr{(A\wedge F_{A})} + \tr{(\Theta\wedge F_{A})}\,.
\end{equation}
Then using $\Sigma = G^{-1}\,(A_0+\Theta)\,G$, we have
\begin{equation}
    \tr{(\Sigma\wedge \Sigma\wedge \Sigma)}= \tr{[(A_0+\Theta)\wedge (A_0+\Theta)\wedge (A_0+\Theta)]}\,.
\end{equation}
Expanding and using invariance of the trace of the wedge product under permutations of the 1-forms, we obtain
\begin{align}
    &\tr{(\Sigma\wedge \Sigma\wedge \Sigma)}\nonumber\\
    &=\tr{(A_0\wedge A_0\wedge A_0)}+3\,\tr{(A_0\wedge A_0\wedge \Theta)}+3\,\tr{(A_0\wedge \Theta\wedge \Theta)}+\tr{(\Theta\wedge \Theta\wedge \Theta)}\,.
    \label{eq:A_cubed_reduction}
\end{align}
Thus it follows that
\begin{align}
    &\tr{(\Sigma\wedge \Sigma\wedge \Sigma)}\wedge\tr{(\mathcal{A}\wedge F_{\mathcal{A}})}\nonumber\\
    &= \tr{(\Theta\wedge \Theta\wedge \Theta)}\wedge\tr{(A\wedge F_{A})} + 3\,\tr{(A_0\wedge \Theta\wedge \Theta)}\wedge\tr{(\Theta\wedge F_{A})}\,,
    \label{eq:red_1_app}
\end{align}
where we used that $\tr{(A\wedge F_{A})}$ is a horizontal 3-form which when wedged with additional horizontal forms gives zero and that the wedge product of more than three $\Theta$ vanishes due to $\mathfrak{su}(2)$ being three-dimensional. Here, we have
\begin{align}
    \tr{(A_0\wedge \Theta\wedge \Theta)}\wedge\tr{(\Theta\wedge F_{A})} = -\frac{1}{8}\,\delta_{pq}\,\varepsilon_{ijk}\,  \Theta^j\wedge \Theta^k\wedge \Theta^p\wedge A^i_0\wedge F_{A}^q\,,
\end{align}
where we have permuted all $\Theta$ to the left producing a minus sign. Using
\begin{equation}
    \Theta^i\wedge  \Theta^j\wedge \Theta^k = \Theta^1\wedge  \Theta^2\wedge \Theta^3\,\varepsilon^{ijk} = -\frac{2}{3}\,\tr{(\Theta\wedge\Theta\wedge \Theta)}\,\varepsilon^{ijk}\,,
    \label{eq:Theta_cubed}
\end{equation}
together with $\varepsilon_{ijk}\epsilon^{jkp} = 2\delta^p_i$, we obtain
\begin{equation}
    \tr{(A_0\wedge \Theta\wedge \Theta)}\wedge\tr{(\Theta\wedge F_{A})} = -\frac{1}{3}\,\tr{(\Theta\wedge \Theta\wedge \Theta)}\wedge\tr{(A_0\wedge F_{A})}
\end{equation}
and \eqref{eq:red_1_app} simplifies to
\begin{equation}
    \tr{(\Sigma\wedge \Sigma\wedge \Sigma)}\wedge\tr{(\mathcal{A}\wedge F_{\mathcal{A}})} = \tr{(\Theta\wedge \Theta\wedge \Theta)}\wedge[\tr{(A\wedge F_{A})} -\tr{(A_0\wedge F_{A})}]\,.
    \label{eq:red_main_1_app}
\end{equation}
Similarly to \eqref{eq:A_cubed_reduction}, we obtain
\begin{equation}
    \tr{(\mathcal{A}\wedge \mathcal{A}\wedge \mathcal{A})}=\tr{(A\wedge A\wedge A)}+3\,\tr{(A\wedge A\wedge \Theta)}+3\,\tr{(A\wedge \Theta\wedge \Theta)}+\tr{(\Theta\wedge \Theta\wedge \Theta)}\,.
\end{equation}
so that
\begin{align}
    &\tr{(\Sigma\wedge \Sigma\wedge \Sigma)}\wedge \tr{(\mathcal{A}\wedge \mathcal{A}\wedge \mathcal{A})}\nonumber\\
    &=\tr{(\Theta\wedge \Theta\wedge \Theta)}\wedge\tr{(A\wedge A\wedge A)} + \tr{(A_0\wedge A_0\wedge A_0)}\wedge\tr{(\Theta\wedge \Theta\wedge \Theta)} \nonumber\\
    &\quad +9\,\tr{(A_0\wedge \Theta\wedge \Theta)}\wedge\tr{(A\wedge A\wedge \Theta)} +9\,\tr{(A_0\wedge A_0\wedge \Theta)}\wedge\tr{(A\wedge \Theta\wedge \Theta)}\,.
    \label{eq:red_2_app}
\end{align}
Here we can write
\begin{equation}
     \tr{(A_0\wedge \Theta\wedge \Theta)}\wedge\tr{(A\wedge A\wedge \Theta)}  = -\frac{1}{16}\,\varepsilon_{ijk}\,\varepsilon_{lmn}\,\Theta^j\wedge  \Theta^k\wedge \Theta^n\wedge A^l\wedge A^m\wedge A_0^i\,,
\end{equation}
where we have permuted all $\Theta$ to the left producing a minus sign. Using \eqref{eq:Theta_cubed} and $\varepsilon_{ijk}\varepsilon^{jkn} = 2\delta^n_i$, we obtain
\begin{equation}
    \tr{(A_0\wedge \Theta\wedge \Theta)}\wedge\tr{(A\wedge A\wedge \Theta)} = -\frac{1}{3}\,\tr{(\Theta\wedge \Theta\wedge \Theta)}\wedge\tr{(A\wedge A\wedge A_0)}\,.
    \label{eq:trace_identity_app}
\end{equation}
A similar calculation with $A\leftrightarrow A_0$ gives
\begin{align}
    \tr{(A_0\wedge A_0\wedge \Theta)}\wedge\tr{(A\wedge \Theta\wedge \Theta)}  &= -\tr{(A\wedge \Theta\wedge \Theta)}\wedge\tr{(A_0\wedge A_0\wedge \Theta)}\nonumber\\
    &= \frac{1}{3}\,\tr{(\Theta\wedge \Theta\wedge \Theta)}\wedge\tr{(A\wedge A_0\wedge A_0)}\,.
\end{align}
Substituting to \eqref{eq:red_2_app}, we obtain
\begin{align}
    &\tr{(\Sigma\wedge \Sigma\wedge \Sigma)}\wedge\tr{(\mathcal{A}\wedge \mathcal{A}\wedge \mathcal{A})}\nonumber\\
    &=\tr{(\Theta\wedge \Theta\wedge \Theta)}\wedge[\tr{(A\wedge A\wedge A)}- \tr{(A_0\wedge A_0\wedge A_0)}]\nonumber\\
    &\quad -3\,\tr{(\Theta\wedge \Theta\wedge \Theta)}\wedge[\tr{(A\wedge A\wedge A_0)}-\tr{(A\wedge A_0\wedge A_0)}]\,.
    \label{eq:red_main_2_app}
\end{align}
Substituting \eqref{eq:red_main_1_app} and \eqref{eq:red_main_2_app} to \eqref{eq:CS_Omega_app}, we obtain
\begin{align}
    &2\,\tr{(\Sigma\wedge \Sigma\wedge \Sigma)}\wedge\text{CS}[\mathcal{A}]\nonumber\\
    &= \tr{(\Theta\wedge \Theta\wedge \Theta)}\wedge\biggl(\tr{(A\wedge F_{A})}-\frac{1}{3}\,\tr{(A\wedge A\wedge A)}\biggr)\nonumber\\
    &\quad + \tr{(\Theta\wedge \Theta\wedge \Theta)}\wedge\tr{\biggl(-A_0\wedge F_{A}+\frac{1}{3}\, A_0\wedge A_0\wedge A_0+A\wedge A\wedge A_0-A\wedge A_0\wedge A_0\biggr)}\,.
    \label{eq:red_midstep_1}
\end{align}
By writing
\begin{equation}
    \tr{(A_0\wedge F_{A})} = \tr{(A_0\wedge \d A)} + \tr{(A_0\wedge A\wedge A)}\,,
\end{equation}
we see that there is a cancellation on the second line in \eqref{eq:red_midstep_1} yielding
\begin{align}
    &2\,\tr{(\Sigma\wedge \Sigma\wedge \Sigma)}\wedge\text{CS}[\mathcal{A}]\nonumber\\
    &= 2\,\tr{(\Theta\wedge \Theta\wedge \Theta)}\wedge\text{CS}[A]\nonumber\\
    &\quad + \tr{(\Theta\wedge \Theta\wedge \Theta)}\wedge\biggl(-\tr{(A_0\wedge \d A)}+\frac{1}{3}\, \tr{(A_0\wedge A_0\wedge A_0)}-\tr{(A\wedge A_0\wedge A_0)}\biggr)\,.
    \label{eq:red_midstep_2}
\end{align}
Using $A_0\wedge A_0 = -\d A_0$ since $F_{A_0} = 0$, we have
\begin{equation}
    -\tr{(A_0\wedge \d A)}-\tr{(A\wedge A_0\wedge A_0)} = -\tr{(A_0\wedge \d A)}+\tr{(A\wedge \d A_0)}= \d \,\tr{(A_0\wedge A)}\,.
\end{equation}
Together with 
\begin{equation}
    \tr{(\Theta\wedge \Theta\wedge \Theta)}\wedge \d\, \tr{(A_0\wedge A)}
    = -\d [\tr{(\Theta\wedge \Theta\wedge \Theta)}\wedge\tr{(A_0\wedge A)}]\,,
\end{equation}
which follows from
\begin{equation}
    \tr{(\d \Theta\wedge \Theta\wedge \Theta)} = \tr{( \Theta\wedge \Theta\wedge \Theta\wedge \Theta)} = 0\,,
\end{equation}
we obtain
\begin{align}
    2\,\tr{(\Sigma\wedge \Sigma\wedge \Sigma)}\wedge\text{CS}[\mathcal{A}] &= \tr{(\Theta\wedge \Theta\wedge \Theta)}\wedge\biggl(2\,\text{CS}[A]+ \frac{1}{3}\,\tr{(A_0\wedge A_0\wedge A_0)}\biggr)\nonumber\\
    &\quad -\d [\tr{(\Theta\wedge \Theta\wedge \Theta)}\wedge\tr{(A_0\wedge A)}]\,.
\end{align}
Using $\epsilon_{S^3} = \frac{1}{12}\,\tr{(\Theta\wedge \Theta\wedge \Theta)}$ and $\Omega = \frac{1}{3}\,\tr{(\Sigma\wedge \Sigma\wedge \Sigma)}$, we obtain
\begin{equation}
    \Omega\wedge\text{CS}[\mathcal{A}]  = 4\,\epsilon_{S^3}\wedge\text{CS}[A] +\frac{2}{3}\,\epsilon_{S^3}\wedge\tr{(A_0\wedge A_0\wedge A_0)}-2\,\d [\epsilon_{S^3}\wedge\tr{(A_0\wedge A)}]\,.
    \label{eq:CS_reduction_final_app_1}
\end{equation}
Since $F_{A_0} = 0$, we have
\begin{equation}
    \mathrm{CS}[A_0] = \frac{1}{2}\,\tr{\biggl(A_0\wedge F_{A_0} - \frac{1}{3}\,A_0\wedge A_0\wedge A_0\biggr)} = -\frac{1}{6}\,\tr{(A_0\wedge A_0\wedge A_0)}
    \label{eq:CS_A_0}
\end{equation}
so that this simplifies to
\begin{equation}
    \Omega\wedge\text{CS}[\mathcal{A}]  = 4\,\epsilon_{S^3}\wedge (\text{CS}[A] -\text{CS}[A_0])-2\,\d [\epsilon_{S^3}\wedge\tr{(A_0\wedge A)}]\,,
    \label{eq:CS_reduction_final_app_0}
\end{equation}
which is the formula used in the main text. Defining $\Delta A \equiv A-A_0$, we have the identity
\begin{align}
\mathrm{CS}[A]
&=
\mathrm{CS}[A_0]
+
\frac{1}{2}\,\tr{\biggl(
2\, \Delta A \wedge F_{A_0}
+
\Delta A \wedge \d_{A_0} (\Delta A)+
\frac{2}{3}\,
\Delta A \wedge \Delta A \wedge \Delta A
\biggr)}\nonumber\\
&\quad-
\frac{1}{2}\,\d\,\tr{(A_0 \wedge \Delta A)}\,.
\end{align}
Using $F_{A_0} = 0$ and $\tr{(A_0\wedge A_0)} = 0$, we obtain
\begin{equation}
    \epsilon_{S^3}\wedge (\text{CS}[A] -\text{CS}[A_0]) = \epsilon_{S^3}\wedge \mathrm{CS}[\Delta A,A_0] + \frac{1}{2}\,\d[\epsilon_{S^3}\wedge\tr{(A_0 \wedge A)}]\,,
\end{equation}
where we used $\d[\epsilon_{S^3}\wedge\tr{(A_0 \wedge \Delta A)}] = -\epsilon_{S^3}\wedge\d\,\tr{(A_0 \wedge \Delta A)}$. Substituting into \eqref{eq:CS_reduction_final_app_0}, we obtain
\begin{equation}
    \Omega\wedge\text{CS}[\mathcal{A}]  = 2\,\epsilon_{S^3}\wedge\tr{\biggl(
\Delta A \wedge \d_{A_0}( \Delta A)+
\frac{2}{3}\,
\Delta A \wedge \Delta A \wedge \Delta A
\biggr)}\,.
\end{equation}

\subsection{Reduction of 6D volume}\label{subapp:volume_reduction}

Let us now consider the dimensional reduction of the 6D volume of the background:
\begin{equation}
    \Omega\wedge \overbar{\Omega} = \frac{1}{9}\,\tr{(\Sigma\wedge \Sigma\wedge \Sigma)}\wedge \tr{(\overbar{\Sigma}\wedge \overbar{\Sigma}\wedge \overbar{\Sigma})}\,.
\end{equation}
Using the formula \eqref{eq:A_cubed_reduction}, we obtain
\begin{align}
    9\,\Omega\wedge \overbar{\Omega}
    &= \tr{(A_0\wedge A_0\wedge A_0)}\wedge  \tr{(\Theta\wedge \Theta\wedge \Theta)} + \tr{(\Theta\wedge \Theta\wedge \Theta)}\wedge \tr{(\overbar{A}_0\wedge \overbar{A}_0\wedge \overbar{A}_0)}\nonumber\\
    &+9\,\tr{(A_0\wedge A_0\wedge \Theta)}\wedge \tr{(\overbar{A}_0\wedge \Theta\wedge \Theta)}+9\,\tr{(A_0\wedge \Theta\wedge \Theta)}\wedge \tr{(\overbar{A}_0\wedge \overbar{A}_0\wedge \Theta)}\,.
\end{align}
Now here we can use the trace identity \eqref{eq:trace_identity_app} to write
\begin{align}
    \tr{(A_0\wedge A_0\wedge \Theta)}\wedge \tr{(\overbar{A}_0\wedge \Theta\wedge \Theta)} &= -\frac{1}{3}\,\tr{(A_0\wedge A_0\wedge \overbar{A}_0)}\wedge \tr{(\Theta\wedge \Theta\wedge \Theta)}\,,\\
    \tr{(A_0\wedge \Theta\wedge \Theta)}\wedge \tr{(\overbar{A}_0\wedge \overbar{A}_0\wedge \Theta)} &= \frac{1}{3}\,\tr{(\overbar{A}_0\wedge \overbar{A}_0\wedge A_0)}\wedge \tr{(\Theta\wedge \Theta\wedge \Theta)}\,.
\end{align}
Thus we obtain
\begin{align}
    9\,\Omega\wedge \overbar{\Omega}
    &= [\tr{(A_0\wedge A_0\wedge A_0)}- \tr{(\overbar{A}_0\wedge \overbar{A}_0\wedge \overbar{A}_0)}]\wedge \tr{(\Theta\wedge \Theta\wedge \Theta)}\nonumber\\
    &-3\,[\tr{(A_0\wedge A_0\wedge \overbar{A}_0)} - \tr{(\overbar{A}_0\wedge \overbar{A}_0\wedge A_0)}]\wedge \tr{(\Theta\wedge \Theta\wedge \Theta)}\,.
\end{align}
Since $F_{A_0} = 0$ and $F_{\overbar{A}_0} = 0$, we obtain
\begin{align}
    \tr{(A_0\wedge A_0\wedge \overbar{A}_0)} - \tr{(\overbar{A}_0\wedge \overbar{A}_0\wedge A_0)} &= -\tr{(\d A_0\wedge \overbar{A}_0)} + \tr{(\d \overbar{A}_0\wedge A_0)}\nonumber
    \\&=-\d\,\tr{(A_0\wedge \overbar{A}_0)}\,.
\end{align}
Thus the final result is
\begin{align}
\Omega\wedge\overbar{\Omega}
    &= \frac{1}{9}\,[\tr{(A_0\wedge A_0\wedge A_0)}- \tr{(\overbar{A}_0\wedge \overbar{A}_0\wedge \overbar{A}_0)}]\wedge \tr{(\Theta\wedge \Theta\wedge \Theta)}\nonumber\\
    &+\frac{1}{3}\,\d[\tr{(A_0\wedge \overbar{A}_0)}\wedge \tr{(\Theta\wedge \Theta\wedge \Theta)}]\,.
\end{align}
Using $\epsilon_{S^3} = \frac{1}{12}\,\tr{(\Theta\wedge \Theta\wedge \Theta)}$ and permuting it to the left producing a minus sign in the first term but not in the second, we obtain
\begin{equation}
    \Omega\wedge\overbar{\Omega} = -\frac{4}{3}\,\epsilon_{S^3}\wedge[\tr{(A_0\wedge A_0\wedge A_0)}- \tr{(\overbar{A}_0\wedge \overbar{A}_0\wedge \overbar{A}_0)}] +4\,\d[\epsilon_{S^3}\wedge\tr{(A_0\wedge \overbar{A}_0)}]\,.
\end{equation}
Using \eqref{eq:CS_A_0}, we obtain
\begin{equation}
    \Omega\wedge\overbar{\Omega} = 8\,\epsilon_{S^3}\wedge(\mathrm{CS}[A_0]- \mathrm{CS}[\overbar{A}_0]) +4\,\d[\epsilon_{S^3}\wedge\tr{(A_0\wedge \overbar{A}_0)}]\,.
\end{equation}

\paragraph{Alternative reduction.} We will provide an alternative dimensional reduction of the volume form which can also be found in~\cite{Herfray:2016std}. We will write
\begin{equation}
    \Sigma = \mathcal{W}_0 + i \mathcal{E}_0\,,\quad \overbar{\Sigma} = \mathcal{W}_0 - i \mathcal{E}_0\,,
\end{equation}
where $\mathcal{W}_0$ and $\mathcal{E}_0$ are $\mathfrak{su}(2)$-valued 1-forms and $\mathcal{E}_0$ is horizontal. Then we obtain
\begin{equation}
    \Omega = \varrho + i\hat{\varrho}\,,\quad \overbar{\Omega} = \varrho - i\hat{\varrho}\,,
\end{equation}
where the real 3-forms
\begin{equation}
    \varrho = \tr{\biggl(\frac{1}{3}\,\mathcal{W}_0\wedge \mathcal{W}_0\wedge \mathcal{W}_0-\mathcal{W}_0\wedge \mathcal{E}_0\wedge \mathcal{E}_0  \biggr)}\,,\quad \hat{\varrho} = \tr{\biggl(\mathcal{W}_0\wedge \mathcal{W}_0\wedge \mathcal{E}_0 - \frac{1}{3}\,\mathcal{E}_0\wedge \mathcal{E}_0\wedge \mathcal{E}_0\biggr)}\,.
\end{equation}
Now we have
\begin{equation}
    \Omega\wedge\overline{\Omega} = -2i\,\varrho\wedge\hat{\varrho}\,,
\end{equation}
which gives
\begin{equation}
    \frac{\Omega\wedge\overline{\Omega} }{2i}=\tr{(\mathcal{W}_0\wedge \mathcal{E}_0\wedge \mathcal{E}_0)}\wedge \tr{(\mathcal{W}_0\wedge \mathcal{W}_0\wedge \mathcal{E}_0)} +\frac{1}{9}\, \tr{(\mathcal{W}_0\wedge \mathcal{W}_0\wedge \mathcal{W}_0)}\wedge \tr{(\mathcal{E}_0\wedge \mathcal{E}_0\wedge \mathcal{E}_0)}\,.
\end{equation}
Using the trace identity
\begin{equation}
    \tr{(\mathcal{W}_0\wedge \mathcal{E}_0\wedge \mathcal{E}_0)}\wedge \tr{(\mathcal{W}_0\wedge \mathcal{W}_0\wedge \mathcal{E}_0)} = \frac{1}{3}\,\tr{(\mathcal{W}_0\wedge\mathcal{W}_0\wedge \mathcal{W}_0)}\wedge \tr{( \mathcal{E}_0\wedge \mathcal{E}_0\wedge \mathcal{E}_0)}\,,
\end{equation}
this reduces to
\begin{equation}
    \Omega\wedge\overline{\Omega} = \frac{8i}{9}\, \tr{(\mathcal{W}_0\wedge \mathcal{W}_0\wedge \mathcal{W}_0)}\wedge \tr{(\mathcal{E}_0\wedge \mathcal{E}_0\wedge \mathcal{E}_0)}\,.
\end{equation}
Decomposing
\begin{equation}
    A_0 = W_0 + i E_0\,,\quad \overbar{A}_0 = W_0 - i E_0\,,
\end{equation}
we obtain
\begin{equation}
    \mathcal{W}_0 = G^{-1}\,(W_0 + \Theta)\,G\,,\quad \mathcal{E}_0 = G^{-1}\,E_0\,G\,,
\end{equation}
so that
\begin{equation}
     \Omega\wedge\overline{\Omega} = \frac{8i}{9}\, \tr{[(W_0 + \Theta)\wedge (W_0 + \Theta)\wedge (W_0 + \Theta)]}\wedge \tr{(E_0\wedge E_0\wedge E_0)}\,.
\end{equation}
Since $\tr{(E_0\wedge E_0\wedge E_0)}$ is a horizontal top-form, we obtain
\begin{equation}
    \Omega\wedge\overline{\Omega} = \frac{8i}{9}\, \tr{(\Theta\wedge\Theta\wedge\Theta)}\wedge \tr{(E_0\wedge E_0\wedge E_0)} = \frac{32i}{3}\,\epsilon_{S^3}\wedge \tr{(E_0\wedge E_0\wedge E_0)}\,.
\end{equation}

\section{Action of 3D Einstein gravity in the Chern--Simons formulation}\label{app:3D_Einstein_CS_form}

In this appendix, we review the rewriting of 3D gravity in the CS formulation in our conventions.

\paragraph{Conventions.} With the conventions \eqref{eq:g_Gamma} and \eqref{eq:E_W_defs}, the curvature of $W$ is
\begin{equation}
    F_W = \d W + W\wedge W \equiv f^i\tau_i,
    \quad
    f^i = \d W^i + \frac{1}{2}\,\varepsilon^i{}_{jk} W^j\wedge W^k \,.
\end{equation}
In these conventions $f^i$ is the dualized spin curvature,
\begin{equation}
    f^i \equiv -\frac{1}{2}\,\varepsilon^i{}_{jk}\, f^{jk},
    \qquad
    f^{ij} = -\varepsilon^{ij}{}_{k} \,f^k\,,
\end{equation}
such that the Riemann tensor of the independent connection $\Gamma$ is
\begin{equation}
    (f^{ij})_{\mu\nu} = R_{\mu\nu}^{\rho\sigma}\,E^i_\rho E^j_\sigma\,.
    \label{eq:f_R_relation}
\end{equation}

\paragraph{Bulk terms in the first-order formulation.} 
The oriented bulk volume form is
\begin{equation}
    \epsilon_g= \sqrt{g}\,\d^3 y
    = \frac{1}{3!}\,\varepsilon_{ijk}\, E^i\wedge E^j\wedge E^k = -\frac{2}{3}\,\tr{(E\wedge E\wedge E)}\,.
    \label{eq:epsilon_g}
\end{equation}
Therefore the cosmological part of the bulk action is
\begin{equation}
    \frac{1}{4} \int_{\mathbb{H}_3}\d^3 y\sqrt{g}\,2 =\frac{1}{2}\int_{\mathbb{H}_3}\epsilon_g
    = -\frac{1}{3}\int_{\mathbb{H}_3}\tr{(E\wedge E\wedge E)}\,.
    \label{eq:cosmo_term}
\end{equation}
We assume that the dreibein $E^i_\mu$ is invertible with inverse $E^\mu_i$ so that we have
\begin{equation}
    \varepsilon_{ijk}\, E^i\wedge f^{jk} = \frac{1}{2}\,\varepsilon_{ijk}\,(f^{jk})_{\mu\nu}\,E^\mu_m\,E^\nu_n\, E^i\wedge E^m\wedge E^n \,,
\end{equation}
where we used $f^{ij} = \frac{1}{2}\,(f^{ij})_{\mu\nu}\,\d y^\mu\wedge \d y^\nu$ and $\d y^{\mu} = E^{\mu}_i\,E^i$. Using
\begin{equation}
    E^i\wedge E^m\wedge E^n  = \varepsilon^{imn}\,E^1\wedge E^2\wedge E^3 = \varepsilon^{imn}\,\epsilon_g
\end{equation}
and $\varepsilon_{ijk}\,\varepsilon^{imn} = \delta^m_j\delta^n_k - \delta^n_j\delta^m_k$, we obtain
\begin{equation}
    \varepsilon_{ijk}\, E^i\wedge f^{jk} = (f^{mn})_{\mu\nu}\,E^\mu_m\,E^\nu_n\,E^1\wedge E^2\wedge E^3 = R\,\epsilon_g\,,
\end{equation}
where we used $R = E^{\mu}_iE^{\nu}_j\,(f^{ij})_{\mu\nu}$ as follows from \eqref{eq:f_R_relation}. Using $f^{ij}=-\varepsilon^{ij}{}_{k}\,f^k$ on the left-hand side, we obtain
\begin{equation}
    \varepsilon_{ijk}\, E^i\wedge f^{jk}
    = -\varepsilon_{ijk}\,\varepsilon^{jk}{}_{l} \,E^i\wedge f^l = -2\delta_{il}\,E^i\wedge f^l = 4\,\tr{(E\wedge F_W)}\,,
\end{equation}
which implies
\begin{equation}
    R\,\epsilon_g = 4\,\tr{(E\wedge F_W)}\,.
\end{equation}
Thus the curvature term becomes
\begin{equation}
    \frac{1}{4}\int_{\mathbb{H}_3}\d^3 y \sqrt{g}\,R =\int_{\mathbb{H}_3} \tr{(E\wedge F_W)}
    \label{eq:curvature_term}
\end{equation}
Combining \eqref{eq:cosmo_term} and \eqref{eq:curvature_term}, we obtain
\begin{equation}
    \frac{1}{4}\int_{\mathbb{H}_3}\d^3y\sqrt{g}\,(R+2) = \int_{\mathbb{H}_3}\tr{\biggl(E\wedge F_W - \frac{1}{3}\,E\wedge E\wedge E\biggr)}\,.
    \label{eq:bulk_term_app}
\end{equation}

\paragraph{Boundary counterterm in the first-order formulation.} Let $\Psi$ be a scalar field such that the boundary is located at some constant value of $\Psi$. Then we define the outward-pointing unit normal vector as
\begin{equation}
    n_\mu = \frac{\partial_\mu \Psi}{\sqrt{g^{\rho\sigma}\,\partial_\rho \Psi\,\partial_\sigma \Psi}}\,,\quad n^\mu = g^{\mu\nu}\,n_\nu\,,
\end{equation}
which satisfies $g_{\mu\nu}\,n^\mu n^\nu = 1$. We define the three scalar fields $n^i = E^i_\mu\,n^\mu$, which satisfy $\delta_{ij}\,n^in^j = 1$, and introduce the $\mathfrak{su}(2)$-valued scalar field $N$ as
\begin{equation}
   N\equiv n^i\tau_i\,,
\end{equation}
which satisfies $-2\,\tr{(N^2)} = 1$. We also obtain $n_i = \delta_{ij}\,n^j = E_i^\mu\,n_\mu$.

Now we define the boundary zweibein as the pull-back $\hat{E}^i_{\hat{\mu}} \equiv \partial_{\hat{\mu}}Q^\nu\,E^i_\nu\vert_{\partial \mathbb{H}_3}$ of the dreibein to the boundary where $Q^\mu = Q^\mu(\hat{y})$ is the embedding of the boundary $\Psi(Q(\hat{y}))=\text{constant}$ and $\hat{y}^{\hat{\mu}}$ with $\hat{\mu} = 1,2$ are boundary coordinates. We introduce the restrictions to the boundary
\begin{equation}
    \hat{n}^i \equiv n^i\vert_{\partial \mathbb{H}_3}\,,\quad \hat{N} \equiv N\vert_{\partial \mathbb{H}_3} = \hat{n}^i\tau_i\,,
    \label{eq:hat_N}
\end{equation}
so that $\hat{N}$ is an $\mathfrak{su}(2)$-valued scalar on the boundary. By definition, the pull-back of the normal vector to the boundary vanishes $\partial_{\hat{\mu}}Q^\nu\,n_\nu\vert_{\partial \mathbb{H}_3} = 0$ which implies that
\begin{equation}
    n_i\hat{E}^i_{\hat{\mu}}\vert_{\partial \mathbb{H}_3} = \partial_{\hat{\mu}}Q^\nu\,n_iE^i_\nu\vert_{\partial \mathbb{H}_3} = \partial_{\hat{\mu}}Q^\nu\,n_\nu\vert_{\partial \mathbb{H}_3} = 0\,.
\end{equation}
Using \eqref{eq:hat_N}, this can also be written as $-2\,\tr{(\hat{N}\hat{E})} = 0$.

Now the induced metric on the boundary is given by
\begin{equation}
    \gamma_{\hat{\mu}\hat{\nu}} = \delta_{ij}\,\hat{E}^i_{\hat{\mu}}\hat{E}^j_{\hat{\nu}}
\end{equation}
with the volume form
\begin{equation}
    \epsilon_\gamma = \sqrt{\gamma}\,\d^2\hat{y} = \frac{1}{2}\,\sqrt{\gamma}\,\varepsilon_{\hat{\mu}\hat{\nu}}\,\d \hat{y}^{\hat{\mu}}\wedge \d \hat{y}^{\hat{\nu}} \,.
    \label{eq:boundary_volume_form}
\end{equation}
Notice that we can write (plus sign is fixed by the orientation)
\begin{equation}
    \hat{n}_i = \hat{m}_i\equiv \frac{1}{\sqrt{\gamma}}\frac{1}{2}\,\varepsilon_{ijk}\,\varepsilon^{\hat{\mu}\hat{\nu}}\,\hat{E}^j_{\hat{\mu}}\,\hat{E}^k_{\hat{\nu}} \,,\label{eq:hat_n_m}
\end{equation}
where $\varepsilon^{\hat{\mu}\hat{\nu}}$ is the Levi--Civita symbol. This follows from the fact that $\hat{m}_i$ satisfies
\begin{equation}
    \hat{m}_i \hat{E}^i_{\hat{\mu}} = 0\,,\quad \delta^{ij}\,\hat{m}_i\hat{m}_j = 1\,,
\end{equation}
where the first equality follows from $\hat{E}^i_{[\hat{\rho}}\hat{E}^j_{\hat{\mu}}\,\hat{E}^k_{\hat{\nu}]} = 0$ in two dimensions. Thus, we obtain
\begin{equation}
    \varepsilon_{ijk}\,\hat{n}^i\,
    \hat{E}^j_{\hat{\mu}}\hat{E}^k_{\hat{\nu}}=
    \frac{1}{2\sqrt{\gamma}}\,
    \varepsilon_{ijk}\varepsilon^{i}{}_{lm}\,
    \varepsilon^{\hat{\rho}\hat{\sigma}}\,
    \hat{E}^l_{\hat{\rho}}\hat{E}^m_{\hat{\sigma}}
    \hat{E}^j_{\hat{\mu}}\hat{E}^k_{\hat{\nu}}
    =
    \frac{1}{2\sqrt{\gamma}}\,
    \varepsilon^{\hat{\rho}\hat{\sigma}}
    \left(
        \gamma_{\hat{\rho}\hat{\mu}}
        \gamma_{\hat{\sigma}\hat{\nu}}
        -
        \gamma_{\hat{\rho}\hat{\nu}}
        \gamma_{\hat{\sigma}\hat{\mu}}
    \right)
    =
    \sqrt{\gamma}\,
    \varepsilon_{\hat{\mu}\hat{\nu}}\,.
    \label{eq:2D_levi_civita}
\end{equation}
Substituting into \eqref{eq:boundary_volume_form} yields
\begin{equation}
    \epsilon_\gamma = \frac{1}{2}\,\varepsilon_{ijk}\,\hat{n}^i\,\hat{E}^j\wedge \hat{E}^k= -2\,\tr{(\hat{N}\hat{E}\wedge \hat{E})}\,.
    \label{eq:epsilon_gamma_midstep}
\end{equation}
Thus the counterterm is
\begin{equation}
    -\frac{1}{2}\int_{\partial \mathbb{H}_3} \d^2\hat{y}\sqrt{\gamma} = -\frac{1}{2}\int_{\partial \mathbb{H}_3} \epsilon_\gamma = \int_{\partial \mathbb{H}_3}\tr{(N E\wedge E)}\,,
    \label{eq:counterterm_first}
\end{equation}
where we keep the hats (the pull-back) implicit.

\paragraph{GHY term in the first-order formulation.} Let us then consider the GHY term. The extrinsic curvature tensor is defined as
\begin{equation}
    K_{\hat{\mu}\hat{\nu}} = \partial_{\hat{\mu}}Q^\rho\,\partial_{\hat{\nu}}Q^\sigma\,\nabla_{\rho}n_\sigma\vert_{\partial \mathbb{H}_3}\,,
\end{equation}
where $\nabla_{\mu}n_\nu = \partial_\mu n_\nu - \Gamma^\rho_{\mu\nu}\,n_\rho$ is the covariant derivative with respect to the independent connection. Now we have
\begin{equation}
    \nabla_\mu n_i \equiv  E_i^\nu\,\nabla_{\mu}n_\nu = \partial_\mu n_i - (\omega^k_{\;\;\,i})_\mu\,n_k \,,
\end{equation}
where we used that the spin connection is
\begin{equation}
    (\omega^i_{\;\;\,j})_\mu = E^i_\rho\, E_j^\nu\,\Gamma^{\rho}_{\mu\nu} - E^\nu_j\,\partial_\mu E^i_\nu\,.
    \label{eq:omega_gamma_formula}
\end{equation}
We obtain
\begin{equation}
    K_{\hat{\mu}\hat{\nu}} = \partial_{\hat{\mu}}Q^\rho\,\hat{E}^j_{\hat{\nu}}\,\nabla_\rho n_j\vert_{\partial \mathbb{H}_3}\,.
    \label{eq:extrinsic_definition}
\end{equation}
Now we notice that
\begin{equation}
    \d_W N = \d N + [W,N]
\end{equation}
takes the component form
\begin{equation}
    (\d_W N)^i_\mu = \partial_\mu n^i + \varepsilon^i{}_{jk}\,W^j_\mu\, n^k = \partial_\mu n^i + (\omega^i_{\;\;\,k})_\mu\,n^k\,,
\end{equation}
where we used \eqref{eq:E_W_defs}. We obtain
\begin{equation}
    \delta_{ij}\,(\d_W N)_{\mu}^j = \partial_\mu n_i + \varepsilon_{ijk}\,W^j_\mu\, n^k = \partial_\mu n_i + (\omega_{ik})_\mu\,n^k = \partial_\mu n_i - (\omega^k_{\;\;\,i})_\mu\,n_k\,.
\end{equation}
so that
\begin{equation}
    \nabla_\mu n_i = \delta_{ij}\,(\d_W N)_{\mu}^j\,.
\end{equation}
Substituting to \eqref{eq:extrinsic_definition}, we obtain
\begin{equation}
    K_{\hat{\mu}\hat{\nu}} = \delta_{ij}\,\hat{E}^i_{\hat{\nu}}\,Q^*(\d_W N)_{\hat{\mu}}^j\,.
\end{equation}
which gives
\begin{equation}
    Q^*(\d_W N)_{\hat{\mu}}^i = K_{\hat{\mu}}^{\hat{\nu}}\,\hat{E}^i_{\hat{\nu}} + P_{\hat{\mu}}\,\hat{n}^i
\end{equation}
where $P_{\hat{\mu}}$ is free since $\hat{E}^i_{\hat{\mu}}\,\hat{n}_i = 0$. However, the second term vanishes by
\begin{equation}
    P_{\hat{\mu}} = \hat{n}_i\,Q^*(\d_W N)_{\hat{\mu}}^i = \hat{n}_i\,\partial_{\hat{\mu}} \hat{n}^i + (\hat{\omega}_{ij})_{\hat{\mu}}\,\hat{n}^i\hat{n}^j = \frac{1}{2}\,\partial_{\hat{\mu}}(\hat{n}_i\hat{n}^i) = 0\,,
\end{equation}
where $\hat{\omega}_{ij} \equiv Q^*\omega_{ij}$ and we used $\hat{\omega}_{ij} = -\hat{\omega}_{ji}$. Thus we obtain
\begin{equation}
    Q^*(\d_W N)_{\hat{\mu}}^i = K_{\hat{\mu}}^{\hat{\nu}}\,\hat{E}^i_{\hat{\nu}}\,.
\end{equation}
Now we can compute
\begin{equation}
    \varepsilon_{ijk}\, \hat{n}^i\, \hat{E}^j\wedge Q^*(\d_W N)^k = \varepsilon_{ijk}\, \hat{n}^i\,\hat{E}^j_{[\hat{\mu}}K_{\hat{\nu}]}^{\hat{\rho}}\,\hat{E}^k_{\hat{\rho}}\;\d\hat{y}^{\hat{\mu}}\wedge \d\hat{y}^{\hat{\nu}}\,,
    \label{eq:einstein_midstep_app}
\end{equation}
where $A_{[\mu}B_{\nu]} \equiv \frac{1}{2}\,(A_\mu B_\nu - A_\nu B_\mu)$. In two dimensions, every antisymmetric tensor has only one component and is proportional to the Levi--Civita symbol
\begin{equation}
    \hat{E}^j_{[\hat{\mu}}K_{\hat{\nu}]}^{\hat{\rho}} = L^{j\hat{\rho}}\,\varepsilon_{\hat{\mu}\hat{\nu}}\,,\quad L^{j\hat{\rho}} \equiv \frac{1}{2}\,\varepsilon^{\hat{\mu}\hat{\nu}}\, \hat{E}^j_{\hat{\mu}}K_{\hat{\nu}}^{\hat{\rho}}\,.
\end{equation}
Substituting into \eqref{eq:einstein_midstep_app}, we obtain
\begin{equation}
    \varepsilon_{ijk}\, \hat{n}^i \hat{E}^j\wedge Q^*(\d_W N)^k = \frac{1}{2}\,\varepsilon_{ijk}\,\varepsilon^{\hat{\sigma}\hat{\lambda}}\, \hat{n}^i\,\hat{E}^j_{\hat{\sigma}}\,\hat{E}^k_{\hat{\rho}}\,K_{\hat{\lambda}}^{\hat{\rho}}\;\varepsilon_{\hat{\mu}\hat{\nu}}\,\d\hat{y}^{\hat{\mu}}\wedge \d\hat{y}^{\hat{\nu}}\,.
\end{equation}
Now using \eqref{eq:2D_levi_civita} and $\varepsilon^{\hat{\sigma}\hat{\lambda}}\varepsilon_{\hat{\sigma}\hat{\rho}} = \delta^{\hat{\lambda}}_{\hat{\rho}}$, we obtain
\begin{equation}
    \varepsilon_{ijk}\, \hat{n}^i \hat{E}^j\wedge Q^*(\d_W N)^k =  \frac{1}{2}
    \sqrt{\gamma}\,
    K_{\hat\rho}^{\hat\rho}\,
    \varepsilon_{\hat\mu\hat\nu}
    \,\d\hat y^{\hat\mu}\wedge \d\hat y^{\hat\nu} = K\epsilon_\gamma\,,
\end{equation}
where $K = K^{\hat{\mu}}_{\hat{\mu}}$. This can be rewritten as
\begin{equation}
    K\epsilon_\gamma =  -4\,\tr{(\hat{N} \hat{E}\wedge Q^*(\d_W N))}\,.
\end{equation}
Therefore the GHY term becomes
\begin{equation}
     \frac{1}{2}\int_{\partial \mathbb{H}_3}\d^2\hat{y}\sqrt{\gamma}\,K
    = -2\int_{\partial \mathbb{H}_3}\tr{(N E\wedge \d_W N)}\,,
    \label{eq:GHY_first_order}
\end{equation}
where we have left the pull-back with $Q$ implicit. Using \eqref{eq:counterterm_first} and \eqref{eq:GHY_first_order}, the full boundary term becomes
\begin{equation}
    \frac{1}{2}\int_{\partial \mathbb{H}_3}\d^2\hat{y}\sqrt{\gamma}\,(K-1) = -2\int_{\partial \mathbb{H}_3}\tr{(NE\wedge \d_W N)}+\int_{\partial \mathbb{H}_3}\tr{(NE\wedge E)}\,.
    \label{eq:boundary_term_app}
\end{equation}

\paragraph{CS formulation of the total action.} Combining \eqref{eq:bulk_term_app} and \eqref{eq:boundary_term_app}, we obtain
\begin{align}
    &\frac{1}{4}\int_{\mathbb{H}_3}d^3y\sqrt{g}\,(R+2)+\frac{1}{2}\int_{\partial \mathbb{H}_3}\d^2\hat{y}\sqrt{\gamma}\,(K-1)\\
    &= \int_{\mathbb{H}_3}\tr{\biggl(E\wedge F_W - \frac{1}{3}\,E\wedge E\wedge E\biggr)} -2\int_{\partial \mathbb{H}_3}\tr{(NE\wedge \d_W N)}+\int_{\partial \mathbb{H}_3}\tr{(NE\wedge E)}\,.\nonumber
    \label{eq:Einstein_first-order_formulation}
\end{align}
By writing $A = W+iE$, we obtain
\begin{align}
    \text{CS}[A] &= \frac{1}{2}\,\tr{\biggl(W\wedge \d W- E\wedge \d E+\frac{2}{3}\,W\wedge W\wedge W-2\,W\wedge E\wedge E\biggr)}\nonumber\\
    &\quad + \frac{i}{2}\,\tr{\biggl(W\wedge \d E+E\wedge \d W+2\,W\wedge W\wedge E-\frac{2}{3}\,E\wedge E\wedge E\biggr)}\,.
\end{align}
Here we can write
\begin{equation}
    \tr{(W\wedge \d E)} = \tr{(E\wedge \d W)} + \d\,\tr{(E\wedge W)}\,,
\end{equation}
so that the imaginary part becomes
\begin{align}
    \frac{1}{2i}\,(\text{CS}[A]-\text{CS}[\overbar{A}])&=\tr{\biggl(E\wedge \d W+E\wedge W\wedge W - \frac{1}{3}\,E\wedge E\wedge E\biggr)}\nonumber\\
    &\quad +\frac{1}{2}\,\d\,\tr{(E\wedge W)}\,.
\end{align}
Substituting into \eqref{eq:Einstein_first-order_formulation}, we obtain
\begin{align}
    &\frac{1}{4}\int_{\mathbb{H}_3}d^3y\sqrt{g}\,(R+2)+\frac{1}{2}\int_{\partial \mathbb{H}_3}\d^2\hat{y}\sqrt{\gamma}\,(K-1)\nonumber\\
    &= \frac{1}{2i}\int_{\mathbb{H}_3}(\text{CS}[A]-\text{CS}[\overbar{A}]) +\int_{\partial \mathbb{H}_3}\tr{\biggl(-\frac{1}{2}\,E\wedge W-2\,NE\wedge \d_W N+NE\wedge E\biggr)}\,.
\end{align}
Here in the boundary term we can write explicitly
\begin{equation}
    \tr{(\hat{N} \hat{E}\wedge Q^*(\d_{W}N))} = \tr{(\hat{N} \hat{E}\wedge [\hat{W},\hat{N}])} + 
    \tr{(\hat{N} \hat{E}\wedge \d \hat{N})}\,,
\end{equation}
where the first term is
\begin{equation}
    \tr{(\hat{N} \hat{E}\wedge [\hat{W},\hat{N}])} = -\frac{1}{4}\,\varepsilon_{ijk}\,\hat{n}^i \hat{E}^j\wedge [\hat{W},\hat{N}]^k = -\frac{1}{4}\,\varepsilon_{ijk}\,\varepsilon_{pq}^{\;\;\;\,k}\,\hat{n}^i \hat{E}^j\wedge \hat{W}^p\hat{n}^q 
\end{equation}
Using $\varepsilon_{ijk}\,\varepsilon_{pq}^{\;\;\;\,k} = \delta_{ip}\delta_{jq} - \delta_{iq}\delta_{jp}$, we obtain
\begin{equation}
    \tr{(\hat{N} \hat{E}\wedge [\hat{W},\hat{N}])} = -\frac{1}{4}\,(\hat{n}_i\hat{n}_j\, \hat{E}^i\wedge \hat{W}^j-\hat{n}_i\hat{n}^i\, \hat{E}_j\wedge \hat{W}^j)\,.
\end{equation}
Since $\hat{n}_i\hat{E}^i = 0$ and $\hat{n}_i\hat{n}^i = 1$, we obtain
\begin{equation}
    \tr{(\hat{N} \hat{E}\wedge [\hat{W},\hat{N}])} = \frac{1}{4}\, \hat{E}_i\wedge \hat{W}^i = -\frac{1}{2}\,\tr{(\hat{E}\wedge \hat{W})} 
\end{equation}
so that
\begin{equation}
    \tr{(\hat{N} \hat{E}\wedge Q^*(\d_{W}N))} = -\frac{1}{2}\,\tr{(\hat{E}\wedge \hat{W})}+ 
    \tr{(\hat{N} \hat{E}\wedge \d \hat{N})}\,.
\end{equation}
Thus we obtain
\begin{align}
    &\frac{1}{4}\int_{\mathbb{H}_3}d^3y\sqrt{g}\,(R+2)+\frac{1}{2}\int_{\partial \mathbb{H}_3}\d^2\hat{y}\sqrt{\gamma}\,(K-1)\nonumber\\
    &= \frac{1}{2i}\int_{\mathbb{H}_3}(\text{CS}[A]-\text{CS}[\overbar{A}]) + \int_{\partial \mathbb{H}_3}\tr{\biggl(\frac{1}{2}\,E\wedge W -2\, NE\wedge \d N + NE\wedge E\biggr)}\,.
\end{align}
Here we can write
\begin{equation}
    \tr{(E\wedge W)} = \frac{1}{4i}\,\tr{((A-\overbar{A})\wedge (A+\overbar{A}))} = \frac{1}{2i}\,\tr{(A\wedge \overbar{A})}
\end{equation}
and similarly
\begin{equation}
    \tr{(N E\wedge E)} =-\frac{1}{4}\,\tr{(N\,(A-\overbar{A})\wedge(A-\overbar{A}))}\,.
\end{equation}
We obtain
\begin{align}
    &\frac{1}{4}\int_{\mathbb{H}_3}d^3y\sqrt{g}\,(R+2)+\frac{1}{2}\int_{\partial \mathbb{H}_3}\d^2\hat{y}\sqrt{\gamma}\,(K-1)\nonumber\\
    &= \frac{1}{2i}\int_{\mathbb{H}_3} (\text{CS}[A]-\text{CS}[\overbar{A}])\nonumber\\
    &+ \frac{1}{2i}\int_{\partial \mathbb{H}_3}\tr{\biggl(\frac{1}{2}\,A\wedge \overbar{A} -2\, N\,(A-\overbar{A})\wedge \d N -\frac{i}{2}\, N\,(A-\overbar{A})\wedge(A-\overbar{A})\biggr)}\,.
\end{align}
Multiplying these actions with $-\frac{1}{4\pi G_{\text{N}}}$ gives the actions used in the main text.

\section{Flatness implies integrability}\label{app:flatness_integrability}

In this appendix, we show explicitly that the $(0,2)$-component of flatness $(F_{\mathcal{A}})_{02} = 0$ of $\mathcal{A} = \Sigma' + \iota_{\alpha'}\Sigma'$ implies the KS equation \eqref{eq:KS_equation} for $\alpha'$. More precisely, we will prove that
\begin{equation}
    \tr{[(F_{\mathcal{A}})_{02}\wedge \Sigma'\wedge \Sigma']} = \eta'\biggl(\overbar{\partial}\alpha'-\frac{1}{2}\,[\alpha',\alpha']\biggr)\,,
    \label{eq:F_20_KS_equation}
\end{equation}
when $(F_{\mathcal{A}})_{20} = 0 = (F_{\mathcal{A}})_{11}$. This is more involved than the analogous statement \eqref{eq:KS_eq_A} for $\alpha$ proven in the main text since $\overbar{\partial}\Sigma' \neq 0$ is not holomorphic. The calculation in this appendix is complementary to other results of this paper and gives an alternative proof for the fact that $F_{\mathcal{A}} = 0$ is equivalent to the KS equation for $\alpha'$ as first shown in~\cite{Erdmenger:2025lvv}.

To prove \eqref{eq:F_20_KS_equation}, we start by defining the $(3,0)$-form
\begin{equation}
    \Omega' = \frac{1}{3}\,\tr{(\Sigma'\wedge\Sigma'\wedge\Sigma')}\,,
\end{equation}
which need not be holomorphic since $\overbar{\partial}\Sigma' \neq 0$. Then we define $\eta'(\alpha') = \iota_{\alpha'}\Omega'$ and $\eta'(\alpha'\wedge\alpha') = \iota_{\alpha'}\iota_{\alpha'}\Omega'$ which are explicitly
\begin{equation}
    \eta'(\alpha') = \tr{(\mathcal{A}_{01}\wedge \Sigma'\wedge \Sigma')}\,,\quad \eta'(\alpha'\wedge \alpha') = 2\,\tr{(\mathcal{A}_{01}\wedge \mathcal{A}_{01}\wedge \Sigma')}\,.
    \label{eq:eta_alpha_app}
\end{equation}
The different components of the curvature are
\begin{align}
    (F_{\mathcal{A}})_{20}&= \partial\Sigma' + \Sigma'\wedge\Sigma'\,,\\
    (F_{\mathcal{A}})_{11} &= \partial\mathcal{A}_{01}+\overbar{\partial}\Sigma'+ \mathcal{A}_{01}\wedge\Sigma'+\Sigma'\wedge\mathcal{A}_{01}\,,\\
    (F_{\mathcal{A}})_{02} &= \overbar{\partial}\mathcal{A}_{01} + \mathcal{A}_{01}\wedge \mathcal{A}_{01}\,.
\end{align}
We will now compute them one-by-one and show the relation to the KS equation.

\paragraph{The $(F_{\mathcal{A}})_{11}$ component.} Wedging $(F_{\mathcal{A}})_{11} = \partial\mathcal{A}_{01}+\overbar{\partial}\Sigma'+ \mathcal{A}_{01}\wedge\Sigma'+\Sigma'\wedge\mathcal{A}_{01}$ with $\Sigma'\wedge \Sigma'$ and taking the trace gives
\begin{align}
    \tr{((F_{\mathcal{A}})_{11}\wedge\Sigma'\wedge \Sigma')} &= \tr{(\overbar{\partial}\Sigma'\wedge\Sigma'\wedge \Sigma')}+ \tr{(\partial\mathcal{A}_{01}\wedge\Sigma'\wedge \Sigma')}\nonumber\\
    &+\tr{(\mathcal{A}_{01}\wedge \Sigma'\wedge\Sigma'\wedge \Sigma')}+\tr{(\Sigma'\wedge\mathcal{A}_{01}\wedge \Sigma'\wedge \Sigma')}
     \label{eq:F11_midstep}
\end{align}
Using $\tr{(\tau_i\tau_j\tau_k\tau_l)} = \frac{1}{8}(\delta_{ij}\delta_{kl}-\delta_{ik}\delta_{jl} +\delta_{il}\delta_{jk})$, we have
\begin{equation}
    \tr{(\mathcal{A}_{01}\wedge \Sigma'\wedge\Sigma'\wedge \Sigma')} = \mathcal{A}_{01}^i\wedge \Sigma'^j\wedge\Sigma'^k\wedge \Sigma'^l\,\tr{(\tau_i\tau_j\tau_k\tau_l)} = 0 \,.
    \label{eq:quad_trace_vanishes}
\end{equation}
This same identity also implies that
\begin{equation}
    \tr{(\Sigma'\wedge\mathcal{A}_{01}\wedge \Sigma'\wedge \Sigma')} = 0\,.
\end{equation}
Thus \eqref{eq:F11_midstep} reduces to
\begin{equation}
    \tr{((F_{\mathcal{A}})_{11}\wedge\Sigma'\wedge \Sigma')} = \tr{(\overbar{\partial}\Sigma'\wedge\Sigma'\wedge \Sigma')}+ \tr{(\partial\mathcal{A}_{01}\wedge\Sigma'\wedge \Sigma')}\,.
     \label{eq:F11_midstep_2}
\end{equation}
Here the first term is simply $\overbar{\partial}\Omega'$ while the second term is
\begin{equation}
    \tr{(\partial\mathcal{A}_{01}\wedge\Sigma'\wedge \Sigma')} = \partial\,\tr{(\mathcal{A}_{01}\wedge\Sigma'\wedge \Sigma')}+\tr{(\mathcal{A}_{01}\wedge\partial\Sigma'\wedge \Sigma')}-\tr{(\mathcal{A}_{01}\wedge\Sigma'\wedge \partial\Sigma')}\,.
\end{equation}
Using \eqref{eq:eta_alpha_app} and combining the last two terms, we obtain
\begin{equation}
    \tr{(\partial\mathcal{A}_{01}\wedge\Sigma'\wedge \Sigma')} = \partial\eta'(\alpha')-2\,\tr{(\mathcal{A}_{01}\wedge \Sigma'\wedge\partial\Sigma')}\,.
\end{equation}
Using further \eqref{eq:quad_trace_vanishes}, this becomes
\begin{equation}
    \tr{(\partial\mathcal{A}_{01}\wedge\Sigma'\wedge \Sigma')} = \partial\eta'(\alpha')-2\,\tr{(\mathcal{A}_{01}\wedge \Sigma'\wedge (F_{\mathcal{A}})_{20})}\,.
\end{equation}
Substituting to \eqref{eq:F11_midstep_2}, we obtain
\begin{align}
    \tr{((F_{\mathcal{A}})_{11}\wedge\Sigma'\wedge \Sigma')} &= \partial\eta'(\alpha')+\overbar{\partial}\Omega'-2\,\tr{(\mathcal{A}_{01}\wedge \Sigma'\wedge (F_{\mathcal{A}})_{20})}\,.
     \label{eq:F11_final_app}
\end{align}

\paragraph{The $(F_{\mathcal{A}})_{02}$ component.} Wedging $(F_{\mathcal{A}})_{02} = \overbar{\partial} \mathcal{A}_{01} +\mathcal{A}_{01}\wedge \mathcal{A}_{01} $ with $\Sigma'\wedge \Sigma'$ and using $(F_{\mathcal{A}})_{20} = \partial \Sigma' +\Sigma'\wedge \Sigma'$, we obtain
\begin{align}
    &\tr{((F_{\mathcal{A}})_{02}\wedge\Sigma'\wedge \Sigma')}\nonumber\\
    &= \tr{(\overbar{\partial} \mathcal{A}_{01} \wedge\Sigma'\wedge \Sigma')} - \tr{(\mathcal{A}_{01}\wedge\mathcal{A}_{01}\wedge \partial\Sigma')}+\tr{(\mathcal{A}_{01}\wedge\mathcal{A}_{01}\wedge (F_{\mathcal{A}})_{20})}\,.
    \label{eq:F02_simplified}
\end{align}
Using \eqref{eq:eta_alpha_app} the first term here is simply
\begin{align}
    \tr{(\overbar{\partial} \mathcal{A}_{01} \wedge\Sigma'\wedge \Sigma')} &= \overbar{\partial}\,\tr{( \mathcal{A}_{01} \wedge\Sigma'\wedge \Sigma')}+\tr{( \mathcal{A}_{01} \wedge\overbar{\partial}\Sigma'\wedge \Sigma')}-\tr{( \mathcal{A}_{01} \wedge\Sigma'\wedge \overbar{\partial}\Sigma')}\nonumber\\
    &= \overbar{\partial}\eta'(\alpha')-2\,\tr{( \overbar{\partial}\Sigma'\wedge \mathcal{A}_{01} \wedge \Sigma')}\,.
    \label{eq:KS_eq_term_1}
\end{align}
Thus we obtain
\begin{align}
    &\tr{((F_{\mathcal{A}})_{02}\wedge\Sigma'\wedge \Sigma')}\nonumber\\
    &= \overbar{\partial}\eta'(\alpha')-2\,\tr{( \overbar{\partial}\Sigma'\wedge \mathcal{A}_{01} \wedge \Sigma')} - \tr{(\mathcal{A}_{01}\wedge\mathcal{A}_{01}\wedge \partial\Sigma')}+\tr{(\mathcal{A}_{01}\wedge\mathcal{A}_{01}\wedge (F_{\mathcal{A}})_{20})}\,.
    \label{eq:F_02_midstep}
\end{align}
Then let us compute $\frac{1}{2}\,\partial\eta'(\alpha' \wedge\alpha')$ using \eqref{eq:eta_alpha_app} as
\begin{equation}
    \frac{1}{2}\,\partial\eta'(\alpha' \wedge\alpha') = 2\,\tr{(\partial\mathcal{A}_{01}\wedge \mathcal{A}_{01}\wedge \Sigma')}+\tr{(\mathcal{A}_{01}\wedge \mathcal{A}_{01}\wedge \partial\Sigma')}\,.
\end{equation}
Using $(F_{\mathcal{A}})_{11} = \partial\mathcal{A}_{01}+\overbar{\partial}\Sigma'+ \mathcal{A}_{01}\wedge\Sigma'+\Sigma'\wedge\mathcal{A}_{01}$, we obtain
\begin{align}
    \frac{1}{2}\,\partial\eta'(\alpha' \wedge\alpha') &= \tr{(\mathcal{A}_{01}\wedge \mathcal{A}_{01}\wedge \partial\Sigma')}-2\,\tr{(\overbar{\partial}\Sigma'\wedge \mathcal{A}_{01}\wedge \Sigma')}-2\,\tr{(\mathcal{A}_{01}\wedge \Sigma'\wedge \mathcal{A}_{01}\wedge \Sigma')}\nonumber\\   &\quad -2\,\tr{(\Sigma'\wedge\mathcal{A}_{01}\wedge  \mathcal{A}_{01}\wedge \Sigma')}+2\,\tr{((F_{\mathcal{A}})_{11}\wedge \mathcal{A}_{01}\wedge \Sigma')}\,.
    \label{eq:partial_eta_alpha_midstep}
\end{align}
Using the identity $\tr{(\tau_i\tau_j\tau_k\tau_l)} = \frac{1}{8}(\delta_{ij}\delta_{kl}-\delta_{ik}\delta_{jl} +\delta_{il}\delta_{jk})$, we obtain
\begin{equation}
    \tr{(\mathcal{A}_{01}\wedge \Sigma'\wedge \mathcal{A}_{01}\wedge \Sigma')} = \frac{1}{8}\,\mathcal{A}_{01}^i\wedge \Sigma'^j\wedge \mathcal{A}_{01}^k\wedge \Sigma'^l\,(\delta_{ij}\delta_{kl}-\delta_{ik}\delta_{jl} +\delta_{il}\delta_{jk}) \,.
\end{equation}
The second term vanishes directly by antisymmetry of the wedge product and symmetricity of the Kronecker delta so that
\begin{equation}
    \tr{(\mathcal{A}_{01}\wedge \Sigma'\wedge \mathcal{A}_{01}\wedge \Sigma')} = \frac{1}{8}\,\mathcal{A}_{01}^i\wedge \Sigma'^j\wedge \mathcal{A}_{01}^k\wedge \Sigma'^l\,(\delta_{ij}\delta_{kl} +\delta_{il}\delta_{jk}) = 0\,,
\end{equation}
where the two terms cancel each other. Thus we obtain
\begin{align}
    \frac{1}{2}\,\partial\eta'(\alpha' \wedge\alpha') &= \tr{(\mathcal{A}_{01}\wedge \mathcal{A}_{01}\wedge \partial\Sigma')}-2\,\tr{(\overbar{\partial}\Sigma'\wedge \mathcal{A}_{01}\wedge \Sigma')}-2\,\tr{(\Sigma'\wedge\mathcal{A}_{01}\wedge  \mathcal{A}_{01}\wedge \Sigma')}\nonumber\\   &\quad +2\,\tr{((F_{\mathcal{A}})_{11}\wedge \mathcal{A}_{01}\wedge \Sigma')}\,.
\end{align}
Now it follows that
\begin{equation}
    \tr{(\Sigma'\wedge\mathcal{A}_{01}\wedge  \mathcal{A}_{01}\wedge \Sigma')} = -\tr{(\mathcal{A}_{01}\wedge  \mathcal{A}_{01}\wedge \Sigma'\wedge \Sigma')}\,.
\end{equation}
Using $(F_{\mathcal{A}})_{20} = \partial \Sigma' +\Sigma'\wedge \Sigma'$, we obtain
\begin{equation}
    \tr{(\Sigma'\wedge\mathcal{A}_{01}\wedge  \mathcal{A}_{01}\wedge \Sigma')} = \tr{(\mathcal{A}_{01}\wedge  \mathcal{A}_{01}\wedge \partial\Sigma')}-\tr{(\mathcal{A}_{01}\wedge  \mathcal{A}_{01}\wedge (F_{\mathcal{A}})_{20})}\,.
\end{equation}
Substituting to \eqref{eq:partial_eta_alpha_midstep}, we obtain
\begin{align}
    \frac{1}{2}\,\partial\eta'(\alpha' \wedge\alpha') &= -2\,\tr{(\overbar{\partial}\Sigma'\wedge \mathcal{A}_{01}\wedge \Sigma')}-\tr{(\mathcal{A}_{01}\wedge \mathcal{A}_{01}\wedge \partial\Sigma')}\nonumber\\   &\quad +2\,\tr{((F_{\mathcal{A}})_{11}\wedge \mathcal{A}_{01}\wedge \Sigma')}+2\,\tr{(\mathcal{A}_{01}\wedge \mathcal{A}_{01}\wedge (F_{\mathcal{A}})_{20})}\,.
\end{align}
Substituting this finally to \eqref{eq:F_02_midstep}, we obtain
\begin{align}
    &\tr{[(F_{\mathcal{A}})_{02}\wedge\Sigma'\wedge \Sigma']}\nonumber\\
    &= \overbar{\partial}\eta'(\alpha')+\frac{1}{2}\,\partial\eta'(\alpha' \wedge\alpha')-2\,\tr{((F_{\mathcal{A}})_{11}\wedge \mathcal{A}_{01}\wedge \Sigma')}-\tr{(\mathcal{A}_{01}\wedge\mathcal{A}_{01}\wedge (F_{\mathcal{A}})_{20})}\,.
\end{align}
Thus setting $(F_{\mathcal{A}})_{20} = 0 = (F_{\mathcal{A}})_{11}$, this reduces to
\begin{equation}
    \tr{[(F_{\mathcal{A}})_{02}\wedge\Sigma'\wedge \Sigma']}= \overbar{\partial}\eta'(\alpha')+\frac{1}{2}\,\partial\eta'(\alpha' \wedge\alpha')\,.
\end{equation}

\paragraph{Removing the primes.} We can write $\Omega' = e^{\varrho}\,\Omega$ where $e^{\varrho} = \frac{\det\Sigma'^i_j}{\det \Sigma^i_j}$ so that $\eta' = e^{\varrho}\,\eta$. Thus we obtain
\begin{equation}
    \overbar{\partial}\eta'(\alpha')+\frac{1}{2}\,\partial\eta'(\alpha'\wedge \alpha') = e^{\varrho}\,\biggl(\overbar{\partial}\eta(\alpha')+\frac{1}{2}\,\partial\eta(\alpha'\wedge \alpha')\biggr)+ e^{\varrho}\,\biggl(\overbar{\partial} \varrho\wedge\eta(\alpha') +\frac{1}{2}\,\partial\varrho\wedge\eta(\alpha'\wedge\alpha')\biggr)\,.
    \label{eq:E_prime}
\end{equation}
Now we can compute
\begin{equation}
    \partial\eta'(\alpha') + \overbar{\partial}\Omega' = e^{\varrho}\,(\partial\eta(\alpha') + \partial\varrho\wedge \eta(\alpha') +  \overbar{\partial}\varrho\wedge \Omega)\,.
    \label{eq:removing_primes_midstep}
\end{equation}
where we used $\overbar{\partial}\Omega = 0$. Contracting with $\alpha'$, we first have
\begin{equation}
    \iota_{\alpha'}( \partial\varrho\wedge \eta(\alpha')) = \iota_{\alpha'}(\partial\varrho)\wedge \eta(\alpha') + \partial\varrho\wedge \eta(\alpha'\wedge\alpha') \,.
\end{equation}
Similarly
\begin{equation}
    \iota_{\alpha'}(\overbar{\partial}\varrho\wedge \Omega) = \overbar{\partial}\varrho\wedge \eta(\alpha')\,,
\end{equation}
where we used $\iota_{\alpha'}(\overbar{\partial}\varrho) = 0$. Thus \eqref{eq:removing_primes_midstep} yields
\begin{equation}
    \iota_{\alpha'}(\partial\eta'(\alpha') + \overbar{\partial}\Omega') = e^{\varrho}\,(\iota_{\alpha'}\partial\eta(\alpha')+\overbar{\partial}\varrho\wedge \eta(\alpha') +\iota_{\alpha'}(\partial\varrho)\wedge \eta(\alpha') + \partial\varrho\wedge \eta(\alpha'\wedge\alpha'))\,.
    \label{eq:removing_primes_midstep_2}
\end{equation}
Now we can use that $\partial\varrho\wedge \Omega = 0$ since this is a $(4,0)$-form. It gives
\begin{equation}
    0 = \iota_{\alpha'}\iota_{\alpha'}(\partial\varrho\wedge \Omega) = \iota_{\alpha'}(\iota_{\alpha'}(\partial\varrho)\wedge\Omega + \partial\varrho\wedge\eta(\alpha')) = 2\,\iota_{\alpha'}(\partial\varrho)\wedge\eta(\alpha') +\partial\varrho\wedge\eta(\alpha'\wedge\alpha')\,.
\end{equation}
so that
\begin{equation}
    \iota_{\alpha'}(\partial\varrho)\wedge\eta(\alpha') = -\frac{1}{2}\,\partial\varrho\wedge\eta(\alpha'\wedge\alpha')\,.
\end{equation}
Substituting to \eqref{eq:removing_primes_midstep_2}, we obtain
\begin{equation}
    \iota_{\alpha'}(\partial\eta'(\alpha') + \overbar{\partial}\Omega') = e^{\varrho}\,\biggl(\iota_{\alpha'}\partial\eta(\alpha')+\overbar{\partial}\varrho\wedge \eta(\alpha')  + \frac{1}{2}\,\partial\varrho\wedge \eta(\alpha'\wedge\alpha')\biggr)\,.
\end{equation}
Substituting to \eqref{eq:E_prime}, we obtain
\begin{equation}
    \overbar{\partial}\eta'(\alpha')+\frac{1}{2}\,\partial\eta'(\alpha'\wedge \alpha') = e^{\varrho}\,\biggl(\overbar{\partial}\eta(\alpha')+\frac{1}{2}\,\partial\eta(\alpha'\wedge \alpha')-\iota_{\alpha'}\partial\eta(\alpha')\biggr) + \iota_{\alpha'}(\partial\eta'(\alpha') + \overbar{\partial}\Omega')\,.
\end{equation}
Now $(F_{\mathcal{A}})_{20} = (F_{\mathcal{A}})_{11} = 0$ imply $\partial\eta'(\alpha') + \overbar{\partial}\Omega' = 0$ so that
\begin{equation}
    \tr{[(F_{\mathcal{A}})_{02}\wedge \Sigma'\wedge \Sigma']} = e^{\varrho}\,\biggl(\overbar{\partial}\eta(\alpha')+\frac{1}{2}\,\partial\eta(\alpha' \wedge\alpha')-\iota_{\alpha'}\partial\eta(\alpha')\biggr)\,.
    \label{eq:F02_2}
\end{equation}
Further using Tian's lemma \eqref{eq:Tians_lemma_text}, we obtain
\begin{equation}
    \tr{[(F_{\mathcal{A}})_{02}\wedge \Sigma'\wedge \Sigma']} = \eta'\biggl(\overbar{\partial}\alpha'-\frac{1}{2}\,[\alpha',\alpha']\biggr)\,.
\end{equation}
Now by the argument in Appendix~\ref{subapp:equivalence_hcs_KS}, $(F_{\mathcal{A}})_{02} = 0$ is equivalent to $\eta'(E_{\alpha'}) = 0$. Thus we find that when $(F_{\mathcal{A}})_{20}  = (F_{\mathcal{A}})_{11} = 0$, we have the equivalence
\begin{equation}
    (F_{\mathcal{A}})_{02} = 0\quad \Leftrightarrow \quad \overbar{\partial}\alpha'-\frac{1}{2}\,[\alpha',\alpha'] = 0\,.
\end{equation}
In particular, full flatness implies integrability:
\begin{equation}
    F_{\mathcal{A}} = 0\quad \Rightarrow \quad \overbar{\partial}\alpha'-\frac{1}{2}\,[\alpha',\alpha'] = 0\,.
    \label{eq:flatness_implies_KS}
\end{equation}
The converse holds when $\mathcal{A}$ is $\mathrm{SU}(2)$-equivariant and normalized $\mathcal{A}(\tau_i^{\#}) = \tau_i$. Under these conditions $F_{\mathcal{A}}$ is horizontal and its vanishing is equivalent to $(F_{\mathcal{A}})_{02}$ by \eqref{eq:F02_F_A}. Consequently,
\begin{equation}
    F_{\mathcal{A}} = 0\quad \Leftrightarrow \quad    (F_{\mathcal{A}})_{02} = 0 \quad \Leftrightarrow \quad\overbar{\partial}\alpha'-\frac{1}{2}\,[\alpha',\alpha'] = 0\,.
\end{equation}
The $\mathrm{SU}(2)$-equivariance and normalization condition for $\mathcal{A}$ are the same as the $\mathrm{SU}(2)$-invariance and no vertical invariant subspace conditions imposed on the almost complex structure $J'$ in~\cite{Erdmenger:2025lvv} to obtain the converse of \eqref{eq:flatness_implies_KS}.

\addcontentsline{toc}{section}{References}
\bibliography{bib.bib}
\bibliographystyle{JHEP}

\end{document}